\documentclass[aps,prd,floats,nofootinbib,preprintnumbers,superscriptaddress]{revtex4}
\usepackage{epsfig}
\usepackage{amsmath}
\usepackage{amssymb}
\usepackage{amsthm}
\usepackage{longtable}

\begin{document}

\preprint{NUHEP-TH/07-10}

\title{A Survey of Lepton Number Violation Via Effective Operators}

\author{Andr\'e de Gouv\^ea}
\affiliation{Northwestern University, Department of Physics \&
Astronomy, 2145 Sheridan Road, Evanston, IL~60208, USA}

\author{James Jenkins}
\affiliation{Northwestern University, Department of Physics \&
Astronomy, 2145 Sheridan Road, Evanston, IL~60208, USA}

\begin{abstract}
We survey 129 lepton number
violating effective operators, consistent with the minimal Standard Model gauge group and
particle content, of mass dimension up to and including
eleven. Upon requiring that each one radiatively generates the
observed neutrino masses, we extract an associated
characteristic cutoff energy scale which we use to calculate other
observable manifestations of these operators for a number of current
and future experimental probes, concentrating on lepton number
violating phenomena.  These include searches for
neutrinoless double-beta decay and rare meson, lepton, and gauge boson
decays. We also consider
searches at hadron/lepton collider facilities in
anticipation of the LHC and the future ILC.
We find that some operators
are already disfavored by current data, while more are ripe to
be probed by next-generation experiments.
We also find that our current understanding of lepton mixing disfavors
a subset of higher dimensional operators.
While neutrinoless double-beta decay is the most promising signature of
lepton number violation for the majority of operators, a handful is  best probed by
other means.  We argue that a combination of constraints from various
independent experimental
sources will help to pinpoint the ``correct'' model of neutrino
mass, or at least aid in narrowing down the set of possibilities.
\end{abstract}

\maketitle

\setcounter{equation}{0} \setcounter{footnote}{0}
\section{Introduction}
\label{sec:intro}

The discovery of neutrino masses via their flavor oscillations over
long baselines constitutes the first solid evidence of physics beyond the
standard model (SM) of particle physics \cite{NeutrinoReview}. While
this is an important first step toward a deeper understanding of
nature, it poses many more questions than it answers.
A number of theoretically well-motivated models have
been proposed and explored to address the origin of the neutrino mass but, strictly
speaking, these represent only a handful out of an infinite set of
possibilities.  The question of how well future experiments can
probe and distinguish different scenarios arises naturally and is quite
relevant given the current state of high energy physics.  The coming
years promise detailed explorations of the terascale with the Large
Hadron Collider (LHC) and the more distant International Linear
Collider (ILC) or variants thereof.  Expectations are that combined information from
these two facilities, coupled with high precision, low energy
results and cosmological observations will shed light on some of the
current mysteries of physics, including that of the neutrino mass.

Here, we concentrate on the possibility that the neutrino masses are generated
at some high energy scale $\Lambda$ where $U(1)_{B-L}$, the only non-anomalous
global symmetry of the standard model, is broken. Such a scenario is
 well-motivated by the observed properties of the light neutrinos including tiny
 masses, large mixings and the fact that neutrinos are the only electrically
 neutral fundamental fermions. More specifically, once $U(1)_{B-L}$ is broken, neutrinos
 are not protected from getting non-zero Majorana masses after electroweak symmetry breaking.
 On the other hand, since the renormalizable minimal standard model\footnote{Throughout, we will
 assume that the weak scale degrees of freedoms are the known standard model fields, plus a minimal Higgs sector. Hence, we assume that there are no gauge singlet ``right-handed neutrino'' fermions or higher $SU(2)_L$ Higgs boson representations, such as Higgs boson triplets.} preserves $U(1)_{B-L}$, $B-L$ breaking
 effects will only manifest themselves at low energies through higher dimensional operators. This
 being the case, one generically expects neutrino masses to be suppressed with respect to charged
 fermion masses by $(v/\Lambda)^n$, $n\ge 1$, where $v$ is the Higgs boson vacuum expectation value.

By further assuming that all new degrees of freedom are much heavier than the weak scale, we are guaranteed
that, regardless of the details of the new physics sector, all phenomena below the weak scale are described
by irrelevant, higher dimensional operators. In this spirit, the observable consequences of all high energy models
that lead to small Majorana neutrino masses can be catalogued by understanding the consequences of
irrelevant operators that break $B-L$ by two units. With this in mind, we will survey all such
non-renormalizable effective operators for phenomenological
signatures at future and current experiments.  We restrict ourselves
to operators that will lead to lepton number violation (LNV),
as these will be directly connected to the existence of small Majorana neutrino masses.
This means that we do not consider operators that conserve $L$ but violate $B$, and
hence also $B-L$, by two units (such operators lead to,
for example, neutron--antineutron oscillations), nor do we include operators that respect $B-L$. Most
of the time, the latter will not mediate any observable consequences for large enough $\Lambda$, except
for operators of dimension-six and above that can mediate proton decay.

To begin, we
systematically name and classify all relevant LNV operators.
Fortunately, this has already been done\footnote{The authors of
\cite{Operators} discuss all possible effective operators of
dimensions up to and including eleven, but only explicitly list
those deemed unique in the sense that they cannot be written as the
product of any previous operator with a Standard Model interaction.
We append their list and naming scheme to include these into our
analysis.} in \cite{Operators} up to and including operators of mass
dimension eleven.\footnote{We will argue later that irrelevant operators with mass dimension thirteen and higher, if related to
neutrino masses, will require new physics below the electroweak scale so that we would have already
observed new physics if neutrino masses were generated in this way. Furthermore, from a model building perspective, it is difficult
to develop models that predominantly yield effective operators of
very high mass dimension. The probability that such scenarios are both
theoretically well-motivated and evade all observations appears to be slim.}
For each operator we then
calculate/estimate the analytic form of the radiatively generated
neutrino mass matrix.  Upon setting this expression equal to the
experimentally measured neutrino masses, we extract the
energy scale $\Lambda$ associated to the new LNV physics. Armed with these
scales, we proceed to calculate each operator's
phenomenological signatures at a variety of experimental settings.
Additionally, having explicitly calculated the operator-induced
neutrino mass \emph{matrices}, we may also verify, under some generic assumptions,
whether one can account for the observable lepton mixing pattern.  After such a general survey, one is
adequately equipped to take a step back and select
phenomenologically/theoretically interesting operators for further
detailed study by ``expanding'' effective vertices to reveal
particular ultraviolet completions.  In this way, one can use the results presented here
as a means of systematically generating renormalizable
models with well-defined experimental predictions.

This paper is organized as follows.  Sec.~\ref{sec:Scale} is devoted to an introduction to the effective
operators and methods.  In Sec.~\ref{subsec:NuMass}, we
derive and comment on the scales $\Lambda$ of new physics that are
used throughout the remainder of the text.  In Sec.
\ref{sec:constraints}, we survey various experimental probes of LNV
for each operator, and address if and when our analysis breaks down
due to added model structure or additional assumptions.
Specifically, we study both current constraints and future prospects
for neutrinoless double-beta decay experiments in Sec.~\ref{subsec:bb0nu}, followed,  in
Sec.~\ref{subsec:OtherProcesses},  by a similar analysis of other rare
decay modes, including those of various mesons and W/Z gauge bosons.
In Sec.~\ref{subsec:Collider}, we present collider
signatures of LNV as they apply to future linear collider facilities
running in the $e^-e^-$ collision mode, and describe extensions of our
analysis to include associated $\gamma\gamma$ collisions. We also comment on searches for LNV
in future hadron machines.  Sec.~\ref{sec:Oscillations} describes
current constraints from neutrino oscillation
phenomenology due to the general structure of the derived neutrino
mass matrices.
In Sec.~\ref{sec:interesting}, we highlight a number of ``interesting'' operators,
defined by low cutoff scales and prominence of
experimental signatures, which are still allowed by current constraints on LNV.
We undertake a slightly more detailed
discussion of their characteristics and signatures and present some sample ultraviolet
completions.
We conclude in Sec.~\ref{sec:conclusion} with
a summary of our assumptions and results, augmented by commentary on
future prospects for LNV searches. Our results are tabulated by operator name
in Table~\ref{tab:AllOps} for easy reference.

We hope that this  analysis will prove useful to various audiences on a number of
distinct levels. In the most superficial sense, the casual reader
should note the general features of LNV as well as the diversity of
model variations.  Such information is best expressed in terms of
the operator distribution histograms scattered throughout the text.
These are color-coded by operator dimension or cutoff scale,
and typically contain additional information, including current
experimental prospects.
On the more technical
side, those interested in specific neutrino mass generating models  will find detailed, operator specific,
information that may be utilized as crude model predictions.
Additionally, as already alluded to, one may even ``hand-pick''
operators for model development based on specific phenomenological
criteria. Finally, we urge experimentalists to search for new physics
in all accessible channels. It
is our ultimate goal to provide motivation for experimental
considerations of non-standard LNV effects, beyond neutrinoless
double-beta decay.

\setcounter{footnote}{0} \setcounter{equation}{0}
\section{The lepton number violating scale}
\label{sec:Scale}

Here we analyze $SU(3)_c \times SU(2)_L \times U(1)_Y$ invariant
$\Delta L = 2$ non-renormalizable effective operators of mass
dimension up to and including eleven.  They are composed of only the
SM field content as all other, presumably heavy, degrees of freedom
are integrated out.
As already emphasized, we do not allow for the
existence of SM singlet states (right-handed neutrinos) or any other
``enablers'' of renormalizable neutrino masses, such as Higgs $SU(2)_L$ triplet
states. We therefore
assume that all lepton number violation originates from new ultraviolet
physics and that neutrino masses are generated at
some order in perturbation theory.

A $d$-dimensional operator ${\cal O}^d$ is suppressed by $d-4$ powers of a mass
scale $\Lambda$ that characterizes the new physics, in addition to a dimensionless
coupling constant $\lambda$:
\begin{equation}
{\cal L}\in \sum_i\frac{\lambda_i{\cal O}_i^d}{\Lambda^{d-4}},
\end{equation}
where we sum over all possible flavor combinations that make up the same ``operator-type,''
as defined below.
For each operator, $\Lambda/\lambda$ is approximately the maximum energy scale below which the
new perturbative ultraviolet physics is guaranteed to reside, and $\Lambda$ is used as a
hard momentum cutoff in the effective field theory. Among all $d$-dimensional operators, we
define $\Lambda$ so that the largest dimensionless coupling $\lambda$ is equal to unity. Unless
otherwise noted, we will assume that all other $\lambda$ are of order one.

In the first two columns of Table \ref{tab:AllOps}, we exhaustively
enumerate all possible lepton number violating operators of mass
dimension less than or equal to eleven.
All together, this amounts to 129 different types of
operators, most of which, 101 to be exact, are of dimension eleven and
consist of six fermion and two Higgs fields.  Remaining are 21, 6
and 1 operator of dimension nine, seven, and five, respectively.  The
dimension-nine operators can be of two different kinds,
as defined by their respective field content.  They either contain four
fermion and three Higgs fields or simply six fermion fields with no
 Higgs field content.  For consistency, we use the notation of
reference \cite{Operators}, where such a listing was first
introduced. Our operator naming scheme is also derived
from the same list, which we trivially extend to include $21$
elements only mentioned in that analysis. These are the dimension-nine
 and dimension-eleven LNV operators that can be constructed from the ``product''
of the previously listed dimension-five and dimension-seven operators with the SM
Yukawa interactions. We individually identify those
operators with the same field content but different $SU(2)_L$ gauge
structure with an additional roman character subscript added onto
the original designation from \cite{Operators}.  This is done in order to render
our discussion of the various operators clearer, since specific gauge structures can
play an important role in the derived energy scale and predictions of
a given operator. Note that we neglect effective operators that
contain SM gauge fields, since, as argued in \cite{Operators},
these are not typically generated by renormalizable models of new physics.

Our notation is as follows. \begin{equation}  L =
\left(
                                                            \begin{array}{c}
                                                              \nu_L \\
                                                              e_L \\
                                                            \end{array}
                                                          \right)~~~
                                                         {\rm and}~~~
                                                         Q=\left(
                                                            \begin{array}{c}
                                                              u_L \\
                                                              d_L \\
                                                            \end{array}
                                                          \right)
                                                          \end{equation}
are the left-handed lepton and quark $SU(2)_L$ doublets,
respectively.
$e^c$,
$u^c$ and $d^c$ are the charge-conjugate of the $SU(2)_L$ singlet right-handed
charged lepton and quark fermion operators, respectivelty.
Conjugate fields are denoted with the
usual ``bar'' notation ($\bar{L}$, $\bar{Q}$,
$\bar{e}^c$).
For simplicity, we are omitting flavor indices, but it is understood that
each matter fermion field comes in three flavors.
All matter fields defined above are to be understood as
flavor eigenstates: all SM gauge interactions, including those of the $W$-boson,
are diagonal. Without loss of generality, we will also define the $L$ and $e^c$ fields so that the
charged-lepton Yukawa interactions are flavor-diagonal.

We take the $SU(2)_L$
doublet Higgs scalar to be
\begin{equation}H = \left(
                                                            \begin{array}{c}
                                                              H^+ \\
                                                              H^0 \\
                                                            \end{array}
                                                          \right),
                                                          \end{equation}
and assume that, after electroweak symmetry breaking, its neutral
component acquires a vacuum expectation value (vev) of
magnitude $v \approx 0.174$~TeV,\footnote{Our numerical value
for $v$ is distinct from many treatments of the SM where $v$ is
taken to be $0.246~\rm{TeV}$.  These are equivalent up to a factor
of $\sqrt{2}$ and are both valid provided a consistent treatment of
the interaction Lagrangian.} thus spontaneously breaking the
electroweak gauge symmetry $SU(2)_L \times U(1)_Y \rightarrow
U(1)_{\rm em}$. In Table~\ref{tab:AllOps}, the components of the
$SU(2)_L$ doublets are explicitly listed and labeled with $i,j,k,\ldots=1,2$. In order to form gauge
singlets, operators are contracted either by the
antisymmetric tensor $\epsilon_{ij}$, defined such that
$\epsilon_{12} = 1$, or by trivial contractions with a conjugate
doublet field.  Different gauge contractions are partially responsible for the wide variety of operator
structures encountered in this study.

In order to avoid unnecessarily messy expressions, several features are missing from the operators as
listed in Table \ref{tab:AllOps}.  To begin, $SU(3)_c$ color indices
are suppressed in these expressions.
 Color contractions are only implied here because $SU(3)_c$ is an unbroken
symmetry of the SM and hence there is no sense in distinguishing the
various quark field components.  We assume that the parent
ultraviolet completion to each operator treats the color gauge
symmetry properly by introducing appropriately chosen heavy colored
particles to render the theory gauge invariant.  Slightly more
serious is the omission of flavor indices to label the fermion
generations.  For most of this analysis, we assume that all new
physics effects are generation universal and thus, flavor
independent.  This is not guaranteed to be the case, as is painfully obvious
within the SM.  One will also note that, depending on the $SU(2)$ structure of the effective operator, different flavor-dependent coefficients will be strictly related. For example, including flavor dependent couplings $\lambda^1_{\alpha\beta}$, ${\cal O}_1$ should read
$\lambda^1_{\alpha\beta}L_{\alpha}^iL_{\beta}^jH^kH^l\epsilon_{ik}\epsilon_{jl}$, where $\lambda^1_{\alpha\beta}=\lambda^1_{\beta\alpha}$ (symmetric) for all $\alpha,\beta=e,\mu,\tau$. On the other hand, ${\cal O}_{3a}$ should read (for fixed $Q$ and $d^c$ flavors)   $\lambda^{3_a}_{\alpha\beta}L_{\alpha}^iL_{\beta}^jQ^kd^cH^l\epsilon_{ij}\epsilon_{kl}$, where $\lambda^{3_a}_{\alpha\beta}=-\lambda^{3_a}_{\beta\alpha}$ (antisymmetric) for all $\alpha,\beta=e,\mu,\tau$.
Large differences among the various flavor
structures of each operator may very well exist. Flavor is an important
facet of LNV phenomenology, and is addressed where relevant within the text.

The final
feature missing from our notation is explicit Lorentz structure.
Each operator must, of course, form a Lorentz scalar, but there are
numerous field configurations that can bring this about. The Higgs
field is a scalar, and as such, transforms trivially under the
Lorentz group and is thus of no relevance to this discussion. The
fermions, however, transform non-trivially and their contractions
must be accounted for in each operator.  Simple
combinatorics dictate that there are at most $45$ such
possibilities for the six-fermion operators that comprise the bulk
of our sample, 3 in the four-fermion case and only $1$ for the
lone dimension five operator. Additionally, each contraction can be
made in a variety of ways, corresponding to the bilinear Dirac
operators $\mathbf{1}$, $\gamma^\mu$ and $\sigma^{\mu\nu} =
\frac{i}{2}[\gamma^\mu,\gamma^\nu]$ of the scalar, vector and
tensor types, respectively.  Since we are dealing with chiral
fields, the addition of the $\gamma_5$ matrix to form the
pseudoscalar and axial-vector bilinears is redundant. While this
helps reduce the number of possibilities, the task of
listing, categorizing, and analyzing all possible Lorentz structures
for each operator is still quite overwhelming and is not undertaken
in this general survey. Fortunately, different Lorentz structures for the
same operator-type lead to the same predictions up to order one effects.
This is  especially true for the
``interesting'' operators characterized by TeV $\Lambda$ scales.
We shall quantify this statement and mention specific structures
when relevant.  That being said, the Lorentz
structure of an effective operator can suggest a lot of information
about its parent renormalizable model. For example, it can suggest
the spin of the heavy intermediate states and the forms of various vertices.

Armed with these operators, we can calculate the amplitude of any
$\Delta L=2$ LNV process.
It is important to emphasize that when addressing the
phenomenological consequences of any particular operator ${\cal O}$, we
assume that it characterizes the dominant tree-level effect of the new heavy physics,
and that all other effects --
also characterized by other LNV effective operators of lower mass dimension --
occur at higher orders in perturbation theory.
Our approach is purely diagrammatic,
in that we begin with an operator-defined vertex and then proceed to
close  loops and add SM interactions as needed to yield the
correct external state particles.  In this sense, special care must
be taken to respect the chiral structure as defined by each
operator.  In order to reach the intended external states, to couple
to particular gauge bosons, or to close fermion loops, one must
often induce a helicity flip with a SM mass insertion.  We express
these inserted fermion masses in terms of the respective Yukawa
couplings, $y_f$ ($f=\ell,u,d$) and the Higgs vev, $v$. The Higgs
field can be incorporated into this procedure in a number of ways.
We treat the two charged and single neutral Nambu-Goldstone Higgs
bosons, $H^\pm$ and $H^0$, respectively, within the Feynman-'t Hooft
gauge as propagating degrees of freedom with electroweak scale
masses. The physical neutral Higgs, $h_0$, can be either chosen to
propagate as a virtual intermediate state, or couple to the vacuum
with amplitude $v$.

In order to avoid the task of explicitly evaluating a huge
number of multiloop Feynnman diagrams, we succumb to approximate LNV
amplitudes based on reasonable assumptions and well-motivated rules.
Our methodology is motivated by exact computations with one-loop,
dimension 7 operators where the work is analytically tractable, as
well as on general theoretical grounds.  For select operators, we
have also checked our assumptions against predictions from
ultraviolet complete models with success.  In order to
perform a particular calculation, we draw the appropriate
diagram(s), taking care that no momentum loop integral vanishes by
symmetry reasons. This step is potentially quite
involved, as multiple diagrams can give sizable amplitude
contributions depending on the characteristic energy transfer in the
system, not to mention the cumbersome Dirac algebra within the
respective loops.  Given the high, often super-TeV, mass scale
associated with our calculations, it is often convenient to work in
the gauge field basis where each boson state is associated with a
single SM group generator, as is natural before electroweak symmetry
breaking.  In a similar sense, all fermions, including those of
the third generation, are taken to be massless to zeroth order. All masses are included perturbatively where needed
via mass insertions. At first guess, it would
seem that our results are only valid in the rather
subjective limit $\Lambda \gg v$. By
direct comparison with other more complete approximations, however, we find that our
predictions are very reasonable at all scales above $0.5~\rm{TeV}$.
Keeping all of this in mind, we apply the following ``rules'' to obtain
 approximate amplitude expressions.

\begin{enumerate}
\item
\textit{Trivial numerical factors}:  A number of numerical factors
can be read off trivially from the Feynman diagrams.  Specifically,
one can extract the presence of the suppression scale
$\Lambda^{-(d-4)}$ directly from the dimension $d$ operator, as well
as the dimensionless coupling constants $\lambda$. Generally,
$\lambda$ is a generation dependent quantity, but for lack of any
experimental evidence to the contrary, we take $\lambda = 1$
universally unless stated otherwise.
In  the case of scenarios already constrained by current data, we will relax
this assumption to ``save'' the operator and comment on the
phenomenological consequences of the change.

Furthermore, various
factors of the electroweak scale $v$ may be extracted from the
operator's Higgs field content, in addition to
fermion/gauge boson mass terms.  In this way, we may also
include the various Yukawa and gauge coupling factors $y_f$ ($f=\ell,u,d$) and
$g_i$, respectively, where $i$ runs over the three SM gauge groups.
For simplicity, we neglect the gauge subscript $i$ in further analytic
expressions. Finally, a color factor of $3$ associated with each
quark loop should also be included in our computations, but can (and will) be
neglected for simplicity from algebraic expressions where order one
factors are irrelevant and only serve to render expressions more cumbersome.
We note that all coupling constants are subject to renormalization
group running.  In particular, those occurring within a loop should
be evaluated at the scale $\Lambda$.  We neglect this order one
effect since it is most important at large $\Lambda$ scales where
operators tend to have less of a phenomenological impact due to the $(1/\Lambda)^n$ suppression.
\item
\textit{Loop factors}:  In all of our calculations, we assume
that each operator defines an effective field theory, characterized
by the scale $\Lambda$. This implies that all momentum integrals are
effectively cut off at $\Lambda$, above which new states will emerge
to regularize the theory. Divergences in such loops tend to cancel
the large scale suppressions inherent to the bare operators, and
thus enhance predicted LNV rates.  Specific divergences can be
determined by simple power counting of momentum factors.  Of course,
multiple loop integrals are often convoluted to the point where
substantial simplification is needed to determine the dominant
divergent term.  Such a complication is in part due to the numerator
of the Dirac propagators, which include single momentum factors and
must therefore be present in pairs to contribute effectively to an ultraviolet divergence.  The process of adding loops to induce $\Lambda$
power law divergences should only be pursued to the point where the
suppression of the induced effective term is no less than
$\Lambda^{-1}$.  Any further divergent contribution must be treated
as a renormalization to lower order terms, and hence, can only add
small finite corrections to the total amplitude.  In any case, those
diagrams with the smallest scale suppressions are not always the
most dominant, as will become clear later when we discuss specific results.

 In addition to power-law divergences, each
loop is also associated with a numerical suppression factor.  This
arises from the proper normalization of the loop four-momentum
integral as a factor of $(2\pi)^{-4}$, the characteristic phase
space ``volume'' of a quantum state.  It allows one to view the
integral as a coherent sum over all possible intermediate
configurations in a consistent way.  Partially evaluating these
integrals for a number of examples, one quickly finds that two
powers of $\pi$ cancel with the four dimensional Euclidian space
solid angle $\int d\Omega_4$.  We introduce a suppression
factor of $(16\pi^2)^{-1} \sim 0.0063$ for each diagram loop, which
tends to offset enhancements from associated divergent
factors.
A quadratically divergent loop diagram is often proportional to the lowest order contribution
times $(1/16\pi^2)(\Lambda/v)^{2}$ to the power $n$ (number of loops in the diagram).
This contribution is larger than the leading order one if
 $\Lambda > 4\pi v \sim
2~\rm{TeV}$ for any number of loops.  The situation is
often more involved, as many loops turn out to be
logarithmically divergent or even convergent.  The important
conclusion is that adding loops is not an efficient way to enhance
LNV rates at the low scales accessible to future experiments.  This
fact is demonstrated by example in Sec.~\ref{sec:constraints}.

Finally, as already alluded to, many
diagram loops will exhibit logarithmic divergences, as is the
standard case in renormalizable theories involving fermion and
vector fields. This occurrence typically reflects the differences
between the two characteristic scales inherent to the system, namely
$\Lambda$ and $v$, and are of the general form
$\xi\log^n(\Lambda/v)$ for some power $n$.
$\xi$ is a small, loop suppressed, dimensionless coupling
coefficient.  Numerically, these logarithms are much softer than their
quadratically divergent counterparts seen elsewhere in the diagrams
and can safely be neglected.
\item
\textit{Intermediate states}:  We treat all virtual
intermediate states, outside of loops, as if they carry the
characteristic momentum of the interaction $Q$ and neglect Dirac
structure, unless stated otherwise.  In particular, goldstone
bosons are assigned the propagator $(Q^2 - M_g^2)^{-1}$ and fermions
are assigned $(Q - M_f)^{-1}$.  In the case of an intermediate
neutrino, this reduces to a simple factor of $Q^{-1}$ for all
realistic $Q$ values.  Hence, for very low energy processes ($Q\ll 100$~MeV), neutrino
exchange diagrams tend to dominate LNV rates.
\item
\textit{Lorentz structure}:  For the purposes of our analysis, we
assume that all Lorentz contractions between fermions are
scalar-like.  As previously mentioned, the absolute magnitude of
most LNV amplitudes is robust under this assumption up to order one
factors.  The only qualitative exception to this occurs in some
cases involving fermion bilinear terms with a tensor Lorentz structure ($\bar{\psi}^{\prime}\sigma_{\mu\nu}\psi$).
This factor, when coupled between two fermions contracted in a loop,
will yield a vanishing rate due to its antisymmetry inside of a
trace, since $Tr(\sigma_{\mu\nu})=0$.  This can be bypassed by
introducing a new momentum vector into the trace, implying the
addition of another loop.  In most cases, this  is most
efficiently accomplished with a new gauge boson line, which is accompanied by a
logarithmic divergence.  The combination of both
factors leads to a marginal amplitude suppression (with respect to the same operator where all
fermion bilinears are Lorentz scalars) for all energies of interest.
\end{enumerate}

With these approximations in hand, it is a simple matter to estimate
the amplitude associated to any given diagram.  Still, one must wonder about
the uncertainty induced onto the calculations by such varied
assumptions.  Can results obtained by such methods supply valid
physical predictions?  The answer, of course, depends on the question
that is being asked. Here, we will only be interested in estimating order
of magnitude effects, including what value of $\Lambda$ is required in order
to explain the observed neutrino masses and, once $\Lambda$ is so constrained,
what is the order of magnitude of other related observable effects.

One may wonder whether a more detailed estimate of the effects of
each individual operator would lead to more reliable results. The answer is negative.
It is easy to show that different renormalizable theories that lead to the same
effective operator at tree-level will mediate different processes at the loop-level with order one
different relative strengths. Furthermore,
the derived cutoff scales inherit the
uncertainty from the absolute value of the heaviest neutrino mass, which is only
loosely bounded between $0.05~\rm{eV}$ and $1~\rm{eV}$ by the
extracted atmospheric mass squared difference \cite{OscBestFit} and
tritium beta decay kinematic measurements \cite{Mainz,Troitsk}.
This is an order of magnitude uncertainty that cannot be avoided even
if one were to perform a detailed computation within a well-defined ultraviolet complete
theory.

In summary, given all approximations and uncertainties,
our results are only valid up to $\pm$ an order of magnitude.
In this spirit, one need not explicitly consider order one factors that will
necessarily yield negligible corrections by these standards. Such a
large error tolerance supplies the need for care when interpreting
results. In particular, one should not place too much emphasis on
any one bound or prediction, unless it is very robust, {\it i.e.}, able to
withstand variations of at least a factor of ten.  Of course, for
those operators constrained by several different independent
sources one can, and should, take more marginal results seriously.

\subsection{Neutrino Masses and the Scale of New Physics}
\label{subsec:NuMass}

Having defined the  set of LNV operators, we now extract
the scale of new physics from the direct comparison of radiatively
generated neutrino mass expressions to their observed values.
Since there are three light neutrino masses, we will use the heaviest of these to
set the overall mass scale.
Neutrino oscillation data, currently providing
the only evidence for neutrino masses, constrain the relative
magnitudes of the mass eigenstates but not the overall scale
\cite{NeutrinoReview}. Such data only supply a lower
bound on the heaviest neutrino mass, derived from the largest observed mass squared difference
$\Delta m^2_{13} = |m_3^2-m_1^2| \approx 0.0025~\rm{eV}^2$, the atmospheric
mass squared difference \cite{OscBestFit}.  At least one
neutrino mass must be greater than
$\sqrt{\Delta m^2_{13}}\approx 0.05$~eV. Neutrino oscillations also teach us that the next-to-heaviest
neutrino weighs at least $\sqrt{\Delta m^2_{12}}\approx 0.009$~eV (the solar mass-squared difference), in such a
way that the ratio of the heaviest to the next-to-heaviest neutrino masses is guaranteed to be larger than, approximately,
0.2. No lower bounds can be placed on the lightest neutrino mass.
An upper bound on the heaviest neutrino mass is provided by
several non-oscillation neutrino probes. Cosmology provides interesting constraints
on the sum of light neutrino masses, but these are quite dependent
on unconfirmed details of the thermal history of the universe and its composition \cite{CosmoSum}.
Most direct are kinematic measurements
of the tritium beta decay electron endpoint spectrum
\cite{TritNuEffMass}.  Both types of probes provide upper bounds near
$1~\rm{eV}$, likely to improve in coming years.
We choose to perform our calculations assuming the mass scale
$m_\nu \approx 0.05$~eV, corresponding to the experimental lower
bound.  In this way, each extracted operator scale $\Lambda$,
inversely related to the neutrino mass, represents a loose upper bound.
Since most rates for LNV observables are proportional to some
inverse power of $\Lambda$, this choice implies the added
interpretation that, all else remaining equal, our results for such rates should
conservatively reflect lower limit predictions.

\begin{figure}
\begin{center}
\includegraphics[scale=1]{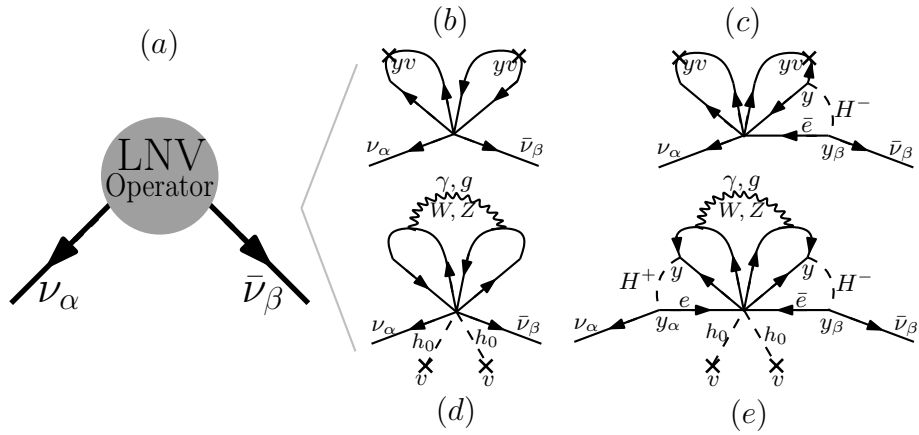}
\caption{Sample diagrams that
radiatively generate Majorana neutrino masses.  Diagram $(a)$ is
representative of all operators that can generate the needed
external state neutrinos.  This usually proceeds via loop
contractions and other couplings, hidden within the light
gray region. Diagrams $(b)-(e)$  help illustrate the
methodology of this analysis. Despite obvious differences, all of
these generate effective dimension-five interactions of
calculable strength. See the text for more details.}
\label{fig:MassDiagrams}
\end{center}
\end{figure}

LNV  neutrino masses are nothing more than
self-energy diagrams evaluated at vanishing momentum transfer.  These must
couple together the left-handed neutrino state $\nu_\alpha$ with the
right-handed anti-neutrino state $\nu_\beta$, as shown schematically in
diagram $(a)$ of Fig.~\ref{fig:MassDiagrams}.  Here the flavor
indices $\alpha$ and $\beta$ can accommodate any of the three lepton flavors ($\alpha,\beta=e,\mu,\tau$).  The
derived Majorana masses $m_{\alpha\beta}=m_{\beta\alpha}$ are generally complex.
 The large grey circle in this diagram represents all
 possible contributions to the neutrino mass.  Specifically, it
 contains the underlying $\Delta L=2$ operator along with all
 modifications needed to yield the correct external state structure.
  This includes such objects as loops, additional gauge boson propagators and SM coupling
  constants.  Generally, several diagrams can contribute to this
  mass generation, but special care must be taken that these are not
  proportional to any positive power of $Q$, the momentum carried by the
  neutrino legs, as this would not lead to a nonzero rest mass
  correction.

Diagrams $(b)-(e)$ of Fig.~\ref{fig:MassDiagrams} are examples
that serve to illustrate some typical features encountered in our
effective operator induced self-energy calculations.  The underlying
LNV operators shown in each diagram contain six fermion fields and
are therefore of dimension nine or eleven, as are the majority of
the analyzed operators.
Each of these diagrams generates an
effective dimension-five interaction
\begin{equation}
{\cal L}_5= \xi_{\alpha\beta}^{(5)} \frac{(L^{\alpha}H)(L^{\beta}H)}{\Lambda},
\label{eq:Gen5}
\end{equation}
where $\xi_{\alpha\beta}^{(5)}=\xi_{\beta\alpha}^{(5)}$ is a generation dependent coupling
constant that is calculable, given the structure of the original
operator.  This is  easily verifiable via direct power
counting, despite differences in dimension, loop number, field
content, and helicity structure.  It turns out that most operators,
especially those characterized by super-TeV scales, possess this
property.

We describe each sample diagram in turn to point out important features. A subset
of the subtleties described below is encountered when estimating the neutrino masses $m_{\alpha\beta}$ for the entire effective operator set.
 Diagram $(b)$ is a simple two-loop
radiatively generated mass term proceeding from dimension-nine
operators, such as
$\mathcal{O}_{11_b}=L^iL^jQ^kd^cQ^ld^c\epsilon_{ik}\epsilon_{jl}$,
containing the fermion structure $f_L f_R^c f_L^\prime f_R^{\prime
c}$, where the fields $f$ and $f^\prime$ are contracted into loops
with mass insertions that supply the needed field chirality flip.
Masses arising from such operators are proportional to two powers of
fermion Yukawa couplings.  Strictly speaking, allowed fermions from
all three generations traverse the closed loops and contribute to
the mass. However, assuming universal new physics coupling
constants, third generation fermions will strongly dominate the
induced neutrino mass.  In cases such as these, one can freely
suppress couplings to the lighter two generations without modifying
the expected value of $\Lambda$. Since diagrams arising from the
majority of our operator set contain at least one loop of this kind,
this property proves quite useful when attempting to avoid low
energy nuclear physics constraints, as will be discussed in more
detail later.

Diagram $(d)$ involves an operator of dimension eleven, such as
$\mathcal{O}_{22}=L^iL^jL^ke^c \bar{L}_k
\bar{e^c}H^lH^m\epsilon_{il}\epsilon_{jm}$, but has a similar
structure to Diagram $(b)$ since both neutral Higgs fields $h_0$
couple to the vacuum, yielding a $v^2$ factor. In this case, the
parent underlying operators contain the fermion structure $f_{L(R)}
f_{L(R)}^c f_{L(R)}^\prime f_{L(R)}^{\prime c}$, or simple variants
thereof. From this, it is clear that such operators will create and
annihilate the \emph{same} field and one can close the fermion loops
without mass insertions. A little thought reveals that such loops,
if left on their own, will vanish by symmetry, since $\int d^4k
[k^\mu/g(k^2)] = 0$ for all functions $g(k^2)$. Hence, non-zero
neutrino masses only appear at a higher order in perturbation theory
({\it i.e.}, we need to add another loop).  To maintain the chiral
structure of the diagram a gauge boson line insertion is always the
most effective. The specific gauge field required in this step
depends critically on the quantum numbers of the fermions $f$ and
$f^\prime$ contained in the operator itself and must be determined
on a case-by-case basis. The absence of Yukawa dependence renders
the estimated value of the cutoff scale $\Lambda$ insensitive to the
values of the dimensionless operator couplings ($\lambda$), given
the way $\Lambda$ is defined.
Notice that this three-loop diagram, like Diagram $(b)$,
predicts an anarchic neutrino Majorana mass matrix, currently allowed by the
neutrino oscillation data \cite{anarchy}. That is, up to
order one corrections, all entries are of the same magnitude, $m_\nu
\approx 0.05$~eV.  This is in contrast to the remaining
sample diagrams ($(c)$ and $(e)$), which both suggest flavor-structured mass matrices.

Dimension-nine operators, such as
$\mathcal{O}_{19}=L^iQ^jd^cd^c\bar{e^c}\bar{u^c}\epsilon_{ij}$,
yielding diagram $(c)$ have the peculiar property that, upon
expanding out the various $SU(2)_L$ contractions in terms of
component fields, no $\nu_\alpha\nu_\beta$ content is present to
form the external legs of a mass diagram. Here the LNV is introduced
via the fermion structure $\nu_\alpha e_{R\beta}$, which annihilates
a left-handed neutrino and creates a left-handed positron.  Hence,
to tie in the needed antineutrino line, one must both flip the
charge lepton helicity and carry away the excess charge with some
bosonic state. Of course, such a charged boson is guaranteed by
charge conservation to be needed elsewhere in the system to close
some $f/f^\prime$ loop.  In this particular example, the process is
illustrated by the exchange of a charged Higgs goldstone boson
$H^-$.
The crucial point is that this mass is necessarily
proportional to a charged lepton Yukawa coupling $y_{\ell_\beta}$, of
the \emph{same} flavor as the external neutrino, since we are working
in the weak eigenbasis where the gauge couplings and the charged-lepton Yukawa couplings are flavor
diagonal.  By symmetry,
the contribution to the $m_{\alpha\beta}$ entry of the neutrino mass matrix
is
proportional to $y_{\ell_\alpha} + y_{\ell_\beta}$, which reduces to
the largest coupling $y_{\ell_\beta}$ in the realistic case of
hierarchial charged lepton Yukawa couplings.

Finally, Diagram $(e)$ yields a five-loop suppressed neutrino
self-energy originating from a dimension-eleven LNV operator such as
$\mathcal{O}_{36} =
\bar{e^c}\bar{e^c}Q^id^cQ^jd^cH^kH^l\epsilon_{ik}\epsilon_{jl}$.
This represents the most complicated structure considered in this
analysis.  As in Diagram $(c)$, no explicit $\nu_\alpha\nu_\beta$
structure is available in the underlying operator, but in this case
all of the LNV arises from $e_{R\alpha}^c e_{R\beta}^c$-type
interactions. Curiously, this interaction already flips helicity as
it annihilates a left-handed positron and creates a right-handed
electron. Unfortunately, being an $SU(2)_L$ singlet, $e_R$ only
couples to the neutrino via a charged Higgs induced Yukawa
interaction; therefore, this amplitude must be proportional to the
product $y_{\ell_\alpha}y_{\ell_\beta}$ to yield a legitimate neutrino
mass contribution.  One might imagine that the Higgs fields
contained in the LNV operator could be used to produce the needed
neutrino legs, but this is not possible since the resulting loop
would have a structure of the form $\int d^4k \frac{k^\mu +
Q^\mu}{g(k^2)} \propto Q^\mu$ which vanishes in the ``rest mass''
limit.  It is clear that both Higgs fields must again couple to the
vacuum and the needed flip must come from the other fermion loops.
The resulting loop integrals can be separated into two convoluted
pieces corresponding to both loop/leg pairs.  A little thought
reveals that each loop set contains three fermion lines whose
associated integral is again proportional to the momentum of the
external neutrino, and thus is not a valid mass correction.  To fix
this last problem without further complicating the chiral
structure, one can add a gauge boson exchange between the fermion
loops, as was also done in Diagram $(d)$.

Despite the dominance of the generated dimension-five interactions
described by Eq.~(\ref{eq:Gen5}) for the majority of the studied LNV
operators, we find that this need not be the case for all of them.
For some operators, the dimension-five neutrino-mass effective
operator Eq.~(\ref{eq:Gen5}) occurs at higher order in perturbation theory than the dimension-seven
neutrino-mass effective operator (schematically, $(LH)^2 H^2$).  For these, the neutrino mass matrix
is generated after electroweak symmetry breaking from
\begin{equation}
{\cal L}_{57} =
\frac{\xi_{\alpha\beta}^{(5)}}{16\pi^2}\frac{(LH)(LH)}{\Lambda} +
\xi_{\alpha\beta}^{(7)}\frac{(LH)(LH)H\bar{H}}{\Lambda^3},
\label{eq:Gen57}
\end{equation}
where $\xi_{\alpha\beta}^{(7)}$ are new  calculable coefficients.  This type of structure is present in
the following operators
\begin{equation}
\mathcal{O}_7,\mathcal{O}_{21_{a,b}},\mathcal{O}_{22},\mathcal{O}_{23},\mathcal{O}_{25},\mathcal{O}_{26_b},\mathcal{O}_{27_b},\mathcal{O}_{29_a}\mathcal{O}_{30_b},\mathcal{O}_{31},\mathcal{O}_{44_c},\mathcal{O}_{57}.
\label{eq:ODim7Mass}
\end{equation}
In general, they are associated with dimension-eleven operators\footnote{This
also occurs with operator $\mathcal{O}_7$, which is of
dimension nine. This operator is the exception, in that it
explicitly contains three Higgs bosons which naturally aids in
building the needed $v^4$ factors in a way similar to that discussed
in the text.} whose mass diagrams are found trivially by connecting
the external fermion loops and coupling the neutral Higgs fields to
the vacuum.  This adds two factors of the electroweak scale to the
mass expressions. Dimensional analysis dictates that the fermion loops must conspire to
yield an addition factor of $v^2$, usually from mass insertions
utilized to flip helicities. For the dimension-seven operators in Eq.~(\ref{eq:Gen57}),
the resulting neutrino mass expression is proportional to
$v^4/\Lambda^3$.  If we assume, as is usually the case, that
most of the dimensionless factors of Eq.~(\ref{eq:Gen57}) are common to
both $\xi_{\alpha\beta}^{(5)}$ and $\xi_{\alpha\beta}^{(7)}$, we find $m_{\alpha\beta} \propto 1/16\pi^2
+ v^2/\Lambda^2$.  In such cases, the dimension-seven contribution is only relevant for operator cutoff
scales $\Lambda\lesssim 4\pi v \approx 2~\rm{TeV}$.  Such low scales are seldom
reached considering that these operators are efficient at mass
generation at low orders and consequently do not possess the
necessary suppression factors. Still, for completeness, we include
these terms when relevant.

\begin{center}
\begin{longtable}{|l|c||c|c|c|c|}
\caption{Dimension-five through dimension-eleven LNV operators
analyzed in this survey.  The first two columns display the operator
name and field structure, respectively. Column three presents the
induced neutrino mass expressions, followed by the
inferred scale of new physics, $\Lambda_{\nu}$.  Column five lists favorable modes of experimental exploration. Column six describes an operator's current status according to the key U (Unconstrained), C (Constrained) and D (Disfavored).  See text for details.} \label{tab:AllOps}   \\
 \hline \\[-2.3ex]
  \multicolumn{1}{c|}{$\mathcal{O}$} &
  \multicolumn{1}{c||}{Operator} &
  \multicolumn{1}{c|}{$m_{\alpha\beta}$} &
  \multicolumn{1}{c|}{$\Lambda_\nu~\rm{(TeV)}$} &
  \multicolumn{1}{c|}{Best Probed} &
  \multicolumn{1}{c|}{Disfavored}
  \\[.1ex] \hline
  \\[-2ex]
\endhead
$1$ & $L^i L^j H^k H^l \epsilon_{ik} \epsilon_{jl}$ & $\frac{v^2}{\Lambda}$ & $6\times 10^{11}$ & $\beta\beta0\nu$ & U \\
$2$ & $L^i L^j L^k e^c H^l \epsilon_{ij} \epsilon_{kl}$ & $\frac{y_\ell }{16\pi^2}\frac{v^2}{\Lambda}$ & $4\times 10^7$ & $\beta\beta0\nu$ & U \\
$3_a$ & $L^i L^j Q^k d^c H^l \epsilon_{ij} \epsilon_{kl}$ & $\frac{y_d g^2}{(16\pi^2)^2}\frac{v^2}{\Lambda}$ & $2\times 10^5$ & $\beta\beta0\nu$ & U \\
$3_b$ & $L^i L^j Q^k d^c H^l \epsilon_{ik} \epsilon_{jl}$ & $\frac{y_d}{16\pi^2}\frac{v^2}{\Lambda}$ & $1\times 10^8$ &$\beta\beta0\nu$& U \\
$4_a$ & $L^i L^j \overline{Q}_i \bar{u^c} H^k \epsilon_{jk}$ & $\frac{y_u}{16\pi^2}\frac{v^2}{\Lambda}$ & $4\times 10^9$ &$\beta\beta0\nu$& U \\
$4_b$ & $L^i L^j \overline{Q}_k\bar{u^c}H^k \epsilon_{ij}$ & $\frac{y_u g^2}{(16\pi^2)^2}\frac{v^2}{\Lambda}$ & $6\times 10^6$ &$\beta\beta0\nu$& U \\
$5$ & $L^i L^j Q^k d^c H^l H^m \overline{H}_i \epsilon_{jl}\epsilon_{km}$ & $\frac{y_d}{(16\pi^2)^2}\frac{v^2}{\Lambda}$ & $6\times 10^5$ &$\beta\beta0\nu$& U \\
$6$ & $L^i L^j \overline{Q}_k\bar{u^c}H^l H^k \overline{H}_i\epsilon_{jl}$ & $\frac{y_u}{(16\pi^2)^2}\frac{v^2}{\Lambda}$ & $2\times 10^7$ &$\beta\beta0\nu$& U \\
$7$ & $L^iQ^j \bar{e^c}\overline{Q}_kH^k H^l H^m\epsilon_{il} \epsilon_{jm}$ & $y_{\ell _\beta}\frac{g^2}{(16\pi^2)^2}\frac{v^2}{\Lambda}\left(\frac{1}{16\pi^2} + \frac{v^2}{\Lambda^2}\right)$ & $4\times 10^2$ &mix& C \\
$8$ & $L^i \bar{e^c} \bar{u^c} d^c H^j \epsilon_{ij}$ & $y_{\ell _\beta}\frac{ y_d y_u}{(16\pi^2)^2}\frac{v^2}{\Lambda}$ & $6\times 10^3$  &mix& C \\
$9$ & $L^i L^j L^k e^c L^l e^c \epsilon_{ij}\epsilon_{kl}$ & $\frac{y_\ell ^2}{(16\pi^2)^2}\frac{v^2}{\Lambda}$ & $3\times 10^3$ &$\beta\beta0\nu$& U \\
$10$  & $L^i L^j L^k e^c Q^l d^c \epsilon_{ij}\epsilon_{kl}$ & $\frac{y_\ell y_d}{(16\pi^2)^2}\frac{v^2}{\Lambda}$ & $6\times 10^3$  &$\beta\beta0\nu$& U \\
$11_a$& $L^i L^j Q^k d^c Q^l d^c \epsilon_{ij} \epsilon_{kl}$ & $\frac{y_d^2 g^2}{(16\pi^2)^3}\frac{v^2}{\Lambda}$ & $30$ &$\beta\beta0\nu$& U \\
$11_b$& $L^i L^j Q^k d^c Q^l d^c \epsilon_{ik}\epsilon_{jl}$ & $\frac{y_d^2}{(16\pi^2)^2}\frac{v^2}{\Lambda}$ & $2\times 10^4$ &$\beta\beta0\nu$& U \\
$12_a$& $L^iL^j\overline{Q}_i\bar{u^c}\overline{Q_j}\bar{u^c}$ & $\frac{y_u^2}{(16\pi^2)^2}\frac{v^2}{\Lambda}$ & $2\times 10^7$ &$\beta\beta0\nu$& U \\
$12_b$& $L^iL^j\overline{Q}_k\bar{u^c}\overline{Q}_l\bar{u^c}\epsilon_{ij}\epsilon^{kl}$ & $\frac{y_u^2 g^2}{(16\pi^2)^3}\frac{v^2}{\Lambda}$ & $4\times 10^4$ &$\beta\beta0\nu$& U \\
$13$ & $L^i L^j \overline{Q}_i \bar{u^c}L^l e^c\epsilon_{jl}$ & $\frac{y_\ell y_u}{(16\pi^2)^2}\frac{v^2}{\Lambda}$ & $2\times 10^5$ &$\beta\beta0\nu$& U \\
$14_a$& $L^iL^j\overline{Q}_k\bar{u^c}Q^kd^c\epsilon_{ij}$ & $\frac{y_dy_ug^2}{(16\pi^2)^3}\frac{v^2}{\Lambda}$ & $1\times 10^3$ &$\beta\beta0\nu$& U \\
$14_b$& $L^i L^j \overline{Q}_i \bar{u^c}Q^ld^c\epsilon_{jl}$ & $\frac{y_dy_u}{(16\pi^2)^2}\frac{v^2}{\Lambda}$ & $6\times 10^5$ &$\beta\beta0\nu$& U \\
$15$ & $L^i L^j L^k d^c \overline{L}_i \bar{u^c}\epsilon_{jk}$ & $\frac{y_dy_ug^2}{(16\pi^2)^3}\frac{v^2}{\Lambda}$ & $1\times 10^3$ &$\beta\beta0\nu$& U \\
$16$ & $L^i L^j e^c d^c \bar{e^c} \bar{u^c}\epsilon_{ij}$ & $\frac{y_dy_ug^4}{(16\pi^2)^4}\frac{v^2}{\Lambda}$ & $2$ &$\beta\beta0\nu$, LHC& U \\
$17$ & $L^i L^j d^c d^c \bar{d^c} \bar{u^c}\epsilon_{ij}$ & $\frac{y_dy_ug^4}{(16\pi^2)^4}\frac{v^2}{\Lambda}$ & $2$ &$\beta\beta0\nu$, LHC& U \\
$18$ & $L^i L^j d^c u^c \bar{u^c} \bar{u^c}\epsilon_{ij}$ & $\frac{y_dy_ug^4}{(16\pi^2)^4}\frac{v^2}{\Lambda}$ & $2$ &$\beta\beta0\nu$, LHC& U \\
$19$ & $L^i Q^j d^c d^c \bar{e^c} \bar{u^c}\epsilon_{ij}$ & $y_{\ell _\beta}\frac{y_d^2y_u}{(16\pi^2)^3}\frac{v^2}{\Lambda}$ & $ 1 $ &$\beta\beta0\nu$, HElnv, LHC, mix & C \\
$20$ & $L^i d^c \overline{Q}_i \bar{u^c} \bar{e^c} \bar{u^c}$ &$y_{\ell _\beta}\frac{y_dy_u^2}{(16\pi^2)^3}\frac{v^2}{\Lambda}$ & $40$ &$\beta\beta0\nu$, mix & C \\
$21_a$& $L^iL^jL^ke^cQ^lu^cH^mH^n\epsilon_{ij}\epsilon_{km}\epsilon_{ln}$ & $\frac{y_\ell y_u}{(16\pi^2)^2}\frac{v^2}{\Lambda}\left(\frac{1}{16\pi^2} + \frac{v^2}{\Lambda^2}\right)$ & $2\times 10^3$ &$\beta\beta0\nu$& U \\
$21_b$& $L^i L^j L^k e^c Q^l u^c H^m H^n \epsilon_{il} \epsilon_{jm}\epsilon_{kn}$ & $\frac{y_\ell y_u}{(16\pi^2)^2}\frac{v^2}{\Lambda}\left(\frac{1}{16\pi^2} + \frac{v^2}{\Lambda^2}\right)$ & $2\times 10^3$ &$\beta\beta0\nu$& U \\
$22$ & $L^i L^j L^k e^c \overline{L}_k \bar{e^c}H^lH^m\epsilon_{il}\epsilon_{jm}$ & $\frac{g^2}{(16\pi^2)^3}\frac{v^2}{\Lambda}$ & $4\times 10^4$  & $\beta\beta0\nu$ & U \\
$23$ & $L^i L^jL^k e^c \overline{Q}_k\bar{d^c}H^lH^m \epsilon_{il} \epsilon_{jm}$ & $\frac{y_\ell y_d}{(16\pi^2)^2}\frac{v^2}{\Lambda}\left(\frac{1}{16\pi^2} + \frac{v^2}{\Lambda^2}\right)$ & $40$  & $\beta\beta0\nu$ & U \\
$24_a$& $L^i L^j Q^k d^c Q^l d^c H^m \overline{H}_i\epsilon_{jk} \epsilon_{lm}$ & $\frac{y_d^2}{(16\pi^2)^3}\frac{v^2}{\Lambda}$ & $1\times 10^2$  & $\beta\beta0\nu$ & U \\
$24_b$& $L^i L^j Q^k d^c Q^l d^cH^m \overline{H}_i \epsilon_{jm} \epsilon_{kl}$ & $\frac{y_d^2}{(16\pi^2)^3}\frac{v^2}{\Lambda}$ & $1\times 10^2$  & $\beta\beta0\nu$ & U \\
$25$ & $L^i L^j Q^k d^c Q^l u^c H^m H^n \epsilon_{im}\epsilon_{jn} \epsilon_{kl}$ & $\frac{y_dy_u}{(16\pi^2)^2}\frac{v^2}{\Lambda}\left(\frac{1}{16\pi^2} + \frac{v^2}{\Lambda^2}\right)$ & $4\times 10^3$  & $\beta\beta0\nu$ & U \\
$26_a$& $L^i L^j Q^k d^c \overline{L}_i \bar{e^c}H^l H^m \epsilon_{jl} \epsilon_{km}$ & $\frac{y_\ell y_d}{(16\pi^2)^3}\frac{v^2}{\Lambda}$ & $40$ & $\beta\beta0\nu$ & U \\
$26_b$& $L^i L^j Q^k d^c \overline{L}_k \bar{e^c} H^lH^m\epsilon_{il}\epsilon_{jm}$ & $\frac{y_\ell y_d}{(16\pi^2)^2}\frac{v^2}{\Lambda}\left(\frac{1}{16\pi^2} + \frac{v^2}{\Lambda^2}\right)$ & $40$ & $\beta\beta0\nu$ & U \\
$27_a$& $L^i L^j Q^k d^c \overline{Q}_i\bar{d^c} H^lH^m \epsilon_{jl} \epsilon_{km}$ & $\frac{g^2}{(16\pi^2)^3}\frac{v^2}{\Lambda}$ & $4\times 10^4$  & $\beta\beta0\nu$ & U \\
$27_b$& $L^i L^j Q^k d^c \overline{Q}_k\bar{d^c} H^l H^m \epsilon_{il} \epsilon_{jm}$ & $\frac{g^2}{(16\pi^2)^3}\frac{v^2}{\Lambda}$ & $4 \times 10^4$  & $\beta\beta0\nu$ & U \\
$28_a$& $L^i L^j Q^k d^c \overline{Q}_j \bar{u^c}H^l \overline{H}_i \epsilon_{kl}$ & $\frac{y_dy_u}{(16\pi^2)^3}\frac{v^2}{\Lambda}$ & $4\times 10^3$  & $\beta\beta0\nu$ & U \\
$28_b$& $L^i L^j Q^k d^c \overline{Q}_k\bar{u^c} H^l \overline{H}_i \epsilon_{jl}$ & $\frac{y_dy_u}{(16\pi^2)^3}\frac{v^2}{\Lambda}$ & $4\times 10^3$  & $\beta\beta0\nu$ & U \\
$28_c$& $L^i L^j Q^k d^c \overline{Q}_l \bar{u^c} H^l\overline{H}_i\epsilon_{jk}$ & $\frac{y_dy_u}{(16\pi^2)^3}\frac{v^2}{\Lambda}$ & $4\times 10^3$  & $\beta\beta0\nu$ & U \\
$29_a$& $L^i L^j Q^k u^c \overline{Q}_k \bar{u^c}H^l H^m \epsilon_{il} \epsilon_{jm}$ & $\frac{y_u^2}{(16\pi^2)^2}\frac{v^2}{\Lambda}\left(\frac{1}{16\pi^2} + \frac{v^2}{\Lambda^2}\right)$ & $2\times 10^5$  & $\beta\beta0\nu$ & U \\
$29_b$& $L^i L^j Q^k u^c \overline{Q}_l \bar{u^c} H^l H^m\epsilon_{ik} \epsilon_{jm}$ & $\frac{g^2}{(16\pi^2)^3}\frac{v^2}{\Lambda}$ & $4\times 10^4$  & $\beta\beta0\nu$ & U \\
$30_a$& $L^i L^j \overline{L}_i \bar{e^c}\overline{Q}_k\bar{u^c} H^k H^l \epsilon_{jl}$ & $\frac{y_\ell y_u}{(16\pi^2)^3}\frac{v^2}{\Lambda}$ & $2\times 10^3$  & $\beta\beta0\nu$ & U \\
$30_b$& $L^i L^j\overline{L}_m \bar{e^c} \overline{Q}_n \bar{u^c} H^k H^l\epsilon_{ik} \epsilon_{jl} \epsilon^{mn}$ & $\frac{y_\ell y_u}{(16\pi^2)^2}\frac{v^2}{\Lambda}\left(\frac{1}{16\pi^2} + \frac{v^2}{\Lambda^2}\right)$ & $2\times 10^3$  & $\beta\beta0\nu$ & U \\
$31_a$& $L^i L^j \overline{Q}_i\bar{d^c}\overline{Q}_k\bar{u^c} H^k H^l \epsilon_{jl}$ & $\frac{y_dy_u}{(16\pi^2)^2}\frac{v^2}{\Lambda}\left(\frac{1}{16\pi^2} + \frac{v^2}{\Lambda^2}\right)$ & $4\times 10^3$  & $\beta\beta0\nu$ & U \\
$31_b$& $L^i L^j \overline{Q}_m\bar{d^c} \overline{Q}_n\bar{u^c}H^k H^l\epsilon_{ik} \epsilon_{jl} \epsilon^{mn}$ & $\frac{y_dy_u}{(16\pi^2)^2}\frac{v^2}{\Lambda}\left(\frac{1}{16\pi^2} + \frac{v^2}{\Lambda^2}\right)$  & $4\times 10^3$  & $\beta\beta0\nu$ & U \\
$32_a$& $L^i L^j \overline{Q}_j \bar{u^c}\overline{Q}_k \bar{u^c} H^k \overline{H}_i$ & $\frac{y_u^2}{(16\pi^2)^3}\frac{v^2}{\Lambda}$ & $2\times 10^5$ & $\beta\beta0\nu$ & U \\
$32_b$& $L^i L^j\overline{Q}_m \bar{u^c} \overline{Q}_n \bar{u^c} H^k\overline{H}_i \epsilon_{jk} \epsilon^{mn}$ & $\frac{y_u^2}{(16\pi^2)^3}\frac{v^2}{\Lambda}$ & $2\times 10^5$ & $\beta\beta0\nu$ & U \\
$33$ & $\bar{e^c} \bar{e^c} L^i L^j e^c e^c H^kH^l \epsilon_{ik} \epsilon_{jl}$ & $\frac{g^2}{(16\pi^2)^3}\frac{v^2}{\Lambda}$ & $4\times 10^4$  & $\beta\beta0\nu$ & U \\
$34$ & $\bar{e^c} \bar{e^c} L^i Q^j e^c d^c H^kH^l \epsilon_{ik} \epsilon_{jl}$ & $y_{\ell _\beta}\frac{y_dg^2}{(16\pi^2)^4}\frac{v^2}{\Lambda}$ & $< 0.5$  & $\beta\beta0\nu$, mix, ILC, LHC & C \\
$35$ & $\bar{e^c} \bar{e^c} L^i e^c\overline{Q}_j\bar{u^c} H^jH^k\epsilon_{ik}$ & $y_{\ell _\beta}\frac{y_ug^2}{(16\pi^2)^4}\frac{v^2}{\Lambda}$ & $2$ & mix, LHC & C \\
$36$ & $\bar{e^c} \bar{e^c} Q^i d^c Q^j d^c H^k H^l \epsilon_{ik} \epsilon_{jl}$ & $y_{\ell _\alpha}y_{\ell _\beta}\frac{y_d^2g^2}{(16\pi^2)^5}\frac{v^2}{\Lambda}$ & $< 0.5$ & $\beta\beta0\nu$, mix, HElnv, ILC, LHC & D \\
$37$ & $\bar{e^c} \bar{e^c} Q^i d^c\overline{Q}_j\bar{u^c} H^j H^k \epsilon_{ik}$ & $y_{\ell _\alpha}y_{\ell _\beta}\frac{y_dy_ug^2}{(16\pi^2)^5}\frac{v^2}{\Lambda}$ & $< 0.5$  & $\beta\beta0\nu$, mix, HElnv, ILC, LHC & D \\
$38$ & $\bar{e^c} \bar{e^c} \overline{Q}_i\bar{u^c}\overline{Q}_j \bar{u^c} H^i H^j$ & $y_{\ell _\alpha}y_{\ell _\beta}\frac{y_u^2g^2}{(16\pi^2)^5}\frac{v^2}{\Lambda}$ & $< 0.5$ & $\beta\beta0\nu$, mix, HElnv, ILC, LHC & D \\
$39_a$& $L^i L^j L^k L^l \overline{L}_i \overline{L}_jH^m H^n \epsilon_{km} \epsilon_{ln}$\footnote{This operator is modified slightly from its original form as given in reference \cite{Operators} where it appeared as $\mathcal{O}_{39}(a) = L^iL^jL^kL^l\overline{L}_i\overline{L}_jH^mH^n\epsilon_{jm}\epsilon_{kl}$.  We corrected this error. } & $\frac{g^2}{(16\pi^2)^3}\frac{v^2}{\Lambda}$ & $8\times 10^4$  & $\beta\beta0\nu$ & U \\
$39_b$& $L^i L^j L^k L^l\overline{L}_m \overline{L}_n H^m H^n \epsilon_{ij} \epsilon_{kl}$ & $\frac{g^2}{(16\pi^2)^3}\frac{v^2}{\Lambda}$ & $4\times 10^4$  & $\beta\beta0\nu$ & U \\
$39_c$& $L^i L^j L^k L^l \overline{L}_i \overline{L}_m H^m H^n\epsilon_{jk} \epsilon_{ln}$ & $\frac{g^2}{(16\pi^2)^3}\frac{v^2}{\Lambda}$ & $4\times 10^4$  & $\beta\beta0\nu$ & U \\
$39_d$& $L^i L^j L^k L^l \overline{L}_p\overline{L}_q H^m H^n \epsilon_{ij} \epsilon_{km} \epsilon_{ln}\epsilon^{pq}$ & $\frac{g^2}{(16\pi^2)^3}\frac{v^2}{\Lambda}$ & $4\times 10^4$  & $\beta\beta0\nu$ & U \\
$40_a$& $L^i L^j L^k Q^l \overline{L}_i \overline{Q}_jH^m H^n \epsilon_{km} \epsilon_{ln}$ & $\frac{g^2}{(16\pi^2)^3}\frac{v^2}{\Lambda}$ & $4\times 10^4$  & $\beta\beta0\nu$ & U \\
$40_b$& $L^i L^j L^k Q^l \overline{L}_i \overline{Q}_l H^m H^n\epsilon_{jm}\epsilon_{kn}$ & $\frac{g^2}{(16\pi^2)^3}\frac{v^2}{\Lambda}$ & $4\times 10^4$  & $\beta\beta0\nu$ & U \\
$40_c$& $L^i L^j L^k Q^l \overline{L}_l \overline{Q}_i H^m H^n\epsilon_{jm}\epsilon_{kn}$ & $\frac{g^2}{(16\pi^2)^3}\frac{v^2}{\Lambda}$ & $4\times 10^4$  & $\beta\beta0\nu$ & U \\
$40_d$& $L^i L^j L^k Q^l \overline{L}_i \overline{Q}_m H^m H^n\epsilon_{jk}\epsilon_{ln}$ & $\frac{g^2}{(16\pi^2)^3}\frac{v^2}{\Lambda}$ & $4\times 10^4$  & $\beta\beta0\nu$ & U \\
$40_e$& $L^i L^j L^k Q^l \overline{L}_i \overline{Q}_m H^m H^n\epsilon_{jl}\epsilon_{kn}$ & $\frac{g^2}{(16\pi^2)^3}\frac{v^2}{\Lambda}$ & $4\times 10^4$  & $\beta\beta0\nu$ & U \\
$40_f$& $L^i L^j L^k Q^l \overline{L}_m \overline{Q}_i H^m H^n\epsilon_{jk}\epsilon_{ln}$ & $\frac{g^2}{(16\pi^2)^3}\frac{v^2}{\Lambda}$ & $4\times 10^4$  & $\beta\beta0\nu$ & U \\
$40_g$& $L^i L^j L^k Q^l \overline{L}_m \overline{Q}_i H^m H^n\epsilon_{jl}\epsilon_{kn}$ & $\frac{g^2}{(16\pi^2)^3}\frac{v^2}{\Lambda}$ & $4\times 10^4$  & $\beta\beta0\nu$ & U \\
$40_h$& $L^i L^j L^k Q^l \overline{L}_m \overline{Q}_n H^m H^n\epsilon_{ij}\epsilon_{kl}$ & $\frac{g^2}{(16\pi^2)^3}\frac{v^2}{\Lambda}$ & $4\times 10^4$ & $\beta\beta0\nu$ & U \\
$40_i$& $L^i L^j L^k Q^l \overline{L}_m \overline{Q}_n H^p H^q\epsilon_{ip}\epsilon_{jq} \epsilon_{kl} \epsilon^{mn}$ & $\frac{g^2}{(16\pi^2)^3}\frac{v^2}{\Lambda}$ & $4\times 10^4$ & $\beta\beta0\nu$ & U \\
$40_j$& $L^i L^j L^k Q^l \overline{L}_m \overline{Q}_n H^p H^q\epsilon_{ip}\epsilon_{lq} \epsilon_{jk} \epsilon^{mn}$ & $\frac{g^2}{(16\pi^2)^3}\frac{v^2}{\Lambda}$ & $4\times 10^4$  & $\beta\beta0\nu$ & U \\
$41_a$& $L^i L^j L^k d^c \overline{L}_i \bar{d^c}H^l H^m \epsilon_{jl} \epsilon_{km}$ & $\frac{g^2}{(16\pi^2)^3}\frac{v^2}{\Lambda}$ & $4\times 10^4$ & $\beta\beta0\nu$ & U \\
$41_b$& $L^i L^j L^k d^c \overline{L}_l \bar{d^c} H^l H^m\epsilon_{ij}\epsilon_{km}$ & $\frac{g^2}{(16\pi^2)^3}\frac{v^2}{\Lambda}$ & $4\times 10^4$ & $\beta\beta0\nu$ & U \\
$42_a$& $L^i L^j L^k u^c \overline{L}_i\bar{u^c} H^l H^m \epsilon_{jl} \epsilon_{km}$ & $\frac{g^2}{(16\pi^2)^3}\frac{v^2}{\Lambda}$ & $4\times 10^4$ & $\beta\beta0\nu$ & U \\
$42_b$& $L^i L^j L^k u^c \overline{L}_l\bar{u^c} H^l H^m \epsilon_{ij} \epsilon_{km}$ & $\frac{g^2}{(16\pi^2)^3}\frac{v^2}{\Lambda}$ & $4\times 10^4$ & $\beta\beta0\nu$ & U \\
$43_a$& $L^i L^j L^k d^c \overline{L}_l\bar{u^c} H^l\overline{H}_i \epsilon_{jk}$ & $\frac{y_dy_ug^2}{(16\pi^2)^4}\frac{v^2}{\Lambda}$ & $6$ & $\beta\beta0\nu$, LHC & U \\
$43_b$& $L^i L^j L^k d^c \overline{L}_j\bar{u^c} H^l\overline{H}_i\epsilon_{kl}$ & $\frac{y_dy_ug^2}{(16\pi^2)^4}\frac{v^2}{\Lambda}$ & $6$ & $\beta\beta0\nu$, LHC & U \\
$43_c$& $L^i L^j L^k d^c \overline{L}_l\bar{u^c} H^m\overline{H}_n\epsilon_{ij} \epsilon_{km} \epsilon^{ln}$ & $\frac{y_dy_ug^2}{(16\pi^2)^4}\frac{v^2}{\Lambda}$ & $6$ & $\beta\beta0\nu$, LHC & U \\
$44_a$& $L^i L^j Q^k e^c \overline{Q}_i \bar{e^c} H^l H^m\epsilon_{jl} \epsilon_{km}$ & $\frac{g^2}{(16\pi^2)^3}\frac{v^2}{\Lambda}$ & $4\times 10^4$ & $\beta\beta0\nu$ & U \\
$44_b$& $L^i L^j Q^k e^c \overline{Q}_k \bar{e^c} H^l H^m\epsilon_{il}\epsilon_{jm}$ & $\frac{g^2}{(16\pi^2)^3}\frac{v^2}{\Lambda}$ & $4\times 10^4$ & $\beta\beta0\nu$ & U \\
$44_c$& $L^i L^j Q^k e^c \overline{Q}_l \bar{e^c} H^l H^m\epsilon_{ij}\epsilon_{km}$ & $\frac{g^4}{(16\pi^2)^4}\frac{v^2}{\Lambda}$ & $60$ & $\beta\beta0\nu$ & U \\
$44_d$& $L^i L^j Q^k e^c \overline{Q}_l \bar{e^c} H^l H^m\epsilon_{ik} \epsilon_{jm}$ & $\frac{g^2}{(16\pi^2)^3}\frac{v^2}{\Lambda}$ & $4\times 10^4$ & $\beta\beta0\nu$ & U \\
$45$ & $L^i L^j e^c d^c \bar{e^c} \bar{d^c} H^k H^l\epsilon_{ik} \epsilon_{jl}$ & $\frac{g^2}{(16\pi^2)^3}\frac{v^2}{\Lambda}$ & $4\times 10^4$ & $\beta\beta0\nu$ & U \\
$46$ & $L^i L^j e^c u^c \bar{e^c} \bar{u^c} H^k H^l\epsilon_{ik} \epsilon_{jl}$ & $\frac{g^2}{(16\pi^2)^3}\frac{v^2}{\Lambda}$ & $4\times 10^4$ & $\beta\beta0\nu$ & U \\
$47_a$& $L^i L^j Q^k Q^l \overline{Q}_i\overline{Q}_j H^mH^n \epsilon_{km} \epsilon_{ln}$ & $\frac{g^2}{(16\pi^2)^3}\frac{v^2}{\Lambda}$ & $4\times 10^4$ & $\beta\beta0\nu$ & U \\
$47_b$& $L^i L^j Q^k Q^l \overline{Q}_i \overline{Q}_k H^m H^n\epsilon_{jm}\epsilon_{ln}$ & $\frac{g^2}{(16\pi^2)^3}\frac{v^2}{\Lambda}$ & $4\times 10^4$ & $\beta\beta0\nu$ & U \\
$47_c$& $L^i L^j Q^k Q^l \overline{Q}_k\overline{Q}_l H^m H^n\epsilon_{im}\epsilon_{jn}$ & $\frac{g^2}{(16\pi^2)^3}\frac{v^2}{\Lambda}$ & $4\times 10^4$ & $\beta\beta0\nu$ & U \\
$47_d$& $L^i L^j Q^k Q^l \overline{Q}_i\overline{Q}_m H^m H^n\epsilon_{jk}\epsilon_{ln}$ & $\frac{g^2}{(16\pi^2)^3}\frac{v^2}{\Lambda}$ & $4\times 10^4$ & $\beta\beta0\nu$ & U \\
$47_e$& $L^i L^j Q^k Q^l \overline{Q}_i\overline{Q}_m H^m H^n\epsilon_{jn}\epsilon_{kl}$ & $\frac{g^2}{(16\pi^2)^3}\frac{v^2}{\Lambda}$ & $4\times 10^4$ & $\beta\beta0\nu$ & U \\
$47_f$& $L^i L^j Q^k Q^l \overline{Q}_k\overline{Q}_m H^m H^n\epsilon_{ij}\epsilon_{ln}$ & $\frac{g^4}{(16\pi^2)^4}\frac{v^2}{\Lambda}$ & $60$ & $\beta\beta0\nu$ & U \\
$47_g$& $L^i L^j Q^k Q^l \overline{Q}_k\overline{Q}_m H^m H^n\epsilon_{il}\epsilon_{jn}$ & $\frac{g^2}{(16\pi^2)^3}\frac{v^2}{\Lambda}$ & $4\times 10^4$ & $\beta\beta0\nu$ & U \\
$47_h$& $L^i L^j Q^k Q^l \overline{Q}_p\overline{Q}_q H^m H^n\epsilon_{ij}\epsilon_{km} \epsilon_{ln} \epsilon^{pq}$ & $\frac{g^4}{(16\pi^2)^4}\frac{v^2}{\Lambda}$ & $60$ & $\beta\beta0\nu$ & U \\
$47_i$& $L^i L^j Q^k Q^l \overline{Q}_p\overline{Q}_q H^m H^n\epsilon_{ik}\epsilon_{jm} \epsilon_{ln} \epsilon^{pq}$ & $\frac{g^2}{(16\pi^2)^3}\frac{v^2}{\Lambda}$ & $4\times 10^4$ & $\beta\beta0\nu$ & U \\
$47_j$& $L^i L^j Q^k Q^l \overline{Q}_p\overline{Q}_q H^m H^n\epsilon_{im}\epsilon_{jn} \epsilon_{kl} \epsilon^{pq}$ & $\frac{g^2}{(16\pi^2)^3}\frac{v^2}{\Lambda}$ & $4\times 10^4$ & $\beta\beta0\nu$ & U \\
$48$ & $L^i L^j d^c d^c \bar{d^c} \bar{d^c} H^kH^l \epsilon_{ik} \epsilon_{jl}$ & $\frac{g^2}{(16\pi^2)^3}\frac{v^2}{\Lambda}$ & $4\times 10^4$ & $\beta\beta0\nu$ & U \\
$49$ & $L^i L^j d^c u^c \bar{d^c} \bar{u^c} H^kH^l \epsilon_{ik} \epsilon_{jl}$ & $\frac{g^2}{(16\pi^2)^3}\frac{v^2}{\Lambda}$ & $4\times 10^4$ & $\beta\beta0\nu$ & U \\
$50$ & $L^i L^j d^c d^c \bar{d^c} \bar{u^c} H^k\overline{H}_i \epsilon_{jk}$ & $\frac{y_dy_ug^2}{(16\pi^2)^4}\frac{v^2}{\Lambda}$ & $6$ & $\beta\beta0\nu$ LHC & U \\
$51$ & $L^i L^j u^c u^c \bar{u^c} \bar{u^c} H^kH^l \epsilon_{ik} \epsilon_{jl}$ & $\frac{g^2}{(16\pi^2)^3}\frac{v^2}{\Lambda}$ & $4\times 10^4$  & $\beta\beta0\nu$ & U \\
$52$ & $L^i L^j d^c u^c \bar{u^c} \bar{u^c} H^k\overline{H}_i \epsilon_{jk}$ & $\frac{y_dy_ug^2}{(16\pi^2)^4}\frac{v^2}{\Lambda}$ & $6$  & $\beta\beta0\nu$, LHC & U \\
$53$ & $L^i L^j d^c d^c \bar{u^c} \bar{u^c}\overline{H}_i \overline{H}_j$ & $\frac{y_d^2y_u^2g^2}{(16\pi^2)^5}\frac{v^2}{\Lambda}$ & $< 0.5$  & $\beta\beta0\nu$, HElnv, ILC, LHC & D \\
$54_a$& $L^i Q^j Q^k d^c \overline{Q}_i \bar{e^c}H^l H^m \epsilon_{jl} \epsilon_{km}$ & $y_{\ell _\beta}\frac{y_dg^2}{(16\pi^2)^4}\frac{v^2}{\Lambda}$ & $< 0.5$ & $\beta\beta0\nu$, mix, HElnv, ILC, LHC & D \\
$54_b$& $L^i Q^j Q^k d^c \overline{Q}_j \bar{e^c} H^l H^m\epsilon_{il}\epsilon_{km}$ & $y_{\ell _\beta}\frac{y_dg^2}{(16\pi^2)^4}\frac{v^2}{\Lambda}$ & $< 0.5$ & $\beta\beta0\nu$, mix, HElnv, ILC, LHC & D \\
$54_c$& $L^i Q^j Q^k d^c \overline{Q}_l \bar{e^c} H^l H^m\epsilon_{im}\epsilon_{jk}$ & $y_{\ell _\beta}\frac{y_dg^2}{(16\pi^2)^4}\frac{v^2}{\Lambda}$ & $< 0.5$ & $\beta\beta0\nu$, mix, ILC, LHC & D \\
$54_d$& $L^i Q^j Q^k d^c \overline{Q}_l \bar{e^c} H^lH^m \epsilon_{ij} \epsilon_{km}$ & $y_{\ell _\beta}\frac{y_dg^2}{(16\pi^2)^4}\frac{v^2}{\Lambda}$ & $< 0.5$ & $\beta\beta0\nu$,mix, HElnv, ILC, LHC & D \\
$55_a$& $L^i Q^j \overline{Q}_i \overline{Q}_k\bar{e^c} \bar{u^c} H^k H^l \epsilon_{jl}$ & $y_{\ell _\beta}\frac{y_ug^2}{(16\pi^2)^4}\frac{v^2}{\Lambda}$ & $2$ & $\beta\beta0\nu$, mix, LHC & C \\
$55_b$& $L^i Q^j \overline{Q}_j \overline{Q}_k \bar{e^c}\bar{u^c}H^k H^l \epsilon_{il}$ & $y_{\ell _\beta}\frac{y_ug^2}{(16\pi^2)^4}\frac{v^2}{\Lambda}$ & $2$ & $\beta\beta0\nu$, mix, LHC & C \\
$55_c$& $L^i Q^j \overline{Q}_m \overline{Q}_n \bar{e^c}\bar{u^c} H^k H^l\epsilon_{ik} \epsilon_{jl} \epsilon^{mn}$ & $y_{\ell _\beta}\frac{y_ug^2}{(16\pi^2)^4}\frac{v^2}{\Lambda}$ & $2$  & $\beta\beta0\nu$, mix, LHC & C \\
$56$ & $L^i Q^j d^c d^c \bar{e^c} \bar{d^c} H^kH^l \epsilon_{ik} \epsilon_{jl}$ & $y_{\ell _\beta}\frac{y_dg^2}{(16\pi^2)^4}\frac{v^2}{\Lambda}$ & $< 0.5$  & $\beta\beta0\nu$, mix, ILC, LHC & C \\
$57$ & $L^i d^c \overline{Q}_j \bar{u^c} \bar{e^c}\bar{d^c} H^j H^k \epsilon_{ik}$ & $y_{\ell _\beta}\frac{y_ug^2}{(16\pi^2)^4}\frac{v^2}{\Lambda} $ & $2$ & $\beta\beta0\nu$, mix, LHC & C \\
$58$ & $L^i u^c \overline{Q}_j \bar{u^c} \bar{e^c}\bar{u^c} H^j H^k \epsilon_{ik}$ & $y_{\ell _\beta}\frac{y_ug^2}{(16\pi^2)^4}\frac{v^2}{\Lambda}$& $2$ & mix, LHC & C \\
$59$ & $L^i Q^j d^c d^c \bar{e^c} \bar{u^c}H^k \overline{H}_i \epsilon_{jk}$ & $y_{\ell _\beta}\frac{y_d^2y_u}{(16\pi^2)^4}\frac{v^2}{\Lambda}$ & $< 0.5$ & $\beta\beta0\nu$, mix, HElnv, ILC, LHC & D \\
$60$ & $L^i d^c \overline{Q}_j \bar{u^c}\bar{e^c}\bar{u^c} H^j \overline{H}_i$ & $y_{\ell _\beta}\frac{y_dy_u^2}{(16\pi^2)^4}\frac{v^2}{\Lambda}$ & $< 0.5$ & $\beta\beta0\nu$, mix, HElnv, ILC, LHC & D \\
$61$ & $L^i L^j H^k H^l L^r e^c \overline{H}_r \epsilon_{ik} \epsilon_{jl}$ & $\frac{y_\ell }{16\pi^2}\frac{v^2}{\Lambda}\left(\frac{1}{16\pi^2} + \frac{v^2}{\Lambda^2}\right)$ & $2\times 10^{5}$ & $\beta\beta0\nu$ & U \\
$62$ & $L^i L^j L^k e^c H^l L^r e^c \overline{H}_r\epsilon_{ij} \epsilon_{kl}$ & $\frac{y_\ell ^2}{(16\pi^2)^2}\frac{v^2}{\Lambda}\left(\frac{1}{16\pi^2} + \frac{v^2}{\Lambda^2}\right)$ & $20$  & $\beta\beta0\nu$ & U \\
$63_a$& $L^i L^j Q^k d^c H^l L^r e^c \overline{H}_r\epsilon_{ij} \epsilon_{kl}$ & $\frac{y_\ell  y_d}{(16\pi^2)^3}\frac{v^2}{\Lambda}$ & $40$ & $\beta\beta0\nu$ & U \\
$63_b$& $L^i L^j Q^k d^c H^l L^r e^c \overline{H}_r\epsilon_{ik} \epsilon_{jl}$ & $\frac{y_\ell  y_d}{(16\pi^2)^2}\frac{v^2}{\Lambda}\left(\frac{1}{16\pi^2} + \frac{v^2}{\Lambda^2}\right)$ & $40$ & $\beta\beta0\nu$ & U \\
$64_a$& $L^i L^j \overline{Q}_i \bar{u^c} H^k L^r e^c \overline{H}_r\epsilon_{jk}$ & $\frac{y_\ell  y_u}{(16\pi^2)^2}\frac{v^2}{\Lambda}\left(\frac{1}{16\pi^2} + \frac{v^2}{\Lambda^2}\right)$ & $2\times 10^3$ & $\beta\beta0\nu$ & U \\
$64_b$& $L^i L^j \overline{Q}_k\bar{u^c}H^k L^r e^c \overline{H}_r\epsilon_{ij}$ & $\frac{y_\ell  y_u}{(16\pi^2)^3}\frac{v^2}{\Lambda}$ & $2\times 10^3$ & $\beta\beta0\nu$ & U \\
$65$ & $L^i \bar{e^c} \bar{u^c} d^c H^j L^r e^c \overline{H}_r\epsilon_{ij}$ & $\frac{y_d y_ug^2}{(16\pi^2)^4}\frac{v^2}{\Lambda}$ & $6$ & $\beta\beta0\nu$, LHC & U \\
$66$ & $L^i L^j H^k H^l \epsilon_{ik} Q^r d^c \overline{H}_r\epsilon_{jl}$ & $\frac{y_d}{16\pi^2}\frac{v^2}{\Lambda}\left(\frac{1}{16\pi^2} + \frac{v^2}{\Lambda^2}\right)$ & $6\times 10^{5}$ & $\beta\beta0\nu$ & U \\
$67$ & $L^i L^j L^k e^c H^l Q^r d^c \overline{H}_r\epsilon_{ij} \epsilon_{kl}$ & $\frac{y_\ell y_d}{(16\pi^2)^2}\frac{v^2}{\Lambda}\left(\frac{1}{16\pi^2} + \frac{v^2}{\Lambda^2}\right)$ & $40$ & $\beta\beta0\nu$ & U \\
$68_a$& $L^i L^j Q^k d^c H^l Q^r d^c \overline{H}_r\epsilon_{ij} \epsilon_{kl}$ & $\frac{y_d^2 g^2}{(16\pi^2)^3}\frac{v^2}{\Lambda}\left(\frac{1}{16\pi^2} + \frac{v^2}{\Lambda^2}\right)$ & $1$ & $\beta\beta0\nu$, LHC & U \\
$68_b$& $L^i L^j Q^k d^c H^l Q^r d^c \overline{H}_r\epsilon_{ik} \epsilon_{jl}$ & $\frac{y_{q_d^2}}{(16\pi^2)^2}\frac{v^2}{\Lambda}\left(\frac{1}{16\pi^2} + \frac{v^2}{\Lambda^2}\right)$ & $1\times 10^2$  & $\beta\beta0\nu$ & U \\
$69_a$& $L^i L^j \overline{Q}_i \bar{u^c} H^k Q^r d^c \overline{H}_r\epsilon_{jk}$ & $\frac{y_dy_u}{(16\pi^2)^2}\frac{v^2}{\Lambda}\left(\frac{1}{16\pi^2} + \frac{v^2}{\Lambda^2}\right)$ & $4\times 10^3$ & $\beta\beta0\nu$ & U \\
$69_b$& $L^i L^j \overline{Q}_k\bar{u^c}H^k Q^r d^c \overline{H}_r\epsilon_{ij}$ & $\frac{y_dy_u g^2}{(16\pi^2)^3}\frac{v^2}{\Lambda}\left(\frac{1}{16\pi^2} + \frac{v^2}{\Lambda^2}\right)$ & $7$ & $\beta\beta0\nu$, LHC & U \\
$70$ & $L^i \bar{e^c} \bar{u^c} d^c H^j Q^r d^c \overline{H}_r\epsilon_{ij}$ & $y_{\ell _\beta}\frac{ y_d^2 y_u}{(16\pi^2)^3}\frac{v^2}{\Lambda}\left(\frac{1}{16\pi^2} + \frac{v^2}{\Lambda^2}\right)$ & $< 0.5$ & $\beta\beta0\nu$, mix, HElnv, ILC, LHC & D \\
$71$ & $L^i L^j H^k H^l Q^r u^c H^s \epsilon_{rs} \epsilon_{ik} \epsilon_{jl}$ & $\frac{y_u}{16\pi^2}\frac{v^2}{\Lambda}\left(\frac{1}{16\pi^2} + \frac{v^2}{\Lambda^2}\right)$ & $2\times 10^{7}$ & $\beta\beta0\nu$ & U \\
$72$ & $L^i L^j L^k e^c H^l Q^r u^c H^s \epsilon_{rs}\epsilon_{ij} \epsilon_{kl}$ & $\frac{y_\ell y_u}{(16\pi^2)^2}\frac{v^2}{\Lambda}\left(\frac{1}{16\pi^2} + \frac{v^2}{\Lambda^2}\right)$ & $2\times 10^3$  & $\beta\beta0\nu$ & U \\
$73_a$& $L^i L^j Q^k d^c H^l Q^r u^c H^s \epsilon_{rs}\epsilon_{ij} \epsilon_{kl}$ & $\frac{y_dy_u g^2}{(16\pi^2)^3}\frac{v^2}{\Lambda}\left(\frac{1}{16\pi^2} + \frac{v^2}{\Lambda^2}\right)$ & $7$ & $\beta\beta0\nu$, LHC & U \\
$73_b$& $L^i L^j Q^k d^c H^l Q^r u^c H^s \epsilon_{rs}\epsilon_{ik} \epsilon_{jl}$ & $\frac{y_dy_u}{(16\pi^2)^2}\frac{v^2}{\Lambda}\left(\frac{1}{16\pi^2} + \frac{v^2}{\Lambda^2}\right)$ & $4\times 10^3$ & $\beta\beta0\nu$ & U \\
$74_a$& $L^i L^j \overline{Q}_i \bar{u^c} H^k Q^r u^c H^s \epsilon_{rs}\epsilon_{jk}$ & $\frac{y_u^2}{(16\pi^2)^2}\frac{v^2}{\Lambda}\left(\frac{1}{16\pi^2} + \frac{v^2}{\Lambda^2}\right)$ & $2\times 10^5$ & $\beta\beta0\nu$ & U \\
$74_b$& $L^i L^j \overline{Q}_k\bar{u^c}H^k Q^r u^c H^s \epsilon_{rs}\epsilon_{ij}$ & $\frac{y_u^2 g^2}{(16\pi^2)^3}\frac{v^2}{\Lambda}\left(\frac{1}{16\pi^2} + \frac{v^2}{\Lambda^2}\right)$ & $2\times 10^2$ & $\beta\beta0\nu$ & U \\
$75$ & $L^i \bar{e^c} \bar{u^c} d^c H^j Q^r u^c H^s \epsilon_{rs}\epsilon_{ij}$ & $y_{\ell _\beta}\frac{ y_d y_u^2}{(16\pi^2)^3}\frac{v^2}{\Lambda}\left(\frac{1}{16\pi^2} + \frac{v^2}{\Lambda^2}\right)$ & $1$ & $\beta\beta0\nu$, mix & C \\
\hline
 \end{longtable}
\end{center}

The third column of Table \ref{tab:AllOps}, labeled
$m_{\alpha\beta}$, presents our estimate for the  operator-induced Majorana
neutrino mass expressions.  These were derived based on the
estimation procedure discussed earlier. Trivial order one
factors, as well as the generation dependent coupling constants $\lambda$ have
been omitted, as already advertised.  Flavor specific charged lepton Yukawa
couplings are explicitly denoted $y_{\ell_\alpha}$ and $y_{\ell_\beta}$ to distinguish
them from $y_\ell$, $y_u$ and $y_d$, meant to represent $\alpha,\beta$-independent
Yukawa couplings.  A summation over all ``internal flavors'' is assumed for each entry.
For order one coupling constants, this sum is
strongly dominated by third generation Yukawa couplings.
Upon setting these
mass expressions equal to the observed scale of light neutrino
masses ($0.05~\rm{eV}$), we extract the required cutoff
scale $\Lambda$ for each operator.  This
quantity, defined to be $\Lambda_{\nu}$, is listed in column four in units
of one TeV. Numerical results were obtained assuming the current
best fit values for all SM parameters. Associated errors are
negligibly small as far as our aspirations are concerned.

\begin{figure}
\begin{center}
\includegraphics[angle=270,scale=.6]{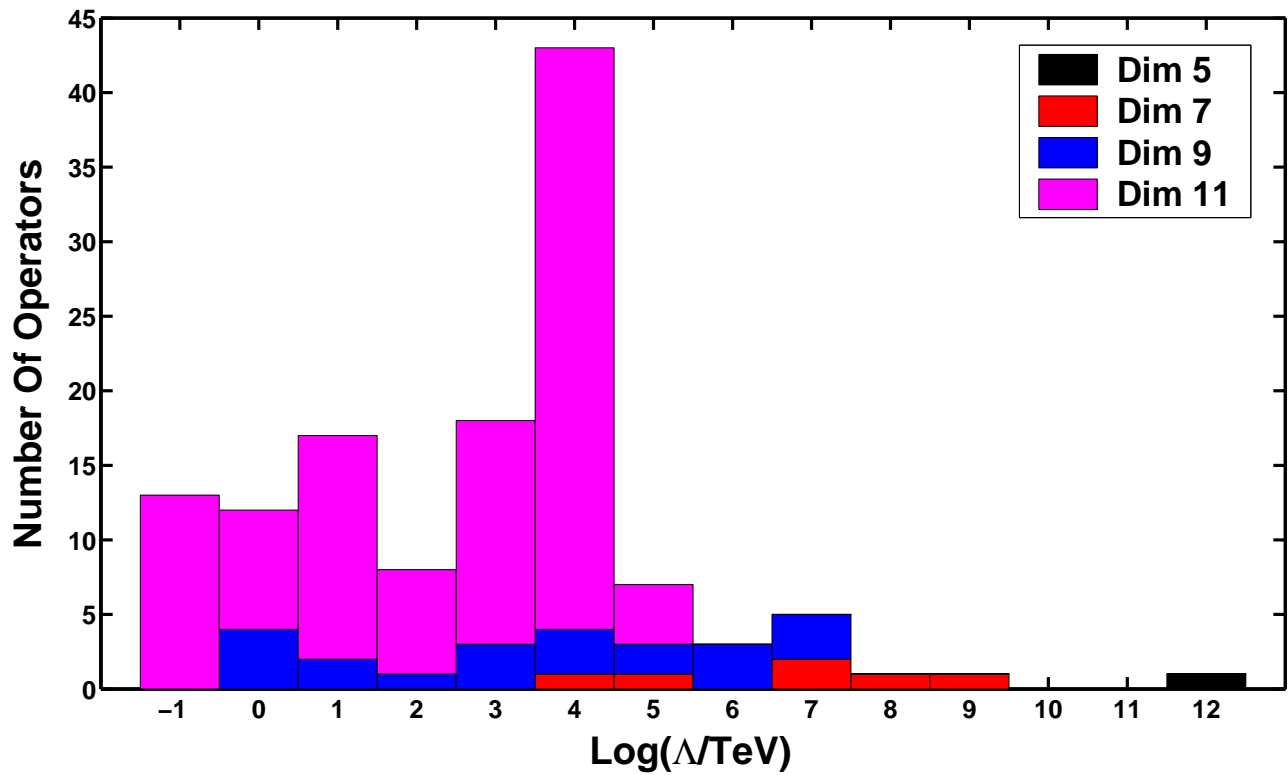}
\caption{A summary histogram of
the scale of new physics $\Lambda_{\nu}$ extracted from the 129 LNV
operators introduced in Table \ref{tab:AllOps}.  We assume a
radiatively generated neutrino mass of $0.05~\rm{eV}$ and universal
order one coupling constants.  The contributions of operators of different mass dimensions
are associated to different colors (shades of gray), as indicated in the caption.} \label{fig:OpSum}
\end{center}
\end{figure}

Fig.~\ref{fig:OpSum} displays the distribution of extracted
cutoff scales, $\Lambda_{\nu}$.
The histogram bars are color coded to reflect the different operator mass
dimensions.  The distribution spans thirteen orders of magnitude,
from the electroweak scale to $10^{12}~\rm{TeV}$.  It is interesting
to note the general trend of operator dimension with scale:  as
expected, higher dimension operators are characterized by lower
ultraviolet scales.  For operators associated with the lowest ultraviolet cutoffs,
the lepton number breaking physics occurs at the same energy scale as electroweak
symmetry breaking. In this case, one needs to revisit some of the
assumptions that go into obtaining the bounds and predictions discussed here.
Regardless, it is fair to say that some of these effective operators should be severely
constrained by other experimental probes, as will be discussed in the next section.

The natural scale for most of the explored
operators is well above $10~\rm{TeV}$, and thus outside the reach of
future experimental efforts except, perhaps, those looking for neutrinoless double-beta decay. The
remainder, however, should yield observable consequences in next-generation
experiments.  This small subset arguably contains the most
interesting cases on purely economic grounds, as they naturally
predict tiny neutrino masses as well as TeV scale new physics, which
is already thought to exist for independent reasons. It is
aesthetically pleasing to imagine that all, or at least most, of
nature's current puzzles can arise from the same source, as opposed
to postulating various solutions at different energy scales.  It is important to note that
one can ``push'' more of the operators into the observable TeV
window by modifying the coupling of the new physics to different
fermion generations.  In particular, since many of the induced
neutrino masses depend upon fermion Yukawa couplings, one
can efficiently reduce scales by simply and uniformly decoupling the
third generation. In most cases this can yield a $\Lambda_{\nu}$ reduction of
several orders of magnitude; a factor that can be further enhanced
by also decoupling the second generation.  Under these conditions,
the resulting distribution, analogous to Fig.~\ref{fig:OpSum}, would show the majority of the operators piled up
near and slightly above the electroweak scale.
A detailed exploration of this possibility would be
impractical and is not pursued further.
We will, however, like to emphasize that this strategy of decoupling the new
physics from the \emph{heavy} fermions is very non-standard. In most cases, one is
tempted to decouple \emph{light} fermions from new physics both because these
lead to the strongest constraints and because one tends to believe that the large Yukawa couplings
of third generation fermions are entangled with the physics of electroweak symmetry breaking.

Not all extracted cutoff scales are subject to a strong dependency on SM Yukawa couplings.
In particular, the $\Lambda_{\nu}$ values for the majority of dimension-eleven
operators in the large histogram bar near $10^4~\rm{TeV}$ would not
shift down at all under this hypothetical decoupling of the third
generation from the new physics.  These are
the operators, as shown in Diagram $(d)$ of
Fig.~\ref{fig:MassDiagrams}, whose induced neutrino mass matrix is independent of the Yukawa sector.
In such cases, $m_{\alpha\beta}$ are only functions of the various gauge couplings.
As such, these constitute the most
robust results of our analysis. These operators all
predict an anarchic Majorana neutrino mass matrix of overall scale
given by $m_{\nu} = g^2/(16\pi^2)^3 v^2/\Lambda$,
implying an energy scale $\Lambda_{\nu}\sim 10^5~\rm{TeV}$.
The only other ``Yukawa invariant'' cutoff scale estimate arises for the
dimension-five operator $\mathcal{O}_1$.
$\mathcal{O}_1$ captures the physics of all versions of the
 seesaw mechanism \cite{SeeSaw}, and is at the heart of
most of the model building currently done within the neutrino
sector. Its ultraviolet completion can precede
in only three distinct ways \cite{Bajc:2006ia}.  These possibilities
are via the exchange of heavy gauge singlet fermions (type I seesaw),
$SU(2)_L$ triplet scalars (type II seesaw) \cite{SeeSaw2}, $SU(2)_L$
triplet fermions (type III seesaw) \cite{SeeSaw3}, or some combination
thereof.  Its popularity is well-founded for a number of reasons,
including its underlying simplicity in structure as well as the
purely empirical fact that it is the ``lowest order means'' of neutrino
mass, and as such is easily generated by a ``generic'' LNV model.
Additionally, the high scale
associated with the seesaw mechanism can be easily incorporated
within existing theoretical models and serves to
help explain the observed baryon antisymmetry of the universe via
leptogenesis \cite{Leptogenesis}.

For the purposes of direct
observation, $\mathcal{O}_1$'s high cutoff scale, nearly
$10^{12}~\rm{TeV}$, places it well outside of the ``detectable region'' ($\Lambda\lesssim 10$~TeV)
and renders it uninteresting for the purposes of our analysis.
Of course, there always remains the possibility that $\mathcal{O}_1$
is generate by very weakly coupled new physics (or very finely-tuned new physics \cite{fine_nus}), in which case we
expect to run into the new ultraviolet degrees of freedom at energies
well below $10^{12}~\rm{TeV}$. In the case of $\mathcal{O}_1$, it has been argued that
new physics at almost any energy scale (from well below the sub-eV realm to well above the weak scale)
will lead to light neutrino masses \cite{LowScaleSeeSaw} without contradicting current experimental results.
Such possibilities -- related to the fact that the new physics is very weakly coupled -- are not being explored here, as
we always assume that the new degrees of freedom are heavier than typical experimentally accessible energy
scales.

Armed with our derived new physics scales $\Lambda_{\nu}$, we proceed to plug
them back into the different irrelevant LNV operators and search for possible
means of future observation as well as already existing constraints.
Generally, those operators that yield the largest experimental
signals have the lowest cutoff scales.
We conclude that, if associated to neutrino masses, the effective cutoff
scale $\Lambda_{\nu}$ of the following effective operators is constrained to be less than  $1~\rm{TeV}$:
\begin{equation}
\mathcal{O}_{34},\mathcal{O}_{36},\mathcal{O}_{37},\mathcal{O}_{38},\mathcal{O}_{53},\mathcal{O}_{54_{a,b,c,d}},\mathcal{O}_{56},\mathcal{O}_{59},\mathcal{O}_{60},\mathcal{O}_{70}.
\label{eq:O1TeV}
\end{equation}
These may lead to observable effects at future high energy accelerator
facilities. Additionally, such low scales may also indirectly lead to
observable effects in ``low energy'' (but high sensitivity) experiments.  There are
more operators associated with slightly higher scales between
$(1-10)~\rm{TeV}$ that may manifest themselves experimentally via virtual
effects.  These are
\begin{equation}
\mathcal{O}_{16},\mathcal{O}_{17},\mathcal{O}_{18},\mathcal{O}_{19},\mathcal{O}_{35},\mathcal{O}_{43_{a,b,c}},\mathcal{O}_{50},\mathcal{O}_{52},\mathcal{O}_{55_{a,b,c}},\mathcal{O}_{57},\mathcal{O}_{58},\mathcal{O}_{65},\mathcal{O}_{68_{a,b}},\mathcal{O}_{73_a},\mathcal{O}_{75}.
\label{eq:O1to10TeV}
\end{equation}
These operators yield finite
predictions for more than one observable, such that experimental
efforts in seemingly unrelated fields can help  constrain the
class of possible LNV models or even help identify the true LNV
model.

\setcounter{footnote}{0} \setcounter{equation}{0}
\section{General operator constraints and predictions}
\label{sec:constraints}

There are, currently, bounds on LNV processes from a number of
independent experimental sources \cite{LNVUpperBounds,PDG}.  Many of
these are presently too mild to constrain the operators listed in Table~\ref{tab:AllOps} once
their ultraviolet cutoffs $\Lambda$ are set to the required value indicated by the presence of
non-zero neutrino masses, $\Lambda_{\nu}$.
The situation, however,  is expected to improve in the next several years with increased
rare decay sensitivities and higher collider energies. Here we
survey the experimental signatures of these operators in terms of
the minimal scenarios described above.  Specifically, we address the
potential of neutrinoless double-beta decay (Sec.~\ref{subsec:bb0nu}), rare meson decays (Sec.~\ref{subsec:OtherProcesses}), and
collider experiments (Sec.~\ref{subsec:Collider}) to constrain the effective operators in question, assuming that, indeed,
they are responsible for the observed non-zero neutrino masses.
As before, we will use the approximations discussed in Sec.~\ref{sec:Scale}, and warn
readers that all the results presented are to be understood as order of magnitude estimates.
The results, however, are useful as far as recognizing the most promising LNV probes and identifying
different scenarios that may be probed by combinations of different LNV searches.

Most of this section will be devoted to  probes of LNV via
simple variants of the following process, which can be written schematically as
\begin{equation}
\ell _\alpha\ell _\beta \leftrightarrow d_\kappa d_\zeta
\bar{u}_\rho \bar{u}_\omega. \label{eq:GoldenCh}
\end{equation}
Greek subscripts run over all different fermion flavors. Given the
assumed democratic models, coupled with our present lack of
experimental information, one would expect that all flavor
combinations are equivalent to zeroth order.  Any
indication to the contrary would signify important deviations from
simple expectations, and thus begin to reveal the flavor structure
of the new physics.  The above selected ``golden modes'' often yield the
largest LNV rates, but this is not always the case. For
example, some operators do not allow tree-level charged dilepton
events, but rather prefer to include neutrino initial or final states.  LNV processes with
initial and final state neutrinos are extremely difficult to identify.
The only hope of such discovery channels is, perhaps, via neutrino scattering
experiments on either electron or nucleon targets, using well
understood neutrino beams.  We point out that any
neutrino/anti-neutrino cross contamination induces ambiguity onto
the total lepton number of the incident beam and would serve as a
crippling source of background for LNV searches. This reasoning
rules out conventional superbeam \cite{SupBeam} facilities as well
as proposed neutrino factories \cite{NuFact}, which contain both
neutrino and anti-neutrino components, but does suggest modest
possibilities for future beta-beams \cite{BetaBeam}. Given projected beta-beam luminosities and energies along
with the derived cutoff scales $\Lambda_{\nu}$, it seems unlikely that LNV
can be observed in such experiments.  Another possible discovery
mode  involves only two external
state quarks and an associated  gauge boson as in
the sample process $\ell_{\alpha}  \ell_{\beta} \rightarrow d \bar{u} + W^-$.
It turns out that the rates for such processes are generally suppressed for the majority of operators involving six
fermion fields, as we are trading a phase space suppression for a
stronger loop suppression. For those operators with only four
fermion fields, the situation is not as straightforward and, in some
cases, the three particle final state is preferred.  Typically, the neutrino mass induced
cutoff scales of those operators are high $(\Lambda_{\nu} \gg 100~\rm{TeV})$, so
it would be quite difficult to observe such effects.  Of course, any $W$-boson final state
will either promptly decay leptonically, yielding missing energy and unknown
total lepton number, or hadronically, reducing the reaction back to
that of the golden mode.

Another possibility is to replace two or
more of the external quark states in Eq.~(\ref{eq:GoldenCh}) with
leptons in such a way as to preserve charge, baryon number, and
$\Delta L = 2$ constraints. While many operators favor
this structure, a little thought reveals that at least one external
neutrino state is always present, which leaves only a missing energy
signature, and little means of lepton number identification in a
detector. Such events would not be clean, but of course, three
final-state charged leptons and missing energy are enough to
extract the existence of at least $\Delta L=1$ LNV, provided that the
number of invisible states is known to be no greater than one. This
last requirement is difficult to achieve in the presence of the large
backgrounds and the limited statistics expected at future
collider facilities, but should still be possible given a concrete
model probed near resonance (see for example \cite{EmEm2munu}).
Therefore, while important and potentially observable, this mode is
not generally the best place to look for LNV and is
neglected in the remainder of our analysis. From this perspective,
the only other relevant channel of LNV discovery is related to $W$
and $Z$ rare decays into final states with non-zero total
lepton number. This possibility is briefly addressed in Sec.~\ref{subsec:OtherProcesses}.

\subsection{Neutrinoless Double-Beta Decay} \label{subsec:bb0nu}

Here we probe the expectations for neutrinoless double-beta decay
$(\beta\beta0\nu)$ for each operator listed in Table
\ref{tab:AllOps}.  $\beta\beta0\nu$ is the LNV ($\Delta L=2$)
process where, within a nucleus, two down quarks convert into two up
quarks with the emission of two electrons but no neutrinos, or in
the language of nuclear physics $(A,Z)\rightarrow (A,Z+2)+e^-e^-$.
See \cite{BB0n_review} and references therein for a comprehensive
review. While precise computations of nuclear matrix elements are essential for
making detailed predictions
 \cite{NuMatrixEl}, the minimal parton-level description
given above is adequate for the purposes of this study. There is a
continuing legacy of cutting edge experiments designed to search for
$\beta\beta0\nu$ with no success to date.\footnote{There is currently
a positive report of $\beta\beta0\nu$ at the $4.2\sigma$ level by a subset of
the Heidelberg-Moscow collaboration \cite{BB0nPosSig}. They report a
measured half-life of $1.74^{+0.18}_{-0.16}\times
10^{21}~\rm{years}$ which maps to $m_{ee}^{\rm eff}\sim
(0.2-0.6)~\rm{eV}$. We choose to neglect this controversial result, which is still
awaiting independent conformation.} Currently the $^{76}$Ge half-life for this
process is bounded to be greater than $1.9\times 10^{25}~\rm{yr}$
and $1.57\times 10^{25}~\rm{yr}$ at $90\%$ confidence by the
Heidelberg-Moscow \cite{HMoscowCurrent} and IGEX \cite{IGEX}
experiments, respectively. Future experiments are poised to improve
these limits (for several different nuclei) by a couple of orders of magnitude within the next five to
ten years \cite{BB0nFuture}.

If one assumes
that $\beta\beta0\nu$ proceeds via the exchange of light Majorana
neutrinos, its amplitude is proportional to the $ee$ element of
the Majorana neutrino mass matrix,
\begin{equation}
m_{ee} = \sum_{i=1}^3 m_i U_{ei}^2, \label{eq:std_Mee}
\end{equation}
where $m_i$  are the neutrino masses and $U_{ei}$ are elements of
the leptonic mixing matrix.  With this, one can extract the
upper bound $m_{ee}<0.35~{\rm eV}$ (90\% confidence level bound, \cite{PDG})
from
current experiments while next-generation experiments are aiming at
$m_{ee}\gtrsim 0.05$~eV\footnote{The parameter
change from half-life to $m_{ee}$ depends heavily on nuclear matrix
element calculations.
 Current calculations induce
  an uncertainty of less then a factor of four on $m_{ee}$ for most parent isotopes \cite{NuMatrixEl}.}
   \cite{BB0nFuture}.
 In general, LNV new physics will lead to additional contributions to $\beta\beta0\nu$, most of
 which are not proportional to $m_{ee}$. However, the amplitude for  $\beta\beta0\nu$
can still be expressed in terms of an effective $m_{ee}$, $m_{ee}^{\rm eff}$, which is an operator-specific
quantity that will be used to analyze  new models of LNV.

\begin{figure}
\begin{center}
\includegraphics[scale=0.8]{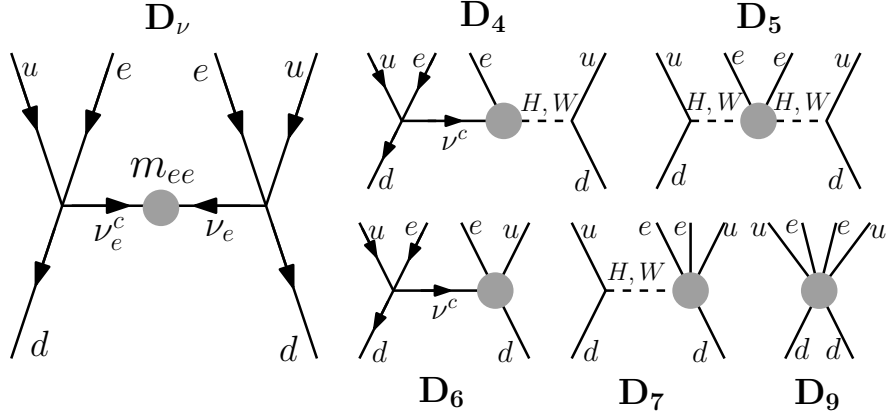}
\caption{The parton level
Feynman diagrams contributing to neutrinoless double-beta decay,
labeled by the dimension of the underlying lepton number violating
interaction, indicated by gray dots.  Each diagram is generated,
at some order in perturbation theory, by all analyzed interactions,
but estimates of their magnitudes depend heavily on the details of
the operators, including their associated scale $\Lambda_{\nu}$, fermion content and
helicity structure.} \label{fig:Bb0nOps}
\end{center}
\end{figure}

Here, we define six different ``classes'' of diagrams one can construct out of LNV irrelevant
operators that
contribute to $\beta\beta 0\nu$ at the parton level. These are illustrated
in Fig.~\ref{fig:Bb0nOps}, and classified by the dimension of the
generated LNV interaction, depicted by large gray dots.
In order to unambiguously separate the different classes, note that the grey circles are
defined in such a way that all fermion and Higgs
legs that come out of it are part of the ``parent'' operator ${\cal O}$
(and not attached on via reducible SM vertices), while all other interactions are SM vertices.
The dots should be viewed as hiding the
underlying LNV interactions. In general, they contain a mixture of
coupling constants and loop factors that must be evaluated
explicitly for each diagram.  It is important to emphasize that the contribution of a  generic operator ${\cal O}$
to $\beta\beta0\nu$ will consist of contributions from all different classes, while usually dominated by one of them.
We show the lepton number conserving
electroweak vertices (point-like) as effective four-fermion
interactions, justified by the low energy scale of nuclear beta
decays.
The dotted lines indicate the exchange of $W$-bosons, labeled by
$W$ and $H$ (charged Higgs goldstone boson).
Helicity arrows are explicitly included where uniquely determined, implying
that the arrowless legs can have any helicity.

${D_\nu}$ describes
the standard scenario of $\beta\beta0\nu$ mediated by light Majorana
neutrinos. It is simply two electroweak vertices held together by a
Majorana mass term on which two neutrinos are annihilated.  The
amplitude for this diagram is proportional to $m_{ee}$, as defined in Eq.~(\ref{eq:std_Mee}).
The dependence on such a neutrino mass is
intuitively clear considering the need for a helicity flip on the
internal neutrino line.  The remaining diagrams are qualitatively different
from this standard case. Most importantly, none of them require
``helicity flips'' and are therefore not directly proportional to neutrino masses.
They are, however, proportional to
inverse powers of the new mass scale $\Lambda_{\nu}$, and hence also suppressed.  These
effects are not entirely independent, since the value of $\Lambda_{\nu}$ was extracted
from the requirement that  neutrino masses are small, but correlations are relaxed
enough to allow nontrivial consequences. It is this partial decoupling
from neutrino masses that allows larger than naively expected contributions to  $\beta\beta0\nu$
from some of the LNV irrelevant operators. Before proceeding, we make
the trivial observation that the amplitudes following from
${D_\nu}$, ${D_4}$ and ${D_5}$ are additionally proportional to two powers of
CKM matrix elements, namely $|V_{ud}|^2$, whereas ${D_6}$
and ${D_7}$ are only proportional to one power of
$V_{ud}$.\footnote{This is true provided that we assume no flavor
structure for the underlying operator, or, equivalently, that all
dimensionless coupling constants are order one.  If one is motivated
by experiment to postulate a minimally flavor violating scenario, to
perhaps ease constraints from flavor changing neutral currents, the
statement must be modified accordingly.}  The tree-level diagram
${D_9}$ has no CKM ``suppression.''  While this is a purely
academic fact in the case of $\beta\beta0\nu$ ($|V_{ud}|\sim 1$!),  it leads to important consequences
for analogous rare decays that depend on the much smaller off-diagonal CKM matrix elements. We will return to these in the next subsection.

For a given operator, the relative size of each diagram's
contribution to the total decay rate depends on many factors
including the operator's dimension, scale, fermion content and
helicity structure.  The dominant contributions must be calculated
on a case by case basis.  Generally, the high scale operators
$(\Lambda_{\nu} \gtrsim 10~\rm{TeV})$ are dominated by the two dimension-four
diagrams ${D_\nu}$ and ${D_4}$ since many factors of $\Lambda_{\nu}$ will
be canceled by divergent loops inside the gray dots thereby
minimizing the $1/\Lambda_{\nu}$ suppression.  All else being equal, ${D_\nu}$ is
the strongest of the pair since it is enhanced by $\sim Q^{-2}$ from
the two propagating neutrino lines as opposed to only $\sim Q^{-1}$
for the one neutrino case shown in diagram $D_4$. For
those operators with no tree-level $\nu\nu$ field content, ${D_4}$
can still be very important, but its dominance is nevertheless
rare. As discussed in Sec.~\ref{sec:Scale}, these are precisely
the operators that have the greatest loop suppressions and
consequently lower energy scales suggesting the need for diagrams
beyond ${D_4}$. The effects of low cutoff scale operators $(\Lambda_{\nu} \lesssim 1~\rm{TeV})$
are not severely suppressed by $1/\Lambda_{\nu}$ (by definition), so the dominant
diagrams will typically be of the highest dimension allowed by the
tree-level structure of the operator.  For such low scales and for
operators of the following schematic form
$dd\bar{u}\bar{u}\bar{e}\bar{e}$ (dimension $9$) or
$dd\bar{u}\bar{u}\bar{e}\bar{e}H_0H_0$ (dimension $11$), ${D_9}$
always dominates the $\beta\beta0\nu$ rate yielding amplitudes
proportional to $1/\Lambda_{\nu}^5$ and $v^2/\Lambda_{\nu}^7$, respectively. For
intermediate scales, and when the operator's field content does not
directly support $\beta\beta0\nu$ due to lack of quark fields, the
situation is not as straight forward and one must perform the
relevant computations to determine the dominant diagrams.  Still, it
should be noted that diagrams containing internally propagating
neutrinos are enhanced by inverse powers of $Q$ and maintain a
slight advantage over their neutrinoless counterparts.  One can thus
generally expect diagram $D_6$ to dominate the decay rates for low
$\Lambda_{\nu}$ scale operators when $D_9$ is suppressed.  The opposite is true for
interactions taking place at higher energies in, for example, next-generation colliders, as discussed in Sec.~\ref{subsec:Collider}.

Since each diagram in Fig.~\ref{fig:Bb0nOps} can have different
external helicity structures, the different  contributions to the total rate
will be added incoherently, thus eliminating the effects of
interference.  There are some case specific coherent
contributions that we neglected in our treatment since most rates
are dominated by a single diagram. Another potential difference among the different
contributions is related to nuclear
matrix element calculations: can the calculations
done assuming $\beta\beta0\nu$ via the standard light Majorana
neutrino exchange scenario of diagram ${D_\nu}$ be applied to the
more general cases encountered here?  We have nothing to add to this
discussion except to naively note that there is no obvious reason
why such rates should be severely suppressed or enhanced relative to
the standard scenario. We therefore assume that all nuclear matrix
elements are identical and can be factored out of the incoherent
sum. We assume that this approximation is not more uncertain than the other sources of
uncertainty inherent to our study (likely a very safe assumption).

As drawn, each diagram $D_i$
contributes to the amplitude that
characterizes $\beta\beta0\nu$.  For example, the amplitude
associated with $D_\nu$ is proportional to
\begin{equation}
\mathcal{A}_{D_\nu} \equiv m_{ee}\frac{|V_{ud}|^2 G_F^2}{Q^2},
 \label{eq:ad}
 \end{equation}
where $G_F$ is the Fermi constant. The remaining diagrams will contribute with $\mathcal{A}_{D_i}\propto\zeta(v,Q)\Lambda^{4-i}$, up to a dimensionless coefficient
containing various numerical/loop factors, as well as general scale
depencies parameterized by some power of the ratio $v/\Lambda$.  The
function $\zeta(v,Q)$ has mass dimension ${i-9}$ so that all ${\cal A}_{D_i}$ have the same mass dimension.  Note that
all aspects of $\mathcal{A}_{D_i}$ are calculable given a LNV
operator and diagram. We can analyze each operator in
terms of an effective $m_{ee}^{\rm eff}$,  defined in
terms of the underlying dimension nine amplitude $\mathcal{A}_{D_i}$ by
\begin{equation}
m_{ee}^{\rm eff} = \frac{Q^2}{G_F^2 |V_{ud}|^2}\sqrt{\sum_i
\mathcal{A}_{D_i}^2},
 \label{eq:MeeEff}
\end{equation}
where $i$ runs over the set $\{\nu,4,5,6,7,9\}$  that labels the
diagrams shown in Fig.~\ref{fig:Bb0nOps}, and $Q\sim 50~\rm{MeV}$
is the typical momentum transfer in $\beta\beta0\nu$. $m_{ee}^{\rm eff}$
can be directly compared with experiment and used to make prediction
for future observations. A few comments are in order regarding this
quantity. First, it is a useful derived object that has no direct
connection to a real neutrino mass and is valid to
arbitrarily large values. Note that in the case of Majorana neutrino exchange,
$m_{ee}^{\rm eff}=m_{ee}$ only if $m_{ee}\ll Q$.  When neutrino masses are
greater than Q, $m_{ee}^{\rm eff}\propto 1/m$.   Our definition of $m_{ee}^{\rm eff}$ also conforms
to the use of large effective masses in \cite{LNVUpperBounds}.  The
second comment is that, unlike the case of $m_{ee}$, which is valid
for any process involving the exchange of electron-like Majorana
neutrinos, $m_{ee}^{\rm eff}$ is case specific.  It must be calculated
separately for each process, as each one, in general, is composed of
different diagrams.  In particular, the calculations of the
effective mass for $\beta\beta0\nu$ expressed here are not directly
applicable to other LNV processes and should not be interpreted as such.
\begin{figure}
\begin{center}
\includegraphics[angle=270,scale=.63]{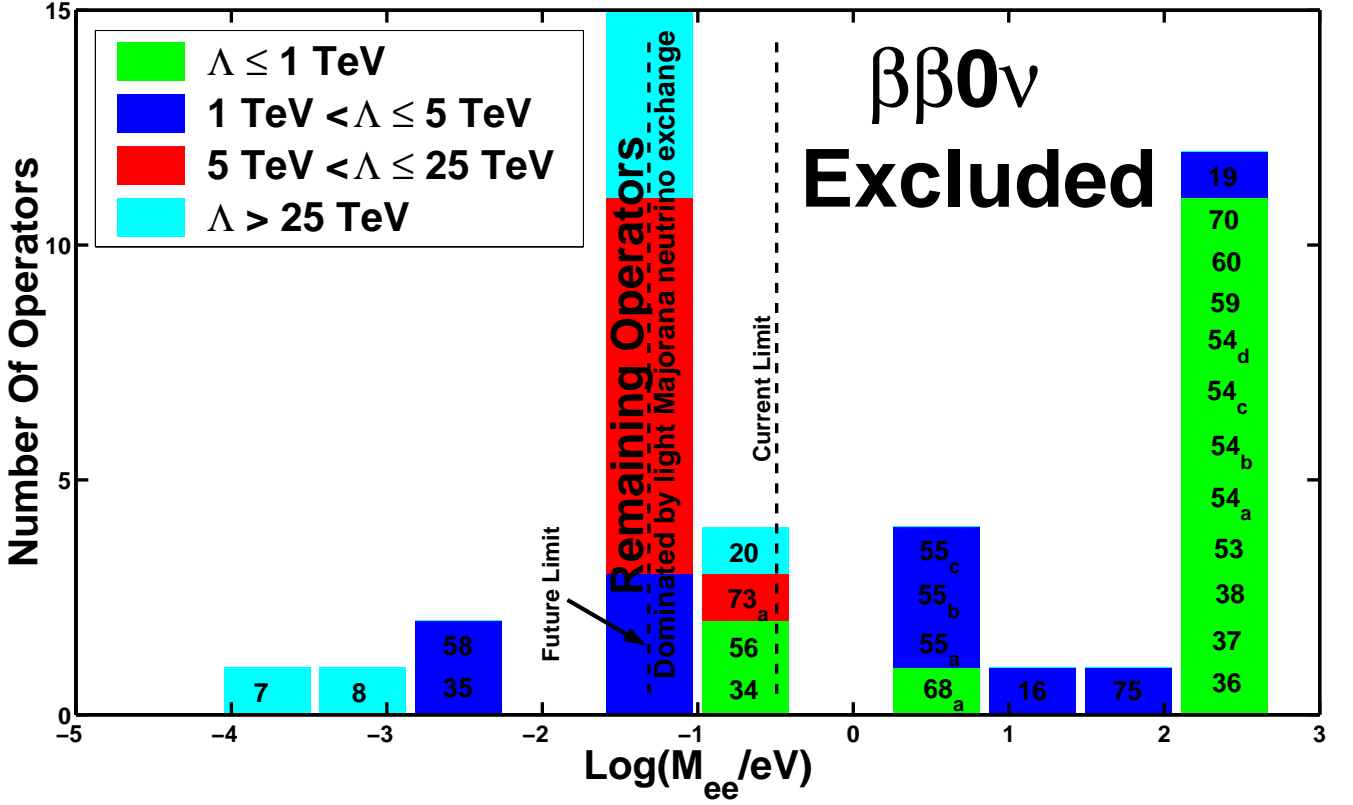}
\caption{$m_{ee}^{\rm eff}$ distribution derived for the neutrinoless double-beta
decay process as described in the text.  The calculations were made
assuming the scales $\Lambda_{\nu}$ derived in Sec.~\ref{sec:Scale}, as
well as universally order one coupling constants. The histogram bars
are labeled explicitly with operator names and color-coded by their
cutoff scales.  Also shown in light gray is the region probed
by next-generation experiments.  The vertical axis is
truncated at 15 operators to best display the relevant features of the plot.}
\label{fig:BBEX}
\end{center}
\end{figure}

The $m_{ee}^{\rm eff}$ distribution extracted from all operators is shown in Fig.~\ref{fig:BBEX}
assuming the scales $\Lambda_{\nu}$ derived in Sec.~\ref{sec:Scale}
and color-coded for convenience within the histogram. Specifically,
we indicate in green the operators that are characterized by sub-TeV
scales and thus accessible to next-generation experiments via direct
production.  The blue and red operators are characterized by scales
between $(1-5)~\rm{TeV}$ and $(5-25)~\rm{TeV}$ respectively, where
virtual effects should be most important for collider searches. The
majority of operators, shown in cyan, are suppressed by scales
greater than $25~\rm{TeV}$ and are hence quite difficult to observe
in other search modes.  We also explicitly label each operator
within the histogram bars for easy identification and comparison.
One should notice the expected general trend that increasing
$\Lambda_{\nu}$ leads to a decrease in $m_{ee}^{\rm eff}$ and vice-versa. The
vertical axis is truncated at $15$ operators, as the
bar near $0.05~\rm{eV}$, dominated by the light Majorana neutrino
exchange described above, would extend to nearly $100$ operators.
With broken vertical lines, we indicate the current 90\% upper bound \cite{PDG},
 $m_{ee}^{\rm eff} = 0.35~\rm{eV}$,
 and the potential reach of future experiments.

This distribution, which spans over six orders of magnitude (from
$10^{-4}$~eV till $10^{2}$~eV), reveals many important features of the
effective operator set.  Beginning at the largest $m_{ee}^{\rm eff}$
values, we find that the twelve operators appearing near
$300~\rm{eV}$ all have the expected common feature of low energy
scales, including $\mathcal{O}_{19}$ with $\Lambda_{\nu}$ only just above
the $1~\rm{TeV}$ mark.  Additionally, the contribution of the majority of these
operators to $\beta\beta0\nu$ is dominated by the tree-level $D_9$ diagram.  The
exceptions are $\mathcal{O}_{54_{c,d}}$ and $\mathcal{O}_{70}$,
all of which are characterized by sub
$0.5~\rm{TeV}$ scales and dominated by diagram $D_6$.  Consequently,
these are subject to a loop and Yukawa/gauge\footnote{As it turns
out these are all suppressed by a single bottom quark Yukawa
coupling as well as two powers of the $SU(2)_L$ gauge coupling $g$,
but this fact cannot be deduced from Fig.~\ref{fig:BBEX} alone.}
suppression relative to their $D_9$ dominated cousins, but the
difference is not visible given the resolution of the figure.  It is interesting to note that these three operators  have the correct quark and lepton content for large
$\beta\beta0\nu$, but their $SU(2)_L$ gauge structures forbid
large tree-level contributions. Similarly, operators
$\mathcal{O}_{16}$, $\mathcal{O}_{55_{a,b,c}}$,
$\mathcal{O}_{68_a}$, and
$\mathcal{O}_{75}$ are also dominated by diagram $D_6$ accompanied by
slightly higher cutoff scales.  This drives down $m_{ee}^{\rm eff}$
significantly considering the leading one-loop scale suppressions of
$\Lambda^{-5}$ and $\Lambda^{-3}$ for the dimension-eleven and dimension-nine
operators respectively.  We point out that operators
$\mathcal{O}_{54_{a,b,c,d}}$ and $\mathcal{O}_{55_{a,b,c}}$ yield almost
identical expressions for their respective $\beta\beta0\nu$
amplitudes (as well as their radiatively generated neutrino mass
expressions) with up and down quark Yukawa couplings exchanged.
While this action enhances most of the $\mathcal{O}_{55}$ $\beta\beta0\nu$
couplings relative to those of $\mathcal{O}_{54}$, it also raises
the $\mathcal{O}_{55}$ $\Lambda_{\nu}$ scale by nearly a factor of four and thus
drives $m_{ee}^{\rm eff}$ down by orders of magnitude.

The remaining
operators all predict $m_{ee}^{\rm eff} < 1~\rm{eV}$, close to current
experimental bounds. The histogram bar near $0.1~\rm{eV}$ is
composed of operators of very different $\Lambda_{\nu}$ scales.  $\mathcal{O}_{34}$
and $\mathcal{O}_{56}$ are both characterized by low cutoff energy
scales around $0.5~\rm{TeV}$, but, due to their fermion and helicity
structure, their contributions to $\beta\beta0\nu$
are dominated by two-loop versions of diagram $D_6$.
The neutrino-mass-required cutoff for $\mathcal{O}_{73_a}$ is around $7~\rm{TeV}$ and its contribution
to $\beta\beta0\nu$ is also dominated by diagram $D_6$. In this case, however, the
two-loop version turns out to be larger than the allowed one-loop amplitude
due to strong scale suppressions (the added loop reduces the
cutoff dependency from $\Lambda^{-5}$ to $\Lambda^{-3}$). This behavior is
characteristic of operators with a larger value of $\Lambda_{\nu}$.  The $\Lambda_{\nu}
= 40~\rm{TeV}$ operator $\mathcal{O}_{20}$, defines the lower edge
of this histogram bar.  It is dominated by the one-loop diagram
$D_6$ enhanced by a top quark Yukawa coupling and, being a dimension
nine operator, is only suppressed by $\Lambda^{-3}$ from the start.
The next bar down contains operators dominated by $D_\nu$.  Most of
these are suppressed by a very high energy scale, but a small subset is
characterized by scales $\Lambda_{\nu} < 25~\rm{TeV}$.  In particular operators
$\mathcal{O}_{17}$, $\mathcal{O}_{18}$ and $\mathcal{O}_{57}$ are
all cutoff at $2~\rm{TeV}$ but, due to their fermion content they
cannot participate in any of the non-standard interactions of
Fig.~\ref{fig:Bb0nOps} at a low enough order in perturbation theory.  Similarly,
the intermediately scaled operators $\mathcal{O}_{43_{a,b,c}}$,
$\mathcal{O}_{50}$, $\mathcal{O}_{52}$, $\mathcal{O}_{62}$,
$\mathcal{O}_{65}$, and $\mathcal{O}_{69_b}$ have either the wrong
fermion content or gauge structure to enhance any of the
$\beta\beta0\nu$ diagrams (other than $D_\nu$) to an observable
level.  These operators are important because their minimal forms
are experimentally unconstrained yet still potentially observable to
both next-generation $\beta\beta0\nu$ and collider experiments. The
remaining histogram bars with $m_{ee}^{\rm eff} < 10^{-2}~\rm{eV}$ are
not accessible to $\beta\beta0\nu$ experiments in the foreseeable future.
Each of these diagrams are dominated by $D_\nu$, either due to high
suppression scales as in the case of $\mathcal{O}_{7}$ and
$\mathcal{O}_{8}$, or, as in $\mathcal{O}_{35}$ and
$\mathcal{O}_{58}$, the operator's fermion content simply disfavors
other contributions to the $\beta\beta0\nu$ amplitude. It
is the general form of the neutrino mass matrix derived in Table
\ref{tab:AllOps}, where we see that $m_{ee}\propto y_e$, that drives
these operators away from their peers near $m_{ee}^{\rm eff} =
0.05~\rm{eV}$.  It is unfortunate that the two ``low'' dimensionality operators
$\mathcal{O}_{7}$ and $\mathcal{O}_{8}$ are cutoff by energy scales $\Lambda_{\nu}$
in excess of $100~\rm{TeV}$ and are hence invisible to any direct
probe. If either of these operators have anything to do with nature,
it is unlikely that LNV will be observed in the foreseeable future
in \emph{any} experiment.  On the other hand,  any observation of
LNV will rule out these types of scenarios. Additionally, as will
become clear shortly in Sec.~\ref{sec:Oscillations}, current
neutrino oscillation data already marginally disfavor such
operators and have ample room to tighten constraints in the near
future.

It is interesting to point out that the lower boundary of the currently
excluded region falls within the $m_{ee}^{\rm eff}$ distribution,
suggesting exciting prospects for the future.  That being
said, one should not read too much into current and future null
results as, for most operators, relatively small cancelations and
order one factors, not accounted for here, can push the relevant
rates below the observable level depending on the underlying
ultraviolet theory.
On the other hand, one is allowed to interpret that
operators that lead to  $m_{ee}^{\rm eff} \gtrsim 10$~eV are severely constrained (if not
ruled out) as proper explanations for neutrino masses if one assumes the new physics
to be flavor ``indifferent'' -- order one factors cannot be evoked to save the scenario.
Once this assumption is dropped, however, it is quite easy to ``fix'' these scenarios, since the large $\beta\beta0\nu$
rate is a
direct consequence of the universal order one couplings and the relatively low cutoff energy
scale $\Lambda_{\nu}$.  For example, one can suppress the
coupling of new physics to first generation fermions (compared to second and third generation fermions), thereby
suppressing the worrisome diagrams of Fig.~\ref{fig:Bb0nOps}. This will have
little effect on the relation between $\Lambda$ and the neutrino masses, discussed in Sec.~\ref{sec:Scale},
since these are either generation independent or highly reliant on
third generation Yukawa couplings. Of course, by combining
$\beta\beta0\nu$ searches with other probes we
can obtain a much better idea of the origins of LNV as well as the
relevant model(s), if any, chosen by nature.

\setcounter{footnote}{0}
\subsection{Other Rare LNV Decay Processes}
\label{subsec:OtherProcesses}

Most of the qualitative discussions of Sec.~\ref{subsec:bb0nu}, devoted to $\beta\beta0\nu$, can be directly
applied to other rare decay processes with the same underlying
kernel interaction described by Eq.~(\ref{eq:GoldenCh}). For such
processes one need only analyze simple variants of the diagrams
listed in Fig.~\ref{fig:Bb0nOps}, using crossing
amplitude symmetries to account for the needed initial
and final state fermions. Other
factors must be added to the various electroweak vertices to
account for quark flavor mixing. The requisite CKM matrix elements
can highly suppress many diagrams for processes involving
cross-generational quark couplings.  In fact, only tree-level $D_9$
diagrams are  safe from such suppressions. Next, and most
importantly, one must include the appropriate characteristic
momentum transfer $Q$ of the new system. Specific rates are highly
dependent on this quantity as effective operator cross-sections
typically grow with some power of $Q$. The particular exponent of
the power law depends on the  diagram, but naive dimensional analysis
dictates that $\Gamma \propto Q^{12}$ for
diagram $D_9$, rendering it highly dependent on a reaction's energy
transfer.  The fact that each diagram varies with $Q$ in a different
way implies that predicting the dominant contributions to a given
process is non-trivial and must be addressed quantitatively.
 Finally, in the cases of hadronic decays, one must also
account for initial/final state matrix elements.  We assume that all
factors can be simply estimated on dimensional grounds.

Unlike the $\beta\beta0\nu$ case, some meson
decay modes proceeding via new LNV tensor interactions are expected
to be suppressed.  Such processes are one
instance in our analysis where an operator's Lorentz structure can
qualitatively affect expected LNV decay rates.  One can understand this by
considering a meson decay mediated by a new tensor particle. The
parton level interaction has the form $(\bar{u}\sigma_{\mu\nu}d)
T^{\mu\nu}$ where the initial state quarks are explicitly shown and
all other fields are contained in the tensor $T^{\mu\nu}$. Following
the standard procedure we factor out the hadronic structure in the
form of a free decay constant and write the amplitude as generally
allowed by Lorentz invariance in terms of the external state's
four-momentum. Due to the antisymmetry of $\sigma_{\mu\nu}$, this amplitude
vanishes to first order.  Non-zero contributions to this decay
mode must necessarily involve individual parton momenta and are
therefore suppressed relative to the usual vector-like decay
calculations. From this, it is clear that models of LNV containing
tensor couplings will often evade the predictions and bounds of this
section. Tensor operators will mediate LNV meson decays into more
complicated final states (one may include, say, initial/final state radiation).
Associated rates are, however, subject to additional gauge coupling and phase space
suppression that  tend to further
reduce the already tiny LNV rates beyond any hope of detection.
\begin{figure}
\begin{center}
\includegraphics[scale=0.8]{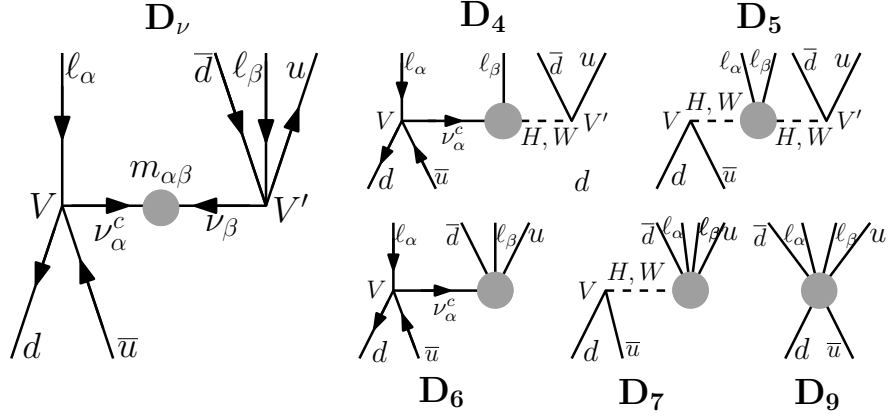}
\caption{The parton level
Feynman diagrams contributing to rare LNV meson decay labeled by the
dimension of the underlying lepton number violating interaction,
indicated by gray dots.  Each diagram is generated, at some order in
perturbation theory, by all analyzed interaction, but estimates of
their magnitudes depend heavily on the details of the operators,
including their associated scale $\Lambda_{\nu}$, fermion content, and helicity
structure.} \label{fig:MesonOps}
\end{center}
\end{figure}

Rare LNV meson decays have been experimentally pursued
for many years \cite{PDG}.  Here, we focus on the $\Delta L=2$
processes $M^\prime \rightarrow M + \ell^{\pm}_{\alpha} \ell^{\pm}_{\beta}$, where
$M^\prime$ and $M$ are the initial and final states mesons
respectively and the $\ell $s represent like-sign lepton pairs of arbitrary flavor.
Electric charge conservation dictates that $M^\prime$
and $M$  have equal and opposite charge.  Here we take each meson to consist of a color singlet
up-type/antidown-type bound state\footnote{For simplicity we assume
that both the process $M^\prime \rightarrow M + \ell _\alpha \ell
_\beta $ and its conjugate have similar amplitudes and therefore
treat them symmetrically.  Large CP-violating effects can invalidate
this assumption.} and factor out all long distance hadronic
effects.  In this way we can view the meson decay process as
$d\bar{u} \rightarrow \ell _\alpha \ell _\beta + \bar{d}u$ for all
up-type and down-type quark flavor combinations.  The effective LNV
diagrams contributing to this process are shown in Fig.
\ref{fig:MesonOps} with the same naming scheme as their analogs in
Fig.~\ref{fig:Bb0nOps}.  Here, $V$ and $V^\prime$ denote potentially distinct elements of the quark mixing matrix.  We additionally point out the
potential dependency on all entries of the Majorana neutrino mass matrix elements
$m_{\alpha\beta}$ in diagram $D_\nu$, as opposed to the $\beta\beta0\nu$ case where $D_{\nu}$
depends only on $m_{ee}$.  These processes probe combinations of the neutrino masses that
are naively unconstrained by $\beta\beta0\nu$ \cite{Flanz:1999ah}.
In general, the varied flavor structures encountered in meson decays allow for experimental probes into new
physics couplings across the fermion generations.
We pointed out earlier that some of the LNV operators lead to unacceptably large rates for $\beta\beta0\nu$ unless first generation
quarks  participate in the new interactions with severely suppressed couplings (compared with second and third generation
quarks). If such a scenario is realized in nature, rare $D$ or $B$ decays may be much more frequent than naive expectations.
 For
this reason, improving rare decay sensitivities to all channels is
essential to completely constrain models of new LNV physics beyond
the minimal framework analyzed here.

\begin{figure}
\begin{center}
\includegraphics[angle=0,scale=.72]{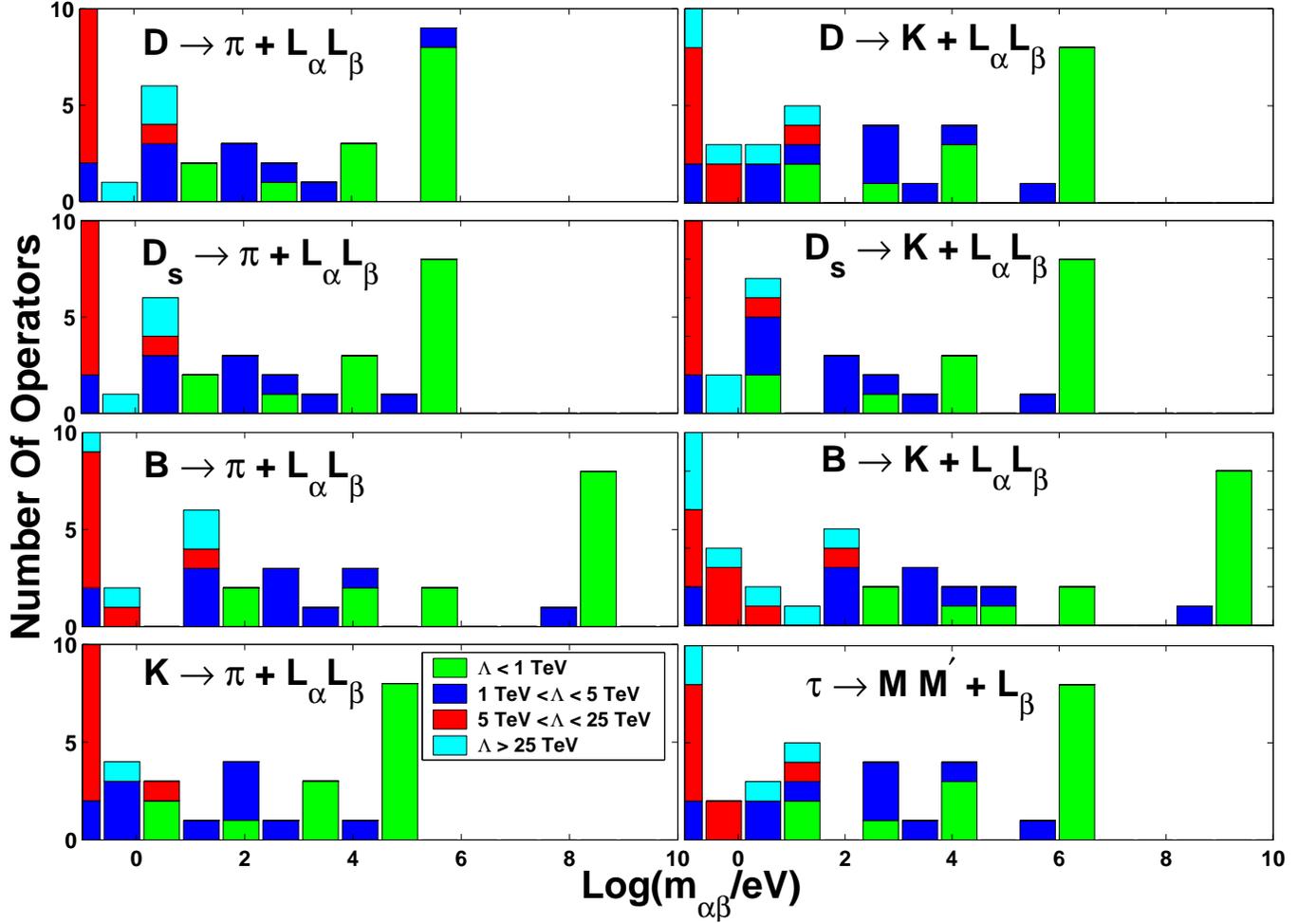}
\caption{$m_{\alpha\beta}^{\rm eff}$ distribution for several rare
LNV meson and $\tau$ decays.  Calculations assumed the charge lepton
flavors $\ell _\alpha\ell _\beta = \mu e$, while the $\tau$ decay
histogram (lower right-hand panel) was obtained assuming the final
state mesons $M M^\prime = KK$.
The histogram bars are color-coded
by suppression scale.  Current bounds on
these processes are typically above $1~\rm{TeV}$ and are
 not visible at these small scales.} \label{fig:Meson}
\end{center}
\end{figure}

Reference \cite{LNVUpperBounds}
summarizes LNV upper bounds on all of these processes in terms of the
effective Majorana neutrino mass matrix element
$m_{\alpha\beta}^{\rm eff}$ that one would extract from observation
assuming that all decay rates are dominated by the light neutrino
exchange shown in $D_\nu$.  Hence, we can compare operator
expectations with current experimental limits in exactly the same
way as was done in Sec.~\ref{subsec:bb0nu}.  For a given LNV meson decay,
$m^{\rm eff}_{\alpha\beta}$ is defined from the contribution of the different classes
of diagrams to the rare meson decay in question, exactly as $m_{ee}^{\rm eff}$ was defined in the
previous subsection (see Eqs.(\ref{eq:ad},\ref{eq:MeeEff})).
Direct estimates for  different  process reveal $m^{\rm eff}$ distributions
similar to that for $m_{ee}^{\rm eff}$ depicted Fig.~\ref{fig:BBEX}, up to ``rescalings'' that reflect the different
kinematics and the presence of small CKM mixing matrix elements. Results are
summarized in Fig.~\ref{fig:Meson} for a representative sample of
charged meson decays. Each histogram is labeled by its
associated decay mode and is color-coded to indicate the neutrino-mass constrained
cutoff scale $\Lambda_{\nu}$ of the different LNV effective operators.  For
simplicity, we refrain from listing operator names on the
individual histogram bars (as opposed to what was done in Fig.~\ref{fig:BBEX}). The ``operator ordering'' is
very similar to that of Fig.~\ref{fig:BBEX}, especially in
the low $\Lambda_{\nu}$ scale, high effective mass regime where decay rate
predictions are particularly important. Note that the horizontal
 axes are relatively fixed for easy comparison and that the vertical direction
is truncated and does not reflect the true ``height'' of the lowest mass
bar (order one hundred operators).

Specifically, we present effective Majorana neutrino mass
distributions for the processes, reading down the panels from left
to right, $D\rightarrow \pi + \ell^{\pm}_{\alpha} \ell^{\pm}_{\beta}$,
$D\rightarrow K + \ell^{\pm}_{\alpha} \ell^{\pm}_{\beta}$, $D_s \rightarrow \pi +
\ell^{\pm}_{\alpha} \ell^{\pm}_{\beta}$, $D_s \rightarrow K + \ell^{\pm}_\alpha \ell^{\pm}
_\beta$, $B\rightarrow \pi + \ell^{\pm}_{\alpha} \ell^{\pm}_{\beta}$,
$B\rightarrow K + \ell^{\pm}_{\alpha} \ell^{\pm}_{\beta}$, $K\rightarrow \pi +
\ell^{\pm}_{\alpha} \ell^{\pm}_{\beta}$, as well as the rare $\tau$ decay, $\tau^{\pm}
\rightarrow M M^\prime + \ell^{\mp}_{\beta}$.\footnote{The actual
calculations displayed in Fig.~\ref{fig:Meson} assumed the charge lepton
flavors $\ell _\alpha\ell _\beta = \mu e$, while the $\tau$ decay
histogram (lower right-hand panel) was produced assuming the final
state mesons $M M^\prime = KK$.}  Here the final state leptons can
be of any flavor allowed by energy conservation. Since, as previously discussed and explicitly
verified numerically, the specific details of the distributions are
mainly dictated by kinematics and CKM matrix elements, these
results are robust under changes in the final state lepton flavors.
The $\tau$ decay distribution shown in the lower right
panel is representative of all possible decay products including
first and second generation charged leptons and light meson states.
One should notice the expected general operator trend within each
histogram as the characteristic cutoff scale is decreased, as well
as the expected peaks near $0.05~\rm{eV}$ dominated by light
Majorana neutrino exchange. Additionally, each distribution is much
``broader'' than the one in Fig.~\ref{fig:BBEX}. This observation
exemplifies the fact that effective mass calculations depend
critically on the underlying process. Indeed, maximum $m_{ee}^{\rm
eff}$ values can reach nearly $10^{10}~\rm{eV}$ for the $B^+
\rightarrow K^- + e^+e^+$ decay but only $10^3~\rm{eV}$ for
$\beta\beta0\nu$. Current upper bounds for $m^{\rm eff}$ from these
processes, mostly well above one TeV, are well beyond the largest
operator predictions here, ranging from $m_{e\mu}^{\rm eff} <
0.09~{\rm TeV}$ for the case of $K^+ \rightarrow \pi^- e^+ \mu^+$ to
$m_{\mu\mu}^{\rm eff} < 1800~{\rm TeV}$ for the case of $B^+
\rightarrow K^- \mu^+ \mu^+$ \cite{LNVUpperBounds}.  It is curious
that the best meson decay bounds come from rare LNV kaon process
but, as can be seen in the lower left panel of Fig.~\ref{fig:Meson},
these yield by far the lowest predictions. Future
experiments have the potential for observing LNV for a select few
operators only provided vast improvements in meson production
luminosities. Current and upgraded B-factories \cite{BFactories} are
expected to provide the most significant improvements, considering
the large derived $B$-meson effective masses shown in
Fig.~\ref{fig:Meson}.  Still, the best cases from the figure yield
only the tiny branching fraction $1.8\times 10^{-17}$ for the case
of the rare decay $B^+ \rightarrow \pi^- e^+ \mu^+$, nearly eleven
orders of magnitude below the current experimental limit of
$1.3\times 10^{-6}$ \cite{PDG}.

Another possible search mode involves the decay of the $Z$-boson into
LNV final states.  The dominant contributions to this process are
generally unrelated to the reactions summarized in
Eq.~(\ref{eq:GoldenCh}) and shown schematically in Figs.~\ref{fig:Bb0nOps} and \ref{fig:MesonOps}.  While there is a
slight connection between them as one can always attach a $Z$-boson to
various fermion lines in each diagram, there are potentially large
lower order contributions arising within the operators themselves.
The latter, when present, can easily overtake the
associated ``golden mode'' counterparts.  In this context, such
processes can be thought of as the decay of the longitudinally
polarized $Z$-boson. Strict bounds exist on such
decays from the LEP-I \cite{LEP} and SLD \cite{SLD} experiments.
Each element of the operator set predicts decays into final state
fermions with total lepton number $L=2$.  The
dilepton pair can be of any flavor and is generally accompanied by
two or four additional fermion states, depending on the dimension of
the operator.  We restrict our discussion to the dimension-eleven
operators comprising the majority of the sample, as these are
typically suppressed by lower cutoff $\Lambda_{\nu}$ scales and, equally important,
explicitly contain Higgs doublets in their field content.  In this case,
tree-level decays result in a six-fermion final state which suffers
from a large phase space suppression and cumbersome multiplicities
that are likely to render even the most sophisticated search ineffective.
The only possibility of this type that yields a charged dilepton
signal is $Z \rightarrow \ell^{\pm}_\alpha\ell^{\pm}_\beta q\bar{q}q\bar{q}$
(quarks of all allowed flavors implied), but many
other possibilities exist involving invisible final state neutrinos. A little
thought also reveals that closing fermion loops in an attempt to
obtain simpler final states and thus render the analysis more
tractable will necessarily result in final state neutrinos. Therefore, the
majority of the $Z$-boson LNV decay channels involve
invisible final states with practically undetermined total lepton
number.  The prospect of direct discovery by these means seems
dismal, but indirect constraints on LNV are still possible from
bounds on the $Z$-boson invisible decay width. There is currently a
statistically insignificant, but nonetheless captivating, $2\sigma$
deviation between the observed invisible decay width and SM
expectations assuming three light neutrino species \cite{PDG}. The
experimentally extracted branching ratio was found to be slightly
\emph{smaller} than its predicted value
so that a new LNV contribution of the form $Z \rightarrow \nu_\alpha \nu_\beta$
would push the invisible branching ratio in the ``wrong'' direction.
From these bounds the decay width of any new contribution to the $Z$-boson
decay is constrained to be less than $2.0~\rm{MeV}$ at the $95\%$
confidence level \cite{PDG,ZInvBound}. A quick estimate reveals that this
constrains the dominant LNV amplitudes $A_Z <
\sqrt{4\pi(2.0~{\rm MeV}/M_Z}) \sim 0.53$.  For the dimension-eleven
operators of interest, the largest possible amplitude is of order
$y^2/(16\pi^2)^2(v/\Lambda)^3$ where $y$
is an arbitrary fermion Yukawa coupling and four powers of the
cutoff scale $\Lambda$ are removed by divergences in the closed
diagram loops. The constraint above translates into
$y^2(v/\Lambda)^3 < 4.1\times 10^2$, which is
easily evaded by even the best case scenario of $y = y_{t} \approx
1$ and $\Lambda \approx v$.  Experimental bounds on $\Gamma_{inv}$
must be improved by a factor of a million before they start significantly
constraining LNV (under the assumptions made here).
This result holds for virtually all possible flavor structures.
We conclude that rare $Z$-boson decays are not
practical discovery modes for the LNV effects considered here, but look to future rare $Z$-boson decay
 studies for more information.

In a similar way, one can also dismiss the case of rare $W$-boson decays
as promising probes of LNV.  As in the $Z$-boson case, the $W$-boson can decay
into a variety of $L = 2$ final states proceeding either through
couplings to left-handed fermion lines or explicit operator
content.  Here, however, there is no six-fermion, same-sign dilepton
final state with no neutrinos due to conservation of charge and weak
isospin, so the lowest order observable mode is already loop
suppressed to $W^- \rightarrow \ell ^-_\alpha\ell ^-_\beta + q\bar{q}$.
Current $W$-boson decay bounds are far too weak
to constrain such suppressed LNV \cite{PDG} and are not likely to
improve to the level implied by the operators under consideration, which predict
the tiny decay rate $\Gamma_{LNV} \leq
m_W(4\pi)/(16\pi^2)^5(v/\Lambda)^{10}
\approx 10^{-5}~\rm{MeV}$ in the best case scenario of electroweak
scale $\Lambda_{\nu}$.  We also point out that, contrary to the $Z$-boson decay
limits, there are no robust, indirect bounds that can be used to
constrain LNV in the case of the $W$-boson.  Note that, despite dismal
prospects for gauge boson decay driven LNV discovery within the
minimal framework of ``natural'' effective operators, one can still
construct theoretically well-motivated models that will yield
observable signals. Particularly, in a weak-scale seesaw mechanism
($\mathcal{O}_1$), the new degrees of freedom,
comprised mostly of Majorana gauge singlet fermions (right-handed
neutrinos), can mediate visible, $\Delta L=2$, $W$-boson mediated processes
 with little or no scale/loop suppression.  This class of model
is analyzed in \cite{WDecay} and is exempt from the discussion
outlined here.

\subsection{Collider LNV Signatures}
\label{subsec:Collider}

If neutrino masses are a consequence of ultraviolet physics related to cutoff scales around the TeV scale,
we expect future high energy collider searches to directly access the new LNV physics.
For example, the direct, resonant, production of new states
could lead to rather spectacular signals of these models. It would also indicate the
breakdown of the effective field theory approach undertaken here. To pursue
such possibilities, one must assume a specific ultraviolet sector and
study its signatures and implications on a case by case basis.  In
the looming shadow of the LHC, \cite{LHC} and the more distant ILC
\cite{ILC}, such an analysis is highly warranted but will not be pursued
here. Instead, we
assume that the masses of new ultraviolet degrees of freedom remain out of the reach
of next-generation accelerator experiments.  Such a situation
can be easily accommodated within the context of the preceding results,
considering the order of magnitude nature of the $\Lambda_{\nu}$ estimates.

We will concentrate on the process
$e^-e^-\rightarrow q\bar{q}q\bar{q}$ (which will usually manifest themselves as jets)
with no missing energy in an
ILC-like environment \cite{ILC} with a center-of-mass energy of
$1~\rm{TeV}$ and an integrated luminosity of
$100~\rm{fb}^{-1}$.  We also make the oversimplifying  assumption
that the detector system has equal acceptance to all quark
flavors, and the ability to efficiently distinguish quarks,
gluons and $\tau$s.  By summing
over all possible quark final states it is simple to estimate the total LNV cross
section for each effective operator, assuming it is responsible for neutrino masses.
Such searches can be complemented  by looking at $e^-e^-\to W^-W^-$,
which have been discussed in detail in the literature \cite{EmEm2WmWm}.  As
discussed in Sec.~\ref{subsec:OtherProcesses}, the different  LNV
operators couple to one or more gauge bosons via an
appropriately closed fermion loop or direct coupling to the Higgs doublet field.

Charge and baryon number conservation dictate that the two quarks in
$e^-e^-\rightarrow q\bar{q}q\bar{q}$ are down-type quarks, while the two antiquarks are up-type antiquarks.
At the parton level, the scattering process is similar to
 $\beta\beta0\nu$, which motivates exploiting simple
variations of the diagrams in Fig.~\ref{fig:Bb0nOps} in order to calculate
the relevant amplitudes, as was done in
Sec.~\ref{subsec:OtherProcesses}.  Here, the
extensions are obvious:  use crossing symmetry to rotate all lepton
lines into the initial state and all quark lines to the final state
taking special care to insert appropriate CKM matrix elements where
needed.  Due to the large characteristic momentum
transfer $Q$ of the $e^-e^-$ scattering, one must also ``expand'' the
electroweak vertices and account for gauge boson
propagation.  With this in mind, the amplitude calculations can be
carried over directly from the previous sections. Specific results are, however,
quite distinct due to the higher
center-of-mass energies involved.  In the language of the underlying
diagrams mediating this reaction, for diagrams characterized by TeV cutoff scales, diagram $D_9$,
if allowed at tree-level, will dominate the rates.  As in the
previous cases, for intermediate to high cutoff scales, general diagram
dominance must be addressed on a case by case basis.
 It is important to appreciate that, since these are non-renormalizable
effective interactions, cross-sections grow with center-of-mass energy. For this reason, we expect many
of the low cutoff scale operators to yield observably large signals at the
ILC.

\begin{figure}
\begin{center}
\includegraphics[angle=270,scale=.63]{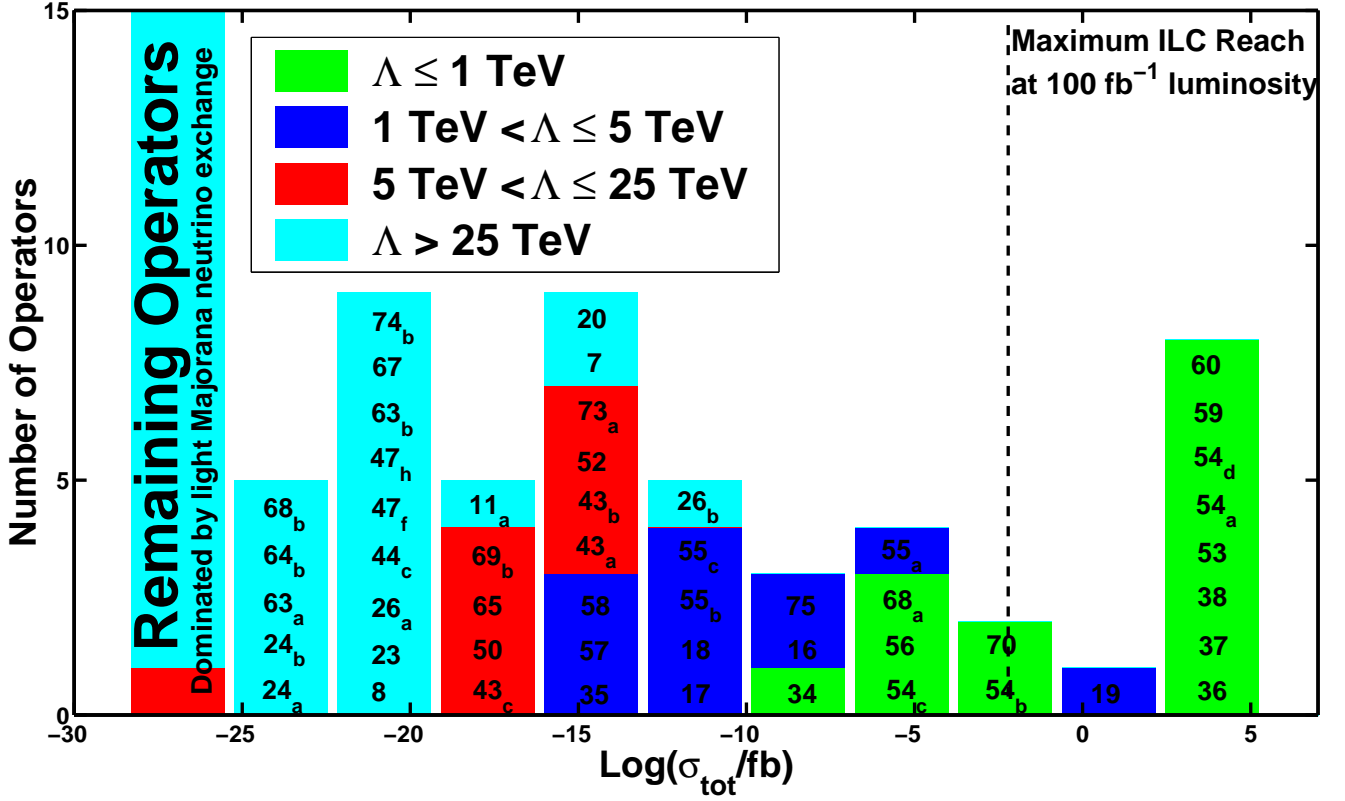}
\caption{Distribution of total cross-section
for the process $e^-e^-\rightarrow q\bar{q}q\bar{q}$ and no missing energy at an $e^-e^-$ collider with
$1~\rm{TeV}$ of center-of-mass energy.  Estimates were obtained
assuming the scales $\Lambda_{\nu}$ derived in Sec.~\ref{sec:Scale}, as
well as order one coupling constants. The histogram bars
are labeled with operator names and color-coded by $\Lambda_{\nu}$ cutoff
scale.  Also shown (broken vertical line) is the reach
of such an experiment assuming $100~\rm{fb}^{-1}$ of integrated
luminosity. The vertical axis is truncated to best
display the relevant features of the plot.} \label{fig:ILC}
\end{center}
\end{figure}

Fig.~\ref{fig:ILC} shows the $e^-e^- \rightarrow q\bar{q}q\bar{q}$ cross-section
distribution, in femtobarns, at the ILC, calculated for all 129 of
the analyzed LNV operators. Once again, the extracted value of the cutoff energy
scale $\Lambda_{\nu}$ assuming constraints from neutrino masses are
color-coded to indicate operators associated with a low ($\Lambda_{\nu}\lesssim 10$~TeV) or high
($\Lambda_{\nu}\gtrsim 10$~TeV)  ultraviolet cutoff.
Each bar is also labeled with the respective constituent
operators, for convenience. Note that the vertical axis is truncated
at fifteen operators (the left-most bin is over 60 operators high) to help clearly display relevant features of
the plot. We also highlight the potential reach (defined as cross-section greater than the inverse of the integrated luminosity) of the ILC with a broken vertical line, assuming
$100~\rm{fb}^{-1}$ of integrated luminosity.
This particular ILC luminosity value should be
considered as a loose lower bound, introduced to give a feeling for
the observable scales involved.  It has recently been
argued, for example, that a realistic machine should be able  to
outperform this estimate by over an order of magnitude \cite{ILC}.

A glance at Fig.~\ref{fig:ILC} reveals that it generally adheres
to the expected correlation of decreasing $\Lambda_{\nu}$ scales with
increasing LNV rates, similar to what is observed for other LNV observable ({\it e.g.},~Fig.~\ref{fig:BBEX}).
The similarities between the different processes
extend beyond mere trends to the specific ordering of the
operators within each histogram. This reflects the common
underlying interactions that drive these
processes.  The operators on the far right of the plot, topping off
the highest cross-sections, are exactly those operators with the
largest $m_{ee}^{\rm eff}$, now ``split'' into three
different bars. The large bar just below $10^5$~fb is
composed of sub-TeV scale operators with tree-level diagram
$D_9$-like fermion content. Slightly smaller are the expectations for $\mathcal{O}_{19}$,
again dominated by diagram $D_9$, but characterized by a slightly
larger $\Lambda_{\nu}$ scale (around one TeV). Moving down in cross-section, this is
followed by the low cutoff scale operators
$\mathcal{O}_{54_{b,c}}$ and $\mathcal{O}_{70}$, dominated by a
combination of diagrams $D_6$ and $D_7$.
On the opposite end of the plot we point out the large bar below
$10^{-25}$~fb, composed mainly of operators associated to high cutoff scales
($\Lambda_{\nu} > 25~\rm{TeV}$).  The contributions of these operators are dominated by
light Majorana neutrino exchange, but their histogram bar contain
far fewer models than their $\beta\beta0\nu$ counterpart, as many of
the latter have been driven up due to new diagram $D_4$ and $D_5$
contributions.  In general, the large
center-of-mass energies tend to magnify differences between
interaction rates that were not relevant in low-energy observables. This naively suggests that high energy probes
have a higher potential for distinguishing different models.

There are eleven operators that lead to an observably  large (as defined earlier) $e^-e^-\to q\bar{q}q\bar{q}$ cross-section at the ILC. Note that all of these were
already ``ruled out'' by current $\beta\beta0\nu$ searches.
As discussed in Sec.~\ref{subsec:bb0nu}, however,  these bounds only effectively limit the
couplings of the new physics to first generation of quarks and leptons, and
hence, if such a scenario is realized in nature, one should still expect large contributions from decay modes that
lead to second and third generation final state quarks.  In fact,
even one such heavy quark is enough to bypass the constraints from $\beta\beta0\nu$
for several effective operators.  Such reasoning implies
that constraints on the new physics flavor structure can be made quite
strong at a linear collider via analyzes of the flavor of the final state quarks.  By identifying and comparing the outgoing quark flavor one can extract individual limits on
quark-lepton coupling constants within the operators.
Additionally, kinematics can be used as a further operator
probe.  For example, one can potentially determine the dominant
underlying LNV diagram (say $D_6$, $D_7$ or $D_9$) by checking whether
the  various kinematic distributions are characteristic of
$W$-boson exchange.

The ILC can cleanly select or discard some LNV scenarios.  This characteristic
is further enhanced by considerations of initial electron
polarization.  Planned linear colliders have the ability to produce
partially polarized beams (80\% polarization for $e^-$, 40\% for $e^+$  \cite{ILC,ILCPol}).
The power of a high energy polarized $e^-e^-$ beam is in model
identification and rejection. Of all operators that yield observably large
cross-sections, the $e_L^-e_L^-$ mode can only probe
$\mathcal{O}_{53}$, and therefore any positive LNV signal cleanly
identifies this as the operator chosen by nature. In a similar way,
the ILC running in its $e_L^-e_R^-$ mode can easily observe LNV from
$\mathcal{O}_{19}$, $\mathcal{O}_{54_a}$, $\mathcal{O}_{54_d}$,
$\mathcal{O}_{59}$ and $\mathcal{O}_{60}$; and to a lesser extent,
operators $\mathcal{O}_{54_b}$ and $\mathcal{O}_{70}$, and possibly
even $\mathcal{O}_{54_c}$. Finally the $e_R^-e_R^-$ mode can probe
operators $\mathcal{O}_{36}$, $\mathcal{O}_{37}$ and
$\mathcal{O}_{38}$.  Within this framework, any LNV detected in one
ILC polarization mode will generally not be seen in the others.
This statement also applies to resonantly enhanced
low scale operators that lie outside the observability window.

While $e^-e^-$ collisions only probe effective operators that ``talk'' to first generation
leptons, there are several lepton collider processes that allow one to explore other members of the
charged lepton family.
Future high energy muon colliders \cite{mumu} could, in principle, also be used to study LNV.  In
this case, all of the preceding  discussions regarding
the ILC are applicable.  Electron linear collider facilities can also be used to study $\gamma e^-$
and $\gamma\gamma$ collisions \cite{gammagamma}. $\gamma e^-$ collisions can be used to probe
$\gamma e^-\to \ell_{\alpha}^++X$ (and hence the ``$e\alpha$'' structure of different LNV operators),
while $\gamma\gamma\to \ell_{\alpha}^{\pm}\ell_{\alpha}^{\pm}+X$ probes all the different $\alpha,\beta$
charged lepton flavors. For $\gamma\gamma$ collisions, for example, considering projected ILC-like
collider parameters, one would expect the same
operator distribution as Fig.~\ref{fig:ILC}, shifted down in
cross-section by, roughly,  a factor of $\alpha^2 \sim 10^{-4}$.  Thus, a handful
of operators should be testable at a future $\gamma\gamma$
collider assuming $100~\rm{fb}^{-1}$ of integrated luminosity.

\begin{figure}
\begin{center}
\includegraphics[scale=.7]{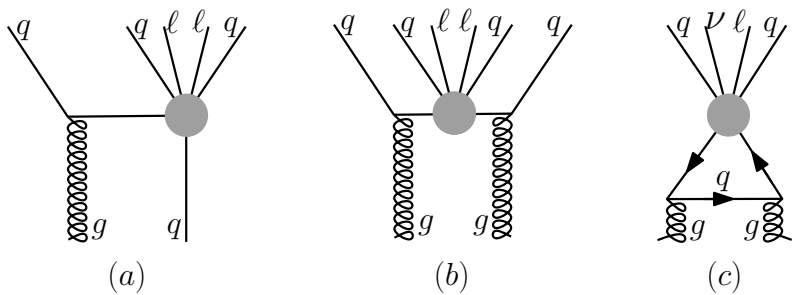}
\caption{Parton level
gluon--gluon and gluon--quark LNV interactions relevant at high energy
hadron colliders.  Each of these yields a same sign dilepton signal
with jets and no missing energy.  Notice that the final state flavor
structure is completely arbitrary under the assumption of random
order one coupling constants.} \label{fig:LHC}
\end{center}
\end{figure}

The preceding analyses carry over to the case of  hadron colliders, such
as the LHC, in a relatively straightforward way.  The LHC, or Large Hadron
Collider, is a proton--proton machine that will operate at a center-of-mass
energy of $14~\rm{TeV}$ and a characteristic integrated
luminosity around $100~\rm{fb}^{-1}$ \cite{LHC} (in its high luminosity mode).
The relevant LNV variants of
Eq.~(\ref{eq:GoldenCh}) are $d d \rightarrow \ell_{\alpha} ^-
\ell_{\beta} ^- uu$ and $u u \rightarrow \ell_{\alpha} ^+
\ell_\beta ^+dd$ with no missing energy. Of course,
at center-of-mass energies well above a TeV, the proton--proton collisions are
 dominated by the gluon content of the proton, so most interactions at the LHC will
be initiated by gluon--gluon and gluon--quark scattering.
The dominant  LNV subprocesses are $q g \rightarrow \ell^{\pm}_\alpha\ell^{\pm} _\beta q\bar{q}q$ and $gg
\rightarrow \ell^{\pm} _\alpha\ell^{\pm}_\beta q\bar{q}q\bar{q}$  and are illustrated in
diagrams $(a)$ and $(b)$ of Fig.~\ref{fig:LHC}, respectively. These are
characterized by similar final states as the quark--quark scattering
reactions but, given that
there is no explicit gauge boson field content in the LNV operators in question
(Table \ref{tab:AllOps}), their amplitudes are  proportional to unimportant order $\alpha_s$ and
$\alpha_s^2$ coefficients,  respectively.  The parton level diagram $(c)$ shows the
related process $gg \rightarrow \ell _\alpha \nu _\beta + q\bar{q}$.  The rate for this process
can be estimated, relative to its four jet cousins, by exchanging a
final state phase space suppression for a single loop suppression.
In all three diagrams depicted in Fig.~\ref{fig:LHC}, the LNV interaction regions
represented by large grey dots  contain all of
the diagrams discussed earlier, meaning that the operator amplitudes
calculated for the ILC can be recycled in this analysis.
While all three
bare diagrams are characterized by rates of the same
order of magnitude
diagram $(c)$ leads to missing transverse energy and potentially undetermined
final-state lepton number, rendering it a less than optimal experimental search mode.
Note that, in all
of these cases, the external, and internal, fermions outside of the
LNV interaction region can be of any flavor. Therefore, hadron
collider experiments have, in principle,  access to \emph{all} LNV
operator parameters. Cleanly identifying and constraining all said
parameters should prove quite difficult for all but the most obvious
signatures.  The above statements regarding
signals at the LHC are also applicable at the Tevatron
with some minor, but important, modifications.  The
Tevatron's $p\overline{p}$ collisions are at a much lower center-of-mass
energy, roughly $2~\rm{TeV}$, while the total expected integrated luminosity, less than
10~fb$^{-1}$ per experiment, is
orders of magnitude smaller.  These
factors lead to much lower amplitudes, reduced by approximately a
factor of $\left(Q_{\rm{Tevatron}}/Q_{LHC}\right)^5 \approx
10^{-5}$.\footnote{Strictly
speaking one must also account for the proton's structure functions
at the Tevatron's energy scale.  Unlike the LHC, where collisions are dominated
by gluon--gluon interactions, proton collisions at the Tevatron are dominated by valence quark
interactions. These considerations do not affect our conclusions.} The smaller center-of-mass energy
also limits the Tevatron's ability to directly produce new physics states.  With this in mind
we conclude  that the Tevatron has little or no chance of discovering LNV (within
this minimal framework).

A detailed set of predictions for the LHC would require a much more
refined analysis, including the effects of parton structure
functions, flux distributions, and backgrounds, and as such is beyond
the scope of this general survey. We would, however, like to point
out that some of the reactions outlined here are subject to large background
rates.  While SM processes are lepton number conserving, many can
fake the LNV signals in the complicated environment of a high energy
hadronic interaction. The requirement of no missing final state
energy is particularly hard to accommodate as some energy is always
lost down the beampipe. As is typically done, one must rely on the less restrictive
conservation of transverse momentum in order to constrain invisible states,
such as neutrinos.  SM same-sign dilepton production
processes arising from, say, $W$-boson pair production, are serious
potential sources of background. Furthermore, it is impossible to
predict correlations among final state jets without selecting a
particular operator and underlying model of new physics, making it
difficult to impose general cuts to reduce other hadronic
backgrounds.  Of course, some of the low scale LNV operators yield
large enough total cross-sections that even crude analyses may suffice to
reveal their existence.  We conclude by pointing out that  a large
amount of recent work has been dedicated to LNV searches at collider
facilities \cite{CollLNV}.  Most of these approach the subject from
the perspective of sub-TeV mass, mostly sterile Majorana neutrinos
that mix with the active neutrinos and are thus  related to light
neutrino masses via the seesaw mechanism \cite{SeeSaw}.  This
amounts to one example that leads to the dimension-five operator
$\mathcal{O}_1$, but where one assumes that the propagating degrees
of freedom are twelve or thirteen orders of magnitude lighter than the
ultraviolet cutoff scale $\Lambda_{\nu}$.\footnote{This can be achieved in
two different ways. Either the new physics is very weakly coupled,
or the new physics -- SM couplings are finely-tuned \cite{fine_nus}. In order to
observe right-handed neutrinos in colliders, the latter must be
realized.}  In this case, LNV interactions are dominated by diagram
$D_\nu$ of Fig. \ref{fig:Bb0nOps} (where heavy (weak scale)
neutrinos are also exchanged),
 and as such one should make use of
specific kinematic cuts to reduce background rates.
These cuts, however, may also remove LNV signals
resulting from many of the scenarios explored here, particularly those whose
rates are dominated by $D_9$ at tree-level.  We urge
experimentalists to account for this possibility while analyzing
future data sets.

\setcounter{footnote}{0} \setcounter{equation}{0}
\section{Neutrino mixing} \label{sec:Oscillations}

Table \ref{tab:AllOps} contains predictions for \emph{all} the entries $m_{\alpha\beta}$,
the Majorana neutrino mass matrix. These are computed in the weak basis where
the weak interactions and the charged-lepton Yukawa couplings are diagonal, so that
the eigenvalues of the neutrino mass matrix are the neutrino masses (bounded by oscillation experiments
and, say, precision measurements of tritium beta-decay  \cite{TritNuEffMass}), while its
eigenvectors determine the neutrino mixing matrix, constrained mostly by oscillation experiments.
Since different LNV effective operators predict different flavor-structures for the neutrino
mass matrix,  there is the possibility to constrain the different scenarios with existing
oscillation data \cite{OscBestFit}.  While we can only predict the values of $m_{\alpha\beta}$
within, at best, an order of magnitude, it is still possible to extract useful
information from the derived large scale structure of the
expressions.  In particular we can test the hypothesis of whether $\lambda$ values
associated to different lepton flavors are allowed to be of the same order of magnitude.
In order to obtain  more accurate predictions and further probe
the fine details of lepton mixing one must succumb to specific models, beyond the scope and
philosophy of this analysis.

The mass matrix for the three light Majorana neutrinos can be reconstructed from nine observables:
three masses $m_1,m_2,m_3$, taken to
be real and positive; three (real) mixing angles
$\theta_{12},\theta_{23},\theta_{13}$; and three CP-violating
phases $\delta,\phi_2,\phi_3$.  Here, $\delta$ is a so-called Dirac phase
that is generally present in the system regardless of the neutrino's
nature (Majorana or Dirac fermion), while $\phi_1,\phi_2$ are so-called Majorana phases, only present if the
neutrinos are Majorana particles (which is the case of all scenarios under consideration here).
Oscillation data determine with relatively good precision
 $\theta_{12}$, $\theta_{23}$,
$\Delta m^2_{12} \equiv m_2^2 - m_1^2$ and $|\Delta m^2_{13}| \equiv |m_3^2 - m_2^2|$.
We define neutrino masses such that $m_1<m_2$  and
$\Delta m^2_{12}<|\Delta m^2_{13}|$, so that the sign of $\Delta m^2_{13}$ remains as an observables which
characterizes the neutrino mass hierarchy (``normal'' for $\Delta m^2_{13}>0$, ``inverted'' for  $\Delta m^2_{13}<0$).
See, for example, \cite{NeutrinoReview} for details.
 As for the third mixing angle, $\sin^2\theta_{13}$ is constrained
to be less than 0.025 (0.058) at $2\sigma$ $(4\sigma)$ from a
three neutrino global oscillation analysis \cite{OscBestFit}.
A considerable amount of
uncertainty remains. In particular we have only
upper bounds on the absolute neutrino mass scale, from kinematical
measurements such as tritium beta decay \cite{Mainz,Troitsk}, plus
cosmological observations \cite{CosmoSum}.  Finally, the three
CP violating phases are completely unconstrained, and we have no information regarding the
neutrino mass hierarchy.

The above experimental results allow for several different ``textures'' for $m_{\alpha\beta}$
in our weak basis of choice (see, for example, \cite{Frigerio:2002fb}). The purpose of this section is to
discuss whether any of the textures predicted by the different LNV effective operators is ``ruled out''
by current observations.
Most of the analyzed operators imply ``anarchic'' \cite{anarchy} neutrino masses.  This
simply means that  all
elements of the neutrino mass matrix are uncorrelated and of the same order of magnitude.
This hypothesis is known to ``fit'' the current data very well \cite{anarchy}.
It will be further challenged by searches for $\theta_{13}$ (the anarchic hypothesis favors large
$\theta_{13}$ values) and probes that may reveal if the neutrino masses are hierarchical or whether
two or three of the masses are almost degenerate (anarchy naively predicts the former).
If future data strongly points towards non-anarchic $m_{\alpha\beta}$, we will be forced to conclude
that there is nontrivial ``leptonic'' structure in the dimensionless coefficients $\lambda$ of most of the LNV operators
considered here.

Many of the operators associated with a low neutrino-mass related cutoff scale ($\Lambda_{\nu} \leq 10~\rm{TeV}$),
on the other hand, naively predict more structured neutrino mass matrices. Operators
\begin{equation}
\mathcal{O}_{7},\mathcal{O}_{8},\mathcal{O}_{19},\mathcal{O}_{20},\mathcal{O}_{34},\mathcal{O}_{35},\mathcal{O}_{54_{a,b,c,d}},\mathcal{O}_{55_{a,b,c}},\mathcal{O}_{56},\mathcal{O}_{57},\mathcal{O}_{58},\mathcal{O}_{59},\mathcal{O}_{60},\mathcal{O}_{70},\mathcal{O}_{75},
\label{eq:lalphaOps}
\end{equation}
which radiatively generate neutrino mass elements proportional to distinct charged lepton Yukawa coupling ($y_{e}, y_{\mu}, y_{\tau}$),  yield mass matrices $m$ such that
\begin{equation}
m \propto \left(
  \begin{array}{ccc}
    y_{e} & y_{\mu} & y_{\tau} \\
    y_{\mu} & y_{\mu} & y_{\tau} \\
    y_{\tau} & y_{\tau} & y_{\tau} \\
  \end{array}
\right). \label{eq:texture1}
\end{equation}
Additionally, models described at low energies by
$\mathcal{O}_{36}$, $\mathcal{O}_{37}$ and $\mathcal{O}_{38}$
generate neutrino masses proportional to both associated charged
Yukawa couplings, such that
\begin{equation}
m \propto \left(
  \begin{array}{ccc}
    y_{e}y_{e} & y_{e}y_{\mu} & y_{e}y_{\tau} \\
    y_{e}y_{\mu} & y_{\mu}y_{\mu} & y_{\mu}y_{\tau} \\
    y_{e}y_{\tau} & y_{\mu}y_{\tau} & y_{\tau}y_{\tau} \\
  \end{array}
\right). \label{eq:texture2}
\end{equation}

The strongly hierarchial nature of the charged lepton masses ($y_e
\ll y_\mu \ll y_\tau$), implies that the $m_{\alpha\beta}$ elements of Eqs.~(\ref{eq:texture1}) and (\ref{eq:texture2}) are
expected to be hierarchical as well. In particular,  the $ee$ matrix element, $m_{ee}$,
proportional to $y_e$ or $y_e^2$ is, for all practical purposes, negligibly small\footnote{Quantitatively, in the scenarios under investigation,
$m_{ee}$ values are, respectively, up to order one corrections, $y_e/y_\tau \sim
10^{-4}$ and $y_e^2/y_\tau^2 \sim 10^{-7}$ times the characteristic
mass scale of the mass matrix.}
in both of these cases.  On the other hand, it is well known that only a normal neutrino mass
hierarchy is consistent with vanishing $m_{ee}$
\cite{Normal0Mee}, so that both  Eqs.~(\ref{eq:texture1}) and (\ref{eq:texture2})
predict the neutrino mass ordering to be normal. In the absence of extra structure,
scenarios characterized by the LNV operators listed in Eq.~(\ref{eq:lalphaOps})
plus $\mathcal{O}_{36}$, $\mathcal{O}_{37}$ and $\mathcal{O}_{38}$ will be ruled out if future  data
favor an inverted mass hierarchy, or if the neutrino masses end up  quasi-degenerate
(regardless of the hierarchy).  As will become clear shortly, Eqs.~(\ref{eq:texture1}) and (\ref{eq:texture2})
predict that the lightest neutrino mass ($m_1$ in this case)
is  small ($\lesssim \sqrt{\Delta m_{12}^2}$).

A more detailed analysis reveals that naive expectations from Eqs.~(\ref{eq:texture1})
are already disfavored, while those from Eqs.~(\ref{eq:texture2}) are virtually excluded.
Assuming the normal hierarchy and very small $m_{ee}$, one can find a relation
between the neutrino mass eigenstates and the oscillation parameters,
thus reducing the number of free parameters in the mass matrix by
one. Consider the diagonalization of the neutrino mass matrix
defined by $m_{\alpha\beta} = UM^DU^T$ with $M^D =
{\rm diag}(m_1,m_2e^{2 i\phi_2},m_3e^{2i\phi_3})$ and $U$
the neutrino mixing matrix, expressed in the PDG parameterization. In this case,
\begin{equation}
m_{ee} = m_1\cos^2\theta_{12}\cos^2\theta_{13} +
m_2\sin^2\theta_{12}\cos^2\theta_{13}e^{2 i \phi_2} +
m_3\sin^2\theta_{13}e^{2 i\left(\phi_3-\delta\right)}.
\end{equation}
Setting $m_{ee}=0$, one can solve for $m_1$ and one of the Majorana phases.
Recalling that, for the normal mass hierarchy, $m_2 = \sqrt{m_1^2 +
\Delta m^2_{12}}$ and $m_3 = \sqrt{m_1^2 + \Delta m^2_{13}}$,  and assuming small $\theta_{13}$
and $\eta \equiv \sqrt{\Delta m^2_{12}/\Delta m^2_{13}}$,
\begin{eqnarray}
 \nonumber \frac{m_1}{\sqrt{\Delta m_{13}^2}} & \approx & \eta \frac{\sin^2\theta_S}{\cos^{1/2}2\theta_S}
 - \theta_{13}^2\frac{\cos^2\theta_S}{\cos
 2\theta_S}\cos [2(\phi_3 - \delta)],\\
 \phi_2 &\approx& \frac{\pi}{2} + \frac{1}{2}\arctan\left(\frac{4\theta_{13}^2}{\eta}\frac{\sqrt{\cos 2\theta_S}}{\sin^2 2\theta_S}\sin [2(\phi_3-\delta)]\right).
\label{eq:m1}
\end{eqnarray}
One can easily obtain approximate expressions for the
other neutrino masses ($m_2, m_3$) and hence all elements $m_{\alpha\beta}$.
Upon substituting the numeric best fit oscillation parameters to
avoid introducing a needlessly cumbersome expression, we get
\begin{eqnarray}
\frac{m_{\alpha\beta}}{\sqrt{\Delta m^2_{13}}} &=&
0.5e^{i2\phi_3}\left(
                      \begin{array}{ccc}
                        0 & 0 & 0 \\
                        0 & 1 & 1 \\
                        0 & 1& 1 \\
                      \end{array}
                    \right) + 0.71\theta_{13}e^{-i(\delta-2\phi_3)}\left(\begin{array}{ccc}
                        0 & 1 & 1 \\
                       1 & 0 & 0 \\
                       1 & 0 & 0 \\
                      \end{array}
                    \right)
                    + 0.45\eta\left(\begin{array}{ccc}
                        0 & -1.3 & 1 \\
                        -1.3 & -1 & 0.61 \\
                        1 & 0.61 & -0.36 \\
                      \end{array}
                    \right) \nonumber \\
                    &+& 0.91\theta_{13}^2\cos [2(\delta-\phi_3)]\left(\begin{array}{ccc}
                        0 & 1 & -0.89 \\
                        1 & 0.12 & 0.02 \\
                        -0.89 & 0.02 & -0.12 \\
                      \end{array}
                    \right)
                     +
                    1.2i\theta_{13}^2\sin [2(\phi_3 - \delta)]\left(\begin{array}{ccc}
                        0 & 1 & -0.67 \\
                        1 & 1.2 & -0.83 \\
                        -0.67 & -0.83 & 0.56 \\
                      \end{array}
                      \right).
                    \label{eq:MassMatrix}
\end{eqnarray}

Eq.~(\ref{eq:MassMatrix})  suggests a clear
hierarchy among the mixing matrix elements. The four, lower
box-diagonal $\mu-\tau$ elements dominate, followed by the
off-diagonal $e\mu$ and $e\mu$ entries, and finally the vanishingly small $m_{ee}$.
Except for the vanishingly small $m_{ee}$, which was required  {\it a priori},
all
of the remaining properties follow directly from the experimentally
determined mixing parameters. Among the dominant $\mu-\tau$ submatrix,
Eq.~(\ref{eq:MassMatrix}) predicts that all entries are equal
 up to small order $\eta$ and $\theta_{13}$ corrections.
The magnitude, and sign, of these
``breaking terms'' can be tuned with the phases $\phi_3$
and $\delta$, and to a lesser extent by varying $\eta$ and
$\theta_{13}$ within their allowed ranges. On the other hand, the relative
sizes of  $m_{e\mu}$ and $m_{e\tau}$ are expected to be similar but not identical, {\it i.e.},
$m_{e\mu}\sim m_{e\tau}\sim( m_{e\mu}-m_{e\tau})$.

While some of the gross features of  Eq.~(\ref{eq:MassMatrix}) are shared by
Eq.~(\ref{eq:texture1}) and Eq.~(\ref{eq:texture2}), a finer analysis reveals
several disagreements.
The major discrepancy lies in the required
relations among the matrix elements.  Eq.~(\ref{eq:texture1}) predicts that all $m_{\alpha\tau}$ elements are
equal, while Eq.~(\ref{eq:texture2}) suggests $m_{e\tau} \ll
m_{\mu\tau} \ll m_{\tau\tau}$.  Both of these contradict, in
different ways, the experimental constraint $m_{e\tau} \ll
m_{\mu\tau} \approx m_{\tau\tau}$.  Additionally, both Eq.~(\ref{eq:texture1}) and Eq.~(\ref{eq:texture2})
predict $m_{ee} \ll m_{\mu\mu} \ll m_{\tau\tau}$, while observations require $m_{ee} \ll
m_{\mu\mu} \approx m_{\tau\tau}$.  Similarly, both sets of operators suggest
$m_{e\mu} \ll m_{e\tau}$ while, experimentally, they are
constrained to be similar.

\begin{figure}
\begin{center}
\includegraphics[scale=1]{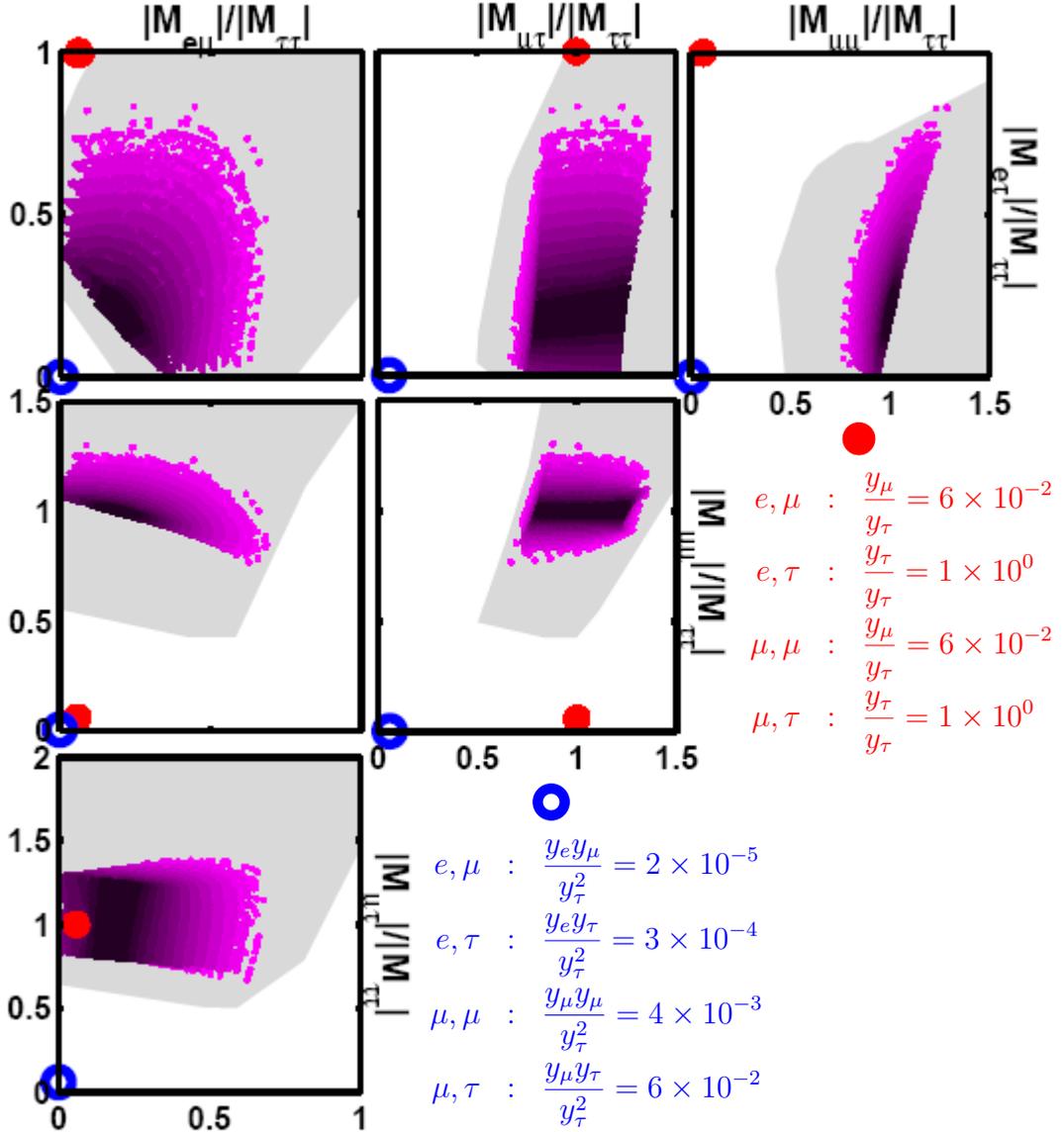}
\caption{Scatter plots of the
symmetric Majorana neutrino mass matrix elements normalized to
$m_{\tau\tau}$.  Each panel is produced assuming the normal mass
hierarchy and parameter constraints insuring that $m_{ee} \leq
10^{-4}~\rm{eV}$.  The light grey region is calculated allowing all
mixing parameters to vary within their respective $95\%$ confidence
intervals.  In the purple (darker) regions, the solar and atmospheric
parameters are held constant while all phases are scanned within
their physical ranges and $\theta_{13}$ is varied between
zero and its 4~$\sigma$ upper bound.
The $\sin^2\theta_{13}$ variation is illustrated by varying the
shading from dark to light.
Also indicated by red (closed) and blue (open) dots are the
expectations derived from Eqs.~(\ref{eq:texture1}) and
(\ref{eq:texture2}), respectively, along with a listing of their
associated coordinate values.} \label{fig:Mn_0ee}
\end{center}
\end{figure}

In order to quantify how much Eq.~(\ref{eq:texture1}) and Eq.~(\ref{eq:texture2})
(dis)agree with our current understanding of neutrino masses and lepton mixing,
we numerically scanned the
allowed mass matrix parameter space assuming the normal neutrino
mass hierarchy and constraining $|m_{ee}| \leq y_e/y_\tau \times 1~\rm{eV} \approx 10^{-4}~\rm{eV}$.
It should be noted that, according to this relation, $m_{ee}$ is
allowed to deviate by nearly a factor of ten above naive
expectations from mass matrix Eq.~(\ref{eq:texture1}), thus accounting for
the possible order of magnitude uncertainties in operator scales and
coupling constants.
This feature is only included for completeness,
as one expects that such $m_{ee}$ excursions from zero will
generally have negligible effect on the mass matrix due to the
robust nature of Eq.~(\ref{eq:MassMatrix}).
Fig.~\ref{fig:Mn_0ee}, a scatter plot of mixing matrix elements,
depicts the result of such a scan.
 Note that  we plot the mass ratios with respect to
assumed-to-be-dominant  $m_{\tau\tau}$ element.
The light grey regions of the plot were produced allowing all
oscillation parameters to vary within their $95\%$ confidance bounds
\cite{OscBestFit} and phases to vary within their entire physical
range subject to the constraints discussed above.  In the purple (dark) region, the
phases and reactor mixing angle $\theta_{13}$ are allowed to vary while
all other mixing parameters are held fixed at their best fit values. We depict the
$\sin^2\theta_{13}$ variation  from zero
to $0.06$ ($4\sigma$ upper bound \cite{OscBestFit}) by varying the
purple shading from dark to light.  It is easy to
check that the numeric (Fig.~\ref{fig:Mn_0ee}) and analytic results (Eq.~(\ref{eq:MassMatrix})) are consistent both
qualitatively and quantitatively.

Fig.~\ref{fig:Mn_0ee} also depicts the
predictions from Eq.~(\ref{eq:texture1}) and Eq.~(\ref{eq:texture2}) with red (closed) and
blue (open) dots, respectively.    As expected, all the predictions from Eq.~(\ref{eq:texture2})
 fall near the origin in each panel and
are safely excluded.  Because expectations  from Eq.~(\ref{eq:texture2})
for all $m_{\alpha\beta}/m_{\tau\tau}$ are much smaller than one,
we also include the dot coordinate values for both textures
within the figure.  In order to render the neutrino mass matrix predicted from
$\mathcal{O}_{36}$, $\mathcal{O}_{37}$ and $\mathcal{O}_{38}$ consistent with
experimental constraints on neutrino masses and lepton mixing, one is required
to choose very hierarchical $\lambda$ coefficients.
In more detail, one needs to choose $\lambda$ values so that
all mixing matrix elements are enhanced relative to the
dominant $m_{\tau\tau}\propto y_\tau y_\tau$ by numerical
factors that range -- for different entries -- from 100 to $10^{5}$.
A possible mechanism for achieving this is to suppress third generation
couplings to new physics, thus driving up the ratio
$m_{\alpha\beta}/m_{\tau\tau}$ along with the required cutoff scale $\Lambda_{\nu}$.
This procedure would have to be accompanied by a more modest reduction of
the couplings of second generation fermions. Basically, we need to impose a flavor
structure that ``destroys'' the naive flavor structure induced by the charged lepton Yukawa coupling
hierarchy. We can safely conclude that
$\mathcal{O}_{36}$, $\mathcal{O}_{37}$, and $\mathcal{O}_{38}$, which
suggest that the neutrino mass matrix has the form  Eq.~(\ref{eq:texture2}),
are strongly disfavored
by current neutrino oscillation data and, if somehow realized in nature, must
be accompanied by a very nontrivial flavor structure.

On the other hand, the operators listed in Eq.~(\ref{eq:lalphaOps}), which predict
Eq.~(\ref{eq:texture1}), are not quite as disfavored.  In
this case the hierarchies among different mass matrix elements are softer,
and one can ask whether the red dots in Fig.~\ref{fig:Mn_0ee} can move toward the experimentally allowed
regions with order 1--10 relative shifts. Many of the predictions are already
in agreement with experimental constraints, or at least close enough to be easily
``nudged'' toward acceptable levels with order one coefficients.
The figure reveals that only $m_{\mu\mu}$ is predicted to be relatively too small. By enhancing it
by a factor of order $y_\tau/y_\mu \sim 20$ one obtains moderately good
agreement between Eq.~(\ref{eq:texture1}) and experimental requirements.
We therefore conclude that
operators listed in Eq.~(\ref{eq:lalphaOps}) are at least marginally allowed by neutrino mixing
phenomenology.

While essential for a complete understanding of
neutrino masses and mixing, improved measurements of the already
determined mixing angles and mass-squared differences will not
 help to further constrain/exclude any of  the LNV scenarios in question.
Considering our  parameter flexibility, only future neutrino
 experiments that provide qualitatively new results can aid in
this endeavor.  In particular, the experimental determination of the
neutrino mass hierarchy is essential in order to properly test  the
scenarios highlighted in this section, as they all predict, in the absence of
very non-trivial flavor structure in the LNV sector,  the normal
hierarchy. Next-generation neutrino oscillation experiments are expected
to provide non-trivial information regarding the neutrino mass hierarchy.
Most rely on a
neutrino/anti-neutrino oscillation asymmetry via Earth matter
effects \cite{NeutrinoReview,NuMassHier0t13}, and depend heavily on a
sufficiently large $\theta_{13}$ mixing angle.  The possibility that
$\theta_{13}$ is vanishingly small, where the standard approach is
ineffective, is addressed in \cite{NuMassHier0t13} considering both
oscillation and non-oscillation probes.  In that case, one can hope
to discern the neutrino mass spectrum in future neutrino factory
\cite{NuFact}/ Superbeam \cite{SupBeam} experiments coupled with
improved constraints on the effective masses extracted from tritium
beta decay \cite{TritNuEffMass} and cosmology
\cite{CosmoSum}.\footnote{One traditionally includes the effective
$\beta\beta0\nu$ mass $m_{ee}$ given by Eq.~(\ref{eq:std_Mee}) in a
neutrino mass hierarchy analysis. However, as discussed in
Sec.~\ref{subsec:bb0nu}, $m_{ee}^{\rm eff}$ is a potentially
convoluted process-dependent quantity that generally has little
(directly) to do with neutrino masses. For this reason,
$\beta\beta0\nu$ constraints cannot be used to determine the
neutrino mass spectrum from the point of this analysis.} Note that
these non-oscillation probes  can be independently used to constrain
LNV models, as they provide information regarding the the magnitude
of the lightest mass eigenstate
($m_1$ [$m_3$] in the case of normal
[inverted] hierarchy).
For example, if either cosmological
observations or tritium beta decay experiments see evidence for
non-zero neutrino masses (in more detail, they constrain
$\Sigma = \sum_i m_i$ and $m_{\nu_e}^2 = \sum_i m_i^2 |U_{ei}|^2$
respectively) such that $\Sigma\gg 0.05$~eV or $m_{\nu_e}\gg 0.01$~eV, one would
conclude, assuming a normal mass hierarchy, that $m_1 \gg
\sqrt{\Delta m^2_{12}}$. This would destroy the possibility of
negligibly small $m_{ee}$, and hence disfavor the operators that
lead to mass matrices of the type Eq.~(\ref{eq:texture1}) and
(\ref{eq:texture2}).  Currently, $\Sigma$ and $m_{\nu_e}$ are bounded to
be below $0.94~{\rm eV}$ and $2.0~{\rm eV}$, respectively, but the
sensitivity to these observable is expected to significantly improve
with next-generation experiments to $0.1~{\rm eV}$ \cite{FutureCosmo} and $0.2~{\rm
eV}$ \cite{Katrin}, respectively.

\setcounter{footnote}{0} \setcounter{equation}{0}
\section{Phenomenologically interesting operators:  Sample Renormalizable Model} \label{sec:interesting}

Having superficially surveyed a large set of LNV operators, we are
now in a position to identify operators with ``interesting'' phenomenological
features for further detailed study.  One  subset of potentially
interesting operators is characterized by those that, when required to ``explain'' the observed neutrino masses, are accompanied by a low cutoff scale of, say, less than several TeV.  Further requiring a small enough
$m_{ee}^{\rm eff}$ in order to evade current $\beta\beta 0 \nu$ constraints, this
set contains only seven elements:
$\mathcal{O}_{17},\mathcal{O}_{18},\mathcal{O}_{34},\mathcal{O}_{35},\mathcal{O}_{56},\mathcal{O}_{57},\mathcal{O}_{58}$.
Of these, all but operators $\mathcal{O}_{35}$ and $\mathcal{O}_{58}$ (which lead to the zeroth-order
neutrino mass matrix Eq.~(\ref{eq:texture1}) and a suppressed $m_{ee}$)
should provided a positive LNV signal in the next round of double-beta decay experiments, baring specific flavor symmetries or finely-tuned
couplings. Furthermore,  $\mathcal{O}_{56}$ leads to a $\beta\beta0\nu$ rate that is higher than what is
naively dictated by the values of the neutrino masses.
Finally, with the possible exception of $\mathcal{O}_{56}$ which may mediate observable LNV processes at high energy
colliders, none of the seven operators above are expect to mediate LNV violating phenomena (as defined here)
at accessible rates.

An ``orthogonal'' subset consists of the higher dimensional
operators already ``excluded'' by $\beta\beta0\nu$. Not
including those operators severely constrained by lepton mixing  in
Sec.~\ref{sec:Oscillations}, this list  contains 11 elements:
$\mathcal{O}_{16},\mathcal{O}_{19},\mathcal{O}_{53},\mathcal{O}_{54a,b,c,d}$,
$\mathcal{O}_{59},\mathcal{O}_{60},\mathcal{O}_{70},\mathcal{O}_{75}$.
Most of these are associated to cutoff scales of order the weak
scale, which are likely to already be constrained by different
searches for new degrees of freedom with masses around
100~GeV. Even if those are considered to be excluded,
$\mathcal{O}_{16},\mathcal{O}_{19},\mathcal{O}_{75}$ are ``safely''
shielded from direct and indirect non-LNV searches,\footnote{Generic
new degrees of freedom at the weak scale are constrained by direct and indirect searches at
high energy colliders
({\it e.g.}, resonances and effective four-fermion interactions, respectively), flavor-violating
({\it e.g.}, $\mu\to e\gamma$), and high precision experiments ({\it
e.g.}, measurements of the anomalous muon magnetic moment).}  while
still mediating potentially observable LNV effects at colliders as
long as the new physics does not couple, to zeroth order, to first
generation quarks (in order to evade the $\beta\beta0\nu$
constraints).

Regardless of whether these different options for the LNV sector lead to observable LNV phenomena,
the low extracted cutoff scale of \emph{all} the operators highlighted above  implies that new degrees of freedom should be
produced and, with a little luck, observed at the LHC or, perhaps, the ILC.
Furthermore, the TeV scale has already been identified as an interesting candidate scale for new physics for
very different reasons, including the dark matter puzzle and the gauge hierarchy problem. The fact that, perhaps, the
physics responsible for neutrino masses also ``lives'' at the TeV scale is rather appealing.

In order to study this new physics,
as already emphasized earlier, ultraviolet complete manifestations of the physics that leads to the effective
operators are required. Here we discuss one concrete example.
Other examples (for different effective
operators) were discussed in \cite{Operators}.
Given a specific LNV operator, it is a simple
matter to write down equivalent renormalizable Lagrangians.
We briefly illustrate this procedure by constructing a renormalizable model that
will lead to the  dimension-eleven
operator $\mathcal{O}_{56}$.  It is
among the interesting LNV effective operators of the sample highlighted above,
since it is currently unconstrained by  $\beta\beta0\nu$ searches regardless of the quark-flavor structure of the operator, while $m_{ee}^{\rm eff}\gg m_{ee}$ for $\beta\beta0\nu$. On the other hand,
$\Lambda_{\nu}$ for $\mathcal{O}_{56}$ is very low (below 500~GeV),
so that the new degrees of freedom may already be constrained by, for example,
Tevatron or LEP data. We will not worry about such constraints henceforth, but will only comment
on possible phenomenological problems.

$\mathcal{O}_{56}$ can be accommodated
by a wide variety of models, as can be seen from its possible
Lorentz structures. In terms of scalar/tensor helicity-violating bilinears $\Gamma_v=1,\sigma_{\mu\nu}$, and vector helicity-conserving
bilinears $\Gamma_c=\gamma_{\mu}$, these are
\begin{eqnarray}
\mathcal{O}_{56} &=&\nonumber \{(L^i\Gamma_v Q^j)(d^c\Gamma_v d^c)(\overline{d^c}\Gamma_v \overline{e^c}),(L^i\Gamma_v Q^j)(d^c\Gamma_c \overline{d^c})(d^c\Gamma_c \overline{e^c}),(L^i\Gamma_v d^c)(Q^j\Gamma_v d^c)(\overline{d^c}\Gamma_v \overline{e^c}),\\
& & \nonumber (L^i\Gamma_v d^c)(Q^j\Gamma_c \overline{d^c})(d^c\Gamma_c \overline{e^c}),(L^i\Gamma_v d^c)(Q^j\Gamma_c \overline{e^c})(d^c\Gamma_c \overline{d^c}),(L^i\Gamma_c \overline{d^c})(Q^j\Gamma_v d^c)(d^c\Gamma_c \overline{e^c}),\\
& & (L^i\Gamma_c \overline{d^c})(Q^j\Gamma_c \overline{e^c})(d^c\Gamma_v d^c),(L^i\Gamma_c \overline{e^c})(Q^j\Gamma_v d^c)(d^c\Gamma_c \overline{d^c}),(L^i\Gamma_c \overline{e^c})(Q^j\Gamma_c
\overline{d^c})(d^c\Gamma_v d^c) \} \times
H^kH^l\epsilon_{ik}\epsilon_{jl}.
\end{eqnarray}

It is clear from the chiral field content that
these operators depend on combination of helicity-conserving
and helicity-violating interactions.  In particular, it is impossible to form any of the
operators in this long list with only
the addition of vector boson states: new heavy scalar and/or tensor particles are probably
required  if $\mathcal{O}_{56}$ is the proper tree-level manifestation
of the LNV physics at low-energies.\footnote{Other possibilities include heavy vector-like fermions.} Furthermore, the
couplings of the new physics fields with one another must be
constrained in order to ``block'' the presence  of lower-dimensional tree-level effective
operators.  This usually implies the
existence of new exact (broken) symmetries to forbid (suppress)
particular interactions.

\begin{figure}
\begin{center}
\includegraphics[scale=.9]{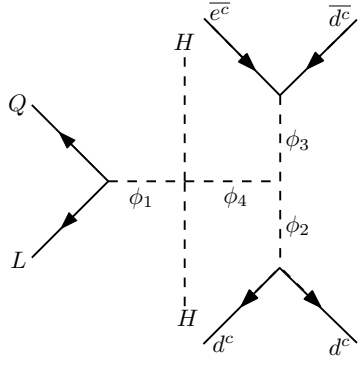}
\caption{Sample scalar
interactions that lead to the ``interesting'' effective operator
$\mathcal{O}_{56}$ with the Lorentz structure $(L^iQ^j)(d^cd^c)(\bar{e^c}\bar{d^c})H^kH^l\epsilon_{ik}\epsilon_{jl}$.} \label{fig:Models}
\end{center}
\end{figure}

Certain Lorentz structures, those containing only $\Gamma_v$ bilinears, can be
realized assuming that the LNV ultraviolet sector contains only heavy \emph{scalar} fields and
we concentrate, for simplicity, on this possibility \cite{ScalarLNV}.
Simple scalar interactions that can lead to $\mathcal{O}_{56}$
are shown in the diagram in Fig.~\ref{fig:Models}.
Specifically, these  yield  the effective operator Lorentz
structure $(L^iQ^j)(d^cd^c)(\bar{e^c}\bar{d^c})H^kH^l\epsilon_{ik}\epsilon_{jl}$
with the introduction of four charged
scalar fields, $\phi_1,\phi_2,\phi_3,\phi_4$.  The gauge structure is such that,  under $(SU(3)_c,SU(2)_L,U(1)_Y)$\footnote{In the case of $U(1)_Y$, `transfoms as $X$' means `has hypercharge $X$'.}, $\phi_1$ transforms as
a $(\bar{3},3,+1/3)$, $\phi_2$ as $(\bar{3},1,-2/3)$, $\phi_3$ as $(3,1,-4/3)$, and $\phi_4$
as $(\bar{3},1,-2/3)$. While $\phi_2$ and $\phi_4$ have identical gauge quantum numbers, they
have different baryon number ($2/3$ versus $-1/3$). $\phi_1$ has baryon number $-1/3$, while $\phi_3$ has baryon number $1/3$. Lepton number cannot be consistently assigned  as it is explicitly violated by two units.

$\phi_4$, which does not couple to any of the SM fermions,
plays an essential role. It acts as a
selective ``insulator'' that connects the various interaction terms in
such a way as to only alow certain tree-level higher dimensional SM effective operators.
All renormalizable theories that lead to only very high dimensional effective operators
contain one or more of these ``hidden sector'' fields.
Note that the
new scalar fields should not acquire vacuum
expectation values in order to avoid the presence of lower dimensional
irrelevant operators that are likely to dominate low-energy phenomenology and -- much more
important -- to
prevent the spontaneous breaking of color or electromagnetic charge.

Given the scalar
field content as well as its  transformation properties
under SM global and local symmetries, it is a simple matter to
write down the minimal interaction Lagrangian density for the system.  A candidate renormalizable Lagrangian is
\begin{equation}
\mathcal{L} = \mathcal{L}^{(SM)} + \sum_i \left(|D_\mu \phi_I|^2 +
M_i|\phi_i|^2\right) +y_1QL \phi_1 + y_2
d^cd^c \phi_2
 + y_3 e^cd^c \phi_3 + \lambda_{14}
\bar{\phi}_1\phi_4 H H + \lambda_{234} M \phi_2\bar{\phi}_3\phi_4 + h.c.~.
 \label{eq:sample}
 \end{equation}
 Each term in Eq.~(\ref{eq:sample}), including those involving covariant derivatives $D_\mu$, is implicitly assumed to
 respect the gauge representations of the associated $\phi_i$
fields, as defined above.  The Yukawa-type couplings $y_i$,  as
well as the $\lambda_i$ scalar vertices are dimensionless, and assumed to be of order one,
while we assume all scalar masses $M_i$ to be of the same order of magnitude.
In this case, $\Lambda\sim M_i$.
In the $\lambda_{234}$ term, an overall mass scale $M$ has been
``factored out'' and is
assumed to be of the same order as the $M_i$. Note that we neglect generation
indicies, which are implied. In the case $\lambda_{234}=0$, lepton number is a classical
global symmetry of Eq.~(\ref{eq:sample}), and one can view this three-scalar coupling
as the source of lepton number
violation. One may even envision a scenario where lepton number is spontaneously broken by the
vacuum expectation value of some SM singlet $\phi_5$ scalar field, $\langle\phi_5\rangle=M$.

Provided all $M_i$ are around 0.5~TeV, as required if this Lagrangian
is to ``explain'' the observed light neutrino masses, LNV is certainly
\emph{not} the only (or even the main) consequence of this model. The $y_1$ and $y_3$ terms
for example, will mediate $\mu\to e$-conversion in nuclei
 at very dangerous levels if their flavor structure is generic. $\phi_2$ can be resonantly
 produced in $dd$-collisions, while $\phi_1$ and $\phi_3$ qualify as scalar lepto-quarks, which
 are constrained by high energy collider experiments, including those at HERA \cite{HERA}, to weigh more than a few hundred GeV \cite{PDG}. For more details, we refer readers to, for example,
 the Particle Data Book \cite{PDG} and references therein.

 We will conclude this discussion by adding that several other effective operator can be realized
 in a very similar way. ${\cal O}_{19}$, for example, if it manifests itself with the Lorentz structure
 $(L^iQ^j)(d^cd^c)(\bar{e^c}\bar{u^c})\epsilon_{ij}$, can be realized by a Lagrangian very similar
 to Eq.~(\ref{eq:sample}) where the $d^c$ field in the $y_3$-coupling interaction is replaced by a $u^c$ field, and
 the $\phi_1$ field is replaced by an $SU(2)_L$ singlet (it is a triplet in Eq.~(\ref{eq:sample})). Of course, hypercharge assignments for the $\phi_i$ also need to be modified in a straight forward way. The
 associated non LNV phenomenology is similar, except for the fact that $\Lambda_{\nu}$ for  ${\cal O}_{19}$ (around 1~TeV) is larger than the one for ${\cal O}_{56}$ and hence ${\cal O}_{19}$ is less constrained by current experimental data. On the other hand, ${\cal O}_{19}$  predicts potentially much larger rates for LNV observables at colliders (see Fig,~\ref{fig:ILC}).

Our definition of ``interesting'' is  arbitrary and
motivated only by the fact that the physics of the ``interesting''
operators highlighted earlier in this section will probably be
explored at next-generation collider and high-precision experiments.  One may argue that
many operators
which lead to the observable neutrino masses for high values of
$\Lambda_{\nu}$ are interesting on their own right, either due to the
theoretically pleasing properties of their associated potential
ultraviolet completions, or by some observational peculiarity. There
are many examples of the first type ranging from the different manifestations of the seesaw
mechanism \cite{SeeSaw,Bajc:2006ia,SeeSaw2,SeeSaw3} to the Zee model
\cite{ZeeModel} and the minimal supersymmetric SM with R-parity
violation \cite{R-parity}. Dedicated analysis of these cases have
been widely pursued in the literature and will not be discussed
here. We would also like to point out that some effective operators,
like  $\mathcal{O}_{7}$ and $\mathcal{O}_{8}$, are, according to our
criteria, very ``uninteresting.'' Both $\mathcal{O}_{7}$ and
$\mathcal{O}_{8}$  predict unobservably suppressed $\beta\beta0\nu$
rates (both predict small $m_{ee}$) and equally hopeless collider
prospects given that they are associated to very high cutoff scales,
$\Lambda_{\nu}\approx4\times 10^2~\rm{TeV}$ and $\Lambda_{\nu}\approx6\times
10^3~\rm{TeV}$, respectively.  If either of these operators are
responsible for the observed tiny neutrino masses, it is quite
possible that we may never directly detect LNV.  It is curious to
consider possible means of indirect detection or other observable
consequences of the different ultraviolet completions of such
scenarios.\footnote{This is very similar to the case of ${\cal
O}_1$. The main redeeming feature of ${\cal O}_1$, other than its
simplicity, is the fact that many of its ultraviolet completions
allow one to explain the matter--antimatter asymmetry of the
universe \cite{Leptogenesis}.} It would also be interesting to ask
whether either of these elusive models has any underlying
theoretical motivation or whether they allow one to solve other
outstanding problems in particle physics.

\setcounter{footnote}{0} \setcounter{equation}{0}
\section{Discussion and Concluding remarks}
\label{sec:conclusion}

If neutrino masses are a consequence of lepton-number violating
physics at a very high energy scale (higher than the scale of
electroweak symmetry breaking), new physics effects -- including the
generation of neutrino Majorana masses -- at low enough energies can
be parameterized in terms of irrelevant operators whose coefficients
are suppressed by inverse powers of an effective cutoff scale
$\Lambda$. As discussed before, $\Lambda$ is, roughly, the energy
scale above which new degrees of freedom must be
observed if the new ultraviolet physics is perturbative (if the new
physics is very weakly coupled, the masses of the new degrees of
freedom can be much smaller than $\Lambda$). We have explored a very
large class of such scenarios through 129 irrelevant operators of
energy dimension less than or equal to eleven that violate lepton number by 2 units.
These are tabulated in first two columns of Table~\ref{tab:AllOps},
along with a summary of our results.

Analyzing each effective operator individually, we estimated the
predicted general form of the Majorana neutrino mass matrix. Our
results are listed in the third column of Table~\ref{tab:AllOps}. By
comparing each such estimate with our current understanding of
neutrino masses, we extracted the cutoff scale $\Lambda_{\nu}$ of
each effective operator, assuming that it provides the dominant
contribution to the observed neutrino masses. These results are
listed in the fourth column of Table~\ref{tab:AllOps} assuming light
neutrino masses equal to 0.05~eV (the square root of the atmospheric
mass-squared difference), and are summarized as follows. Depending
on the field content and dimension of the irrelevant operator, the
``lepton number breaking scale'' $\Lambda_{\nu}$ is predicted to be
anywhere from the weak scale ($\sim 0.1$~TeV) all the way up to
$10^{12}$~TeV (see Fig.~\ref{fig:OpSum}). This means that, depending
on how lepton number is violated and communicated to the SM, the
mass of the associated new degrees of freedom is predicted to be
anywhere between 100~GeV and $10^{12}$~TeV, \emph{even if all new
physics couplings are order one}. We note that in the case of all
variations of the seesaw mechanism (${\cal O}_1$),  neutrino physics
constrains $\Lambda_{\nu}=10^{12}$~TeV such that the new degrees of
freedom are either unobservably heavy, extremely weakly coupled, or
their couplings to the SM degrees of freedom are  finely-tuned. It
is fair to say that this behavior is not characteristic of all LNV
ultraviolet physics. One sample ultraviolet theory that leads to
dimension-eleven LNV effective operators was discussed in
Sec.~\ref{sec:interesting}. Other examples can be found in
\cite{Operators}, and include supersymmetry with trilinear R-parity
violation and the Zee model.

Assuming that a particular operator is responsible for nonzero
neutrino masses, it is straight forward to ask whether it leads to
other observable consequences. Here, we concentrated on several LNV
observables, and included future LNV searches at the LHC and future
lepton machines (like the ILC), along with their ability to directly
produce (and hopefully observe) new physics states lighter than
several hundred GeV. In column five of Table~\ref{tab:AllOps}, we
list the most favorable modes of experimental observation for each
operator.  The different relevant probes are:  neutrinoless double-beta decay ($\beta\beta0\nu$), neutrino oscillation and mixing
(mix), direct searches for new particles at the LHC (LHC) and ILC
(ILC), and virtual LNV effects at collider facilities (HElnv).  We
find it  unlikely that other probes of LNV, including rare meson
decays, should yield a positive signal in the forseeable future.
This conclusion is strongly based on the fact that, for all of our
analysis, we assume that \emph{all} new physics degrees of freedom
are heavier than the weak scale. While the vast majority of
operators is most sensitive to searches for neutrinoless double-beta
decay, that is not true of all operators. Some lead to relatively
suppressed rates for $\beta\beta0\nu$ (mostly because they lead to
mass matrices with a very small $m_{ee}$) even if they are
associated to $\Lambda_{\nu}<1$~TeV, indicating that, for these scenarios,
we are more likely to observe the physics behind neutrino masses
directly at colliders than to see a finite lifetime for
$\beta\beta0\nu$. Other scenarios naively lead to $\beta\beta0\nu$
rates orders of magnitude higher than what is currently allowed by
data. If these are responsible for the generation of neutrino
masses, the new physics is constrained to be somewhat decoupled from
first generation quarks (for example). In this case, there is hope
that LNV phenomena at colliders, which are not restricted to first
generation quarks, occur with non-negligible rates.

The sixth column of Table~\ref{tab:AllOps} lists the current
``status'' of the operator as either experimentally unconstrained
(U), constrained (C), or disfavored (D).  Such labels are assigned
based only on the experimental probes reviewed in this work.  By
arbitrary convention, an `unconstrained' operator can safely
accommodate all existing data even if one assumes \emph{all} its
flavor-dependent dimensionless coefficients to be of order one. A
`constrained' operator can accommodate all existing data after one
allows some of the different flavor-dependent dimensionless
coefficients to be suppressed with respect to the dominant ones by a
factor of 100 or so (as described above).  `Disfavored' operators
can only accommodate all data  only if ``tuned'' much more severely
than the `constrained' ones, and are usually in trouble with more
than one ``type'' of constraint.  A glance at column six reveals
that 11 out of the 129 operators are disfavored by current
data.  The most stringent constraints come from  $\beta\beta0\nu$,
while all `disfavored' operators are associated to cutoffs at or
below $1~\rm{TeV}$. Three of the `disfavored' operators,
$\mathcal{O}_{36}$, $\mathcal{O}_{37}$, and $\mathcal{O}_{38}$, are
also in disagreement with the  neutrino oscillation data (see
Sec.~\ref{sec:Oscillations}).

Our results illustrate that, as far as ``explaining'' neutrino masses, the model-building scene is wide open even if one postulates that neutrino masses arise as a consequence of lepton number violating, ``heavy'' physics. Significant progress will only be achieved once more experimental information becomes available. The observation that neutrinoless double-beta decay occurs with a nonzero rate will help point us in the right direction, but will certainly not reveal much about the mechanism behind neutrino masses. A more complete picture can only arise from combined information from several observables, including other LNV observables and the search for new physics at the electroweak scale. Other important experimental searches, not discussed here,  include all lepton-number conserving ``leptonic'' probes, such as precision measurements of the anomalous magnetic moment of the muon, searches for leptonic electric dipole moments, searches for charged-lepton flavor violation, and precision measurements of neutrino--nucleon and neutrino--lepton scattering.

\section*{Acknowledgments}
JJ would like to thank Michael Schmitt for useful insight into the
experimental feasibility of some of these analyses, as well as the members of
the Argonne National Laboratory theory group for help in identifying
limitations to our assumed operator set.  AdG is indebted to Kai Zuber for discussions on the
relevance of probes of lepton number violation beside $\beta\beta0\nu$.
This work is sponsored in part by the
US Department of Energy Contract DE-FG02-91ER40684.

 \end{document}